\documentclass[12pt]{article}

\usepackage{axodraw}

\usepackage{array}

\def\square{\kern1pt\vbox{\hrule height 1.2pt\hbox{\vrule width 1.2pt\hskip 3pt
   \vbox{\vskip 6pt}\hskip 3pt\vrule width 0.6pt}\hrule height 0.6pt}\kern1pt}

\def \g{\overline{g}}

\begin{document}

\begin{titlepage}

\begin{flushright}
UFIFT-QG-10-08
\end{flushright}

\vspace{1.5cm}

\begin{center}
{\bf Scalar Contribution to the Graviton Self-Energy during Inflation}
\end{center}

\vspace{.5cm}

\begin{center}
Sohyun Park$^{\dagger}$ and R. P. Woodard$^{\ddagger}$
\end{center}

\vspace{.5cm}

\begin{center}
\it{Department of Physics \\
University of Florida \\
Gainesville, FL 32611}
\end{center}

\vspace{1cm}

\begin{center}
ABSTRACT
\end{center}
We use dimensional regularization to evaluate the one loop contribution
to the graviton self-energy from a massless, minimally coupled scalar
on a locally de Sitter background. For noncoincident points our result
agrees with the stress tensor correlators obtained recently by
Perez-Nadal, Roura and Verdaguer. We absorb the ultraviolet divergences
using the $R^2$ and $C^2$ counterterms first derived by 't Hooft and
Veltman, and we take the $D=4$ limit of the finite remainder. The
renormalized result is expressed as the sum of two transverse, 4th order
differential operators acting on nonlocal, de Sitter invariant structure
functions. In this form it can be used to quantum-correct the linearized
Einstein equations so that one can study how the inflationary production
of infrared scalars affects the propagation of dynamical gravitons and
the force of gravity.

\vspace{.5cm}

\begin{flushleft}
PACS numbers:  04.62.+v, 98.80.Cq, 04.60.-m
\end{flushleft}

\vspace{1.5cm}
\begin{flushleft}
$^{\dagger}$ e-mail: spark@phys.ufl.edu \\
$^{\ddagger}$ e-mail: woodard@phys.ufl.edu
\end{flushleft}
\end{titlepage}

\section{Introduction}

The linearized equations for all known force fields do two things:
\begin{itemize}
\item{They give the linearized force fields induced by sources; and}
\item{They describe the propagation of dynamical particles which carry
the force but are, in principle, independent of any source.}
\end{itemize}
This is the classic distinction between the constrained and unconstrained
parts of a force field. In electromagnetism it amounts to the Coulomb
potential versus photons. In gravity there is the Newtonian potential,
plus its three relativistic partners, versus gravitons.

Quantum corrections to the linearized field equations derive from how
the 0-point fluctuations of various fields in whatever background is
assumed, respond to the linearized force fields. These quantum corrections
do not change the dichotomy between constrained and unconstrained fields
but they can, of course, modify classical results. Around flat space
background there is no effect, after renormalization, on the propagation
of dynamical photons or gravitons but there are small corrections to the
Coulomb and Newtonian potentials. As might be expected, the long distance
effects are greatest for the 0-point fluctuations of massless particles
and they take the form required by perturbation theory and dimensional
analysis \cite{EMGR,us},
\begin{equation}
\Bigl( \frac{\Delta \Phi}{\Phi}\Bigr)_{\rm Coul.} \sim -\frac{e^2}{\hbar c}
\, \ln\Bigl(\frac{r}{r_0}\Bigr) \qquad , \qquad \Bigl(
\frac{\Delta \Phi}{\Phi}\Bigr)_{\rm Newt.} \sim -\frac{\hbar G}{c^3 r^2} \; ,
\end{equation}
where $r$ is the distance to the source, $r_0$ is the point at which
the renormalized charge is defined, and the other constants have their
usual meanings.

Schr\"odinger was the first to suggest that the expansion of spacetime
can lead to particle production by ripping the virtual particles
(which are implicit in 0-point fluctuations) out of the vacuum \cite{Schr}.
Following early work by Imamura \cite{TI}, the first quantitative
results were obtained by Parker \cite{Parker1}. He found that the effect
is maximized during accelerated expansion, and for massless particles
which are not conformally invariant \cite{Parker2}, such as massless,
minimally coupled (MMC) scalars and (as noted by Grishchuk \cite{Grishchuk})
gravitons.

The de Sitter geometry is the most highly accelerated expansion
consistent with classical stability. For de Sitter background with
Hubble constant $H$ and scale factor $a(t) = e^{H t}$ it is simple
to show that the number of MMC scalars, or either polarization of
graviton, created with wave vector $\vec{k}$ is \cite{PW1},
\begin{equation}
N(t,\vec{k}) = \Bigl( \frac{H a(t)}{2 c \Vert \vec{k} \Vert} \Bigr)^2 \; .
\end{equation}
It is these particles which comprise the scalar and tensor
perturbations produced by inflation \cite{SMC}, the scalar
contribution of which has been imaged \cite{WMAP}. Of course the
same particles also enter loop diagrams to cause an enormous
strengthening of the quantum effects caused by MMC scalars and gravitons.
A number of analytic results have been obtained for one loop corrections
to the way various particles propagate on de Sitter background and also
to how long range forces act:
\begin{itemize}
\item{In MMC scalar quantum electrodynamics, infrared photons
behave as if they had an increasing mass \cite{vacpol}, and the
charge screening very quickly becomes nonperturbatively strong
\cite{HDW}, but there is no big effect on scalars \cite{KW1};}
\item{For a MMC scalar which is Yukawa-coupled to a massless fermion,
infrared fermions behave as if they had an increasing mass \cite{fermself}
but the associated scalars experience no large correction \cite{LDW};}
\item{For a MMC scalar with a quartic self-interaction, infrared
scalars behave as if they had an increasing mass (which persists to
two loop order) \cite{scalself};}
\item{For quantum gravity minimally coupled to a massless fermion,
the fermion field strength grows without bound \cite{Miao}; and}
\item{For quantum gravity plus a MMC scalar, the scalar shows no
secular effect but its field strength may acquire a momentum-dependent
enhancement \cite{Kahya}.}
\end{itemize}
The great omission from this list is how inflationary scalars and
gravitons affect gravity, both as regards the propagation of dynamical
gravitons and as regards the force of gravity. This paper represents a
first step in completing the list.

One includes quantum corrections to the linearized field equation by
subtracting the integral of the appropriate one-particle-irreducible
(1PI) 2-point function up against the linearized field. For example,
a MMC scalar $\varphi(x)$ in a background metric $g_{\mu\nu}(x)$ whose
1PI 2-point function is $-iM^2(x;x')$, would have the linearized
effective field equation,
\begin{equation}
\partial_{\mu} \Bigl[ \sqrt{g} g^{\mu\nu} \partial_{\nu} \varphi(x)\Bigr]
- \int \!\! d^4x' \, M^2(x;x') \varphi(x') = 0 \; . \label{effeqn}
\end{equation}
To include gravity on the list we must therefore compute the
graviton self-energy, either from MMC scalars or from gravitons,
and then use it to correct the linearized Einstein equation. In
this paper we shall evaluate the contribution from MMC scalars;
a subsequent paper will solve the linearized effective field
equations to determine quantum corrections to the propagation of
gravitons and the gravitational response to a point mass.

It should be noted that the vastly more complicated contribution from
gravitons was derived some time ago \cite{TW1}. However, that result
is not renormalized, and is therefore only valid for noncoincident points.
To use the graviton self-energy in an effective field equation such
as (\ref{effeqn}), where the integration carries $x^{\prime \mu}$
over $x^{\mu}$, one must extract differential operators until the
remaining structure functions are integrable. That is the sort of
form we will derive, using dimensional regularization to control the
divergences and BPHZ counterterms to subtract them.

This paper contains five sections. In section 2 we give those of the
Feynman rules which are needed for this computation, and we describe
the geometry of our $D$-dimensional, locally de Sitter background.
Section 3 derives the relatively simple form for the $D$-dimensional
graviton self-energy with noncoincident points. We show that this
version of the result agrees with the flat space limit \cite{FW} and
with the de Sitter stress tensor correlators recently derived by
Perez-Nadal, Roura and Verdaguer \cite{PNRV}. Section 4 undertakes
the vastly more difficult reorganization which must be done to
isolate the local divergences for renormalization. At the end we
subtract off the divergences with the same counterterms originally
computed for this model in 1974 by 't Hooft and Veltman \cite{HV},
and we take the unregulated limit of $D = 4$. Our discussion
comprises section 5.


\section{Feynman Rules}

In this section we derive Feynman rules for the computation. We
start by expressing the full metric as
\begin{equation}
g_{\mu\nu} = \overline{g}_{\mu\nu} + \kappa h_{\mu\nu} \; ,
\label{full_metric}
\end{equation}
where $\overline{g}_{\mu\nu}$ is the background metric,
$h_{\mu\nu}$ is the graviton field whose indices are
raised and lowered with the background metric, and
$\kappa^2 \equiv 16 \pi G$ is the loop counting parameter of quantum gravity.
Expanding the MMC scalar Lagrangian around the background metric
we get interaction vertices between the scalar and dynamical
gravitons. We take the $D$-dimensional locally de Sitter space
as our background and introduce de Sitter invariant bi-tensors
which will be used throughout the calculation. We close this
section by providing the MMC scalar propagator on the de Sitter background.

\subsection{Interaction Vertices}

The Lagrangian which describes pure gravity and the interaction
between gravitons and the MMC scalar is,
\begin{equation}
     \mathcal{L} =
     \frac{1}{16\pi G}\Bigl[ R - (D\!-\!1)(D\!-\!2)H^2 \Bigr]
     -\frac12 \partial_{\mu} \varphi \partial_{\nu}\varphi g^{\mu\nu} \sqrt{-g} \;.
\label{Lagrangian}
\end{equation}
where $R$ is Ricci scalar, $G$ is Newton's constant and $H$ is the Hubble constant.

Computing the one loop scalar contributions to the graviton self-energy
consists of summing the 3 Feynman diagrams depicted in Figure 1.

\begin{center}
\begin{picture}(360,100)(0,0)
\Photon(15,50)(40,50){2}{5} \Photon(80,50)(105,50){2}{5}
\Vertex(40,50){3} \Vertex(80,50){3} \Text(36,40)[b]{$x$}
\Text(84,40)[b]{$x'$} \CArc(60,50)(20,0,360) \Text(120,50)[]{$+$}
\Photon(135,50)(225,50){2}{15} \Vertex(180,50){3}
\CArc(180,68)(17.6,-90,270) \Text(180,40)[b]{$x$}
\Text(240,50)[]{$+$} \Photon(255,50)(345,50){2}{15}
\Vertex(300,50){3} \Text(301,50)[]{\LARGE $\times$}
\Text(300,40)[b]{$x$}
\end{picture}
\end{center}
\vspace{-1cm}
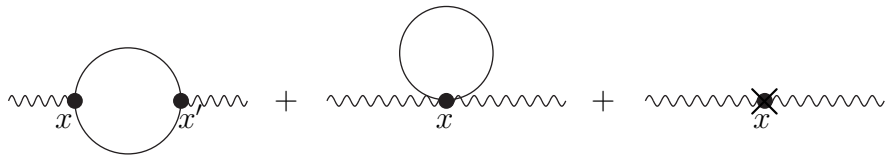
{\noindent \rm \small Figure 1: The one loop graviton
self-energy from MMC scalars.}
\vspace{0.5cm}

\noindent The sum of these three diagrams has the following analytic form:
\begin{eqnarray}\label{anal_graviton_self_energy}
     \lefteqn{-i[^{\mu\nu}\Sigma^{\rho\sigma}](x;x') } \nonumber \\
& & = \frac{1}{2}\sum_{I=1}^{2}T_{I}^{\mu\nu\alpha\beta}(x) \sum_{J=1}^{2}
T_{J}^{\rho\sigma\gamma\delta}(x') \times \partial_{\alpha}
\partial'_{\gamma}i\triangle(x;x') \times
\partial_{\beta}\partial'_{\delta}i\triangle(x;x') \nonumber \\
&& \hspace{2cm} + \frac{1}{2}\sum_{I=1}^{4}F_{I}^{\mu\nu\rho\sigma
\alpha\beta}(x) \times \partial_{\alpha}\partial'_{\beta}i
\triangle(x;x')\times \delta^{D}(x-x') \nonumber \\
&& \hspace{4cm} + 2 \sum_{I=1}^2 C_{I}^{\mu\nu\rho\sigma}(x)
\times \delta^{D}(x-x') \; . \qquad
\end{eqnarray}
The 3-point and 4-point vertex factors $T_I^{\mu\nu\alpha\beta}$ and
$F_I^{\mu\nu\rho\sigma\alpha\beta}$ derive from expanding the MMC scalar
Lagrangian using (\ref{full_metric}),
\begin{eqnarray}
\lefteqn{
-\frac12 \partial_{\mu} \varphi \partial_{\nu} \varphi g^{\mu\nu} \sqrt{-g} } \\
& & \hspace{-0.5cm}= -\frac12 \partial_{\mu} \varphi \partial_{\nu} \varphi
\overline{g}^{\mu\nu} \sqrt{-\overline{g}} -\frac{\kappa}2
\partial_{\mu} \varphi \partial_{\nu} \varphi \Bigl(\frac12 h
\overline{g}^{\mu\nu} - h^{\mu\nu}\Bigr) \sqrt{-\overline{g}} \nonumber \\
& & \hspace{0cm} -\frac{\kappa^2}2 \partial_{\mu} \varphi \partial_{\nu}
\varphi \Biggl\{ \Bigl[\frac18 h^2 \!-\! \frac14 h^{\rho\sigma}
h_{\rho\sigma}\Bigr] \overline{g}^{\mu\nu} \!-\! \frac12 h h^{\mu\nu}
\!+\! h^{\mu}_{~\rho} h^{\rho \nu} \Biggr\} \sqrt{-\overline{g}}
+ O(\kappa^3) \; .
\end{eqnarray}
The resulting 3-point and 4-point vertex factors are given in the Tables 1
and 2, respectively. The procedure to get the counterterm vertex operators
$C_I^{\mu\nu\rho\sigma}(x)$ is given in section 4.

\begin{table}[h]

\vbox{\tabskip=0pt \offinterlineskip
\def\tablerule{\noalign{\hrule}}
\halign to390pt {\strut#& \vrule#\tabskip=1em plus2em& \hfil#\hfil&
\vrule#& \hfil#\hfil&  \hfil#\hfil& & \vrule#\tabskip=0pt\cr
\tablerule \omit&height4pt&\omit&&\omit&&\cr
\omit&height2pt&\omit&&\omit&&\cr &&\omit\hidewidth $ I $ \hidewidth
&& $T_{I}^{\mu\nu\alpha\beta} $ &&\cr
\omit&height4pt&\omit&&\omit&&\cr \tablerule
\omit&height2pt&\omit&&\omit&&\cr && 1 &&
$-\frac{i\kappa}{2}\sqrt{-\overline{g}}\;\overline{g}^{\mu\nu}\overline{g}^{\alpha\beta}
$ && \cr \omit&height2pt&\omit&&\omit&&\cr \tablerule
\omit&height2pt&\omit&&\omit&&\cr && 2 &&
$+i\kappa\sqrt{-\overline{g}}\;\overline{g}^{\mu(\alpha}\overline{g}^{\beta)\nu}
$ && \cr \omit&height2pt&\omit&&\omit&&\cr \tablerule}}

\caption{3-point vertices $T_{I}^{\mu\nu\alpha\beta}$ where
$\overline{g}_{\mu\nu}$ is the de Sitter background metric,
$\kappa^2 \equiv 16\pi G$ and parenthesized indices are
symmetrized.} \label{3vertices}
\end{table}

\begin{table}[h]

\vbox{\tabskip=0pt \offinterlineskip
\def\tablerule{\noalign{\hrule}}
\halign to390pt {\strut#& \vrule#\tabskip=1em plus2em& \hfil#\hfil&
\vrule#& \hfil#\hfil&  \hfil#\hfil& & \vrule#\tabskip=0pt\cr
\tablerule \omit&height4pt&\omit&&\omit&&\cr
\omit&height2pt&\omit&&\omit&&\cr &&\omit\hidewidth $ I $ \hidewidth
&& $F_{I}^{\mu\nu\rho\sigma\alpha\beta} $  &&\cr
\omit&height4pt&\omit&&\omit&&\cr \tablerule
\omit&height2pt&\omit&&\omit&&\cr && 1 &&
$-\frac{i\kappa^2}{4}\sqrt{-\overline{g}}\;\overline{g}^{\mu\nu}
\overline{g}^{\rho\sigma}\overline{g}^{\alpha\beta} $  &&\cr
\omit&height2pt&\omit&&\omit&&\cr \tablerule
\omit&height2pt&\omit&&\omit&&\cr && 2
&&$+\frac{i\kappa^2}{2}\sqrt{-\overline{g}}\;
\overline{g}^{\mu(\rho}\overline{g}^{\sigma)\nu}\overline{g}^{\alpha\beta}$
&&\cr \omit&height2pt&\omit&&\omit&&\cr \tablerule
\omit&height2pt&\omit&&\omit&&\cr && 3 &&
$+\frac{i\kappa^2}{2}\sqrt{-\overline{g}}\;
\bigg[\overline{g}^{\mu(\alpha}\overline{g}^{\beta)\nu}\overline{g}^{\rho\sigma}
+ \overline{g}^{\mu\nu}\overline{g}^{\rho(\alpha}
\overline{g}^{\beta)\sigma} \bigg] $  &&\cr
\omit&height2pt&\omit&&\omit&&\cr \tablerule
\omit&height2pt&\omit&&\omit&&\cr && 4 &&
$-2i\kappa^2\sqrt{-\overline{g}}\;
\overline{g}^{\alpha(\mu}\overline{g}^{\nu)(\rho}\overline{g}^{\sigma)\beta}$
&& \cr \omit&height2pt&\omit&&\omit&&\cr \tablerule}}

\caption{4-point vertices $F_{I}^{\mu\nu\rho\sigma\alpha\beta}$
where $\overline{g}_{\mu\nu}$ is the de Sitter background metric,
$\kappa^2 \equiv 16\pi G$ and parenthesized indices are
symmetrized.}

\label{4vertices}

\end{table}

These interaction vertices are valid for any background metric
$\overline{g}_{\mu\nu}$. In the next two subsections we specialize
to a locally de Sitter background and give the scalar propagator
$i\triangle(x;x')$ on it.

\subsection{Working on de Sitter Space}

We specify our background geometry as the open conformal coordinate submanifold of $D$-dimensional de Sitter space.
A spacetime point $x^{\mu} = (\eta, x^i)$ takes values in the ranges
\begin{equation}
-\infty < \eta < 0 \qquad {\rm and} \qquad  -\infty < x^i < +\infty \; .
\end{equation}
In these coordinates the invariant element is,
\begin{equation}
ds^2 \equiv \g_{\mu\nu} dx^{\mu} dx^{\nu} = a^2 \eta_{\mu\nu}
dx^{\mu} dx^{\nu}\; ,
\end{equation}
where $\eta_{\mu\nu}$ is the Lorentz metric and $a = -1/H\eta$ is
the scale factor. The Hubble parameter $H$ is constant for the de Sitter space.
So in terms of $\eta_{\mu\nu}$ and $a$ our background metric is
\begin{equation}
  \overline{g}_{\mu\nu} \equiv a^2 \eta_{\mu\nu}\;.
\end{equation}

De Sitter space has the maximum number of space-time symmetries in a
given dimension. For our $D$-dimensional conformal coordinates the
$\frac{1}{2}D(D+1)$ de Sitter transformations can be decomposed as
follows:
\begin{itemize}
 \item{Spatial transformations - $(D-1)$ transformations.}
\begin{equation}
 \eta' = \eta\;, ~~ x'^{i} = x^{i} + \epsilon^{i} \;. \label{sym1}
\end{equation}
 \item{Rotations - $\frac{1}{2}(D-1)(D-2)$ transformations.}
\begin{equation}
 \eta' = \eta\;, ~~ x'^{i} = R^{ij}x^{j}\;. \label{sym2}
\end{equation}
 \item{Dilation - 1 transformation.}
\begin{equation}
 \eta' = k\eta\;, ~~ x'^{i} = kx^{j}\;. \label{sym3}
\end{equation}
 \item{Spatial special conformal transformations - $(D-1)$ transformations.}
\begin{equation}
 \eta' = \frac{\eta}{1 - 2\vec{\theta}\cdot\vec{x}
+ \parallel\! \vec{\theta} \!\parallel^2 x\cdot x}\;, ~~ x' =
\frac{x^{i} - \theta^{i}x\cdot x}{1 - 2\vec{\theta}\cdot\vec{x} +
\parallel\! \vec{\theta} \!\parallel^2 x\cdot x}\;. \label{sym4}
\end{equation}
\end{itemize}

It turns out that the MMC scalar contribution to the graviton self-energy
is de Sitter invariant. This suggests to express it in terms of the de
Sitter length function $y(x;x')$,
\begin{equation}
y(x;x') \equiv a a' H^2 \Biggl[ \Bigl\Vert \vec{x} \!-\! \vec{x'}
\Bigr\Vert^2 - \Bigl(\vert \eta \!-\! \eta'\vert \!-\! i
\epsilon\Bigr)^2 \Biggr]\; . \label{ydef}
\end{equation}
Except for the factor of $i\epsilon$ (whose purpose is to enforce
Feynman boundary conditions) the function $y(x;x')$ is closely
related to the invariant length $\ell(x;x')$ from $x^{\mu}$ to $x'^{\mu}$,
\begin{equation}
y(x;x') = 4 \sin^2\Bigl( \frac12 H \ell(x;x')\Bigr) \; .
\end{equation}

With this de Sitter invariant quantity $y(x;x')$, we can form a
convenient basis of de Sitter invariant bi-tensors. Note that
because $y(x;x')$ is de Sitter invariant, so too are covariant
derivatives of it. With the metrics $\g_{\mu\nu}(x)$ and
$\g_{\mu\nu}(x')$, the first three derivatives of $y(x;x')$ furnish
a convenient basis of de Sitter invariant bi-tensors \cite{KW1},
\begin{eqnarray}
\frac{\partial y(x;x')}{\partial x^{\mu}} & = & H a \Bigl(y
\delta^0_{\mu}
\!+\! 2 a' H \Delta x_{\mu} \Bigr) \; , \label{dydx} \\
\frac{\partial y(x;x')}{\partial x'^{\nu}} & = & H a' \Bigl(y
\delta^0_{\nu}
\!-\! 2 a H \Delta x_{\nu} \Bigr) \; , \label{dydx'} \\
\frac{\partial^2 y(x;x')}{\partial x^{\mu} \partial x'^{\nu}} & = &
H^2 a a' \Bigl(y \delta^0_{\mu} \delta^0_{\nu} \!+\! 2 a' H \Delta
x_{\mu} \delta^0_{\nu} \!-\! 2 a \delta^0_{\mu} H \Delta x_{\nu}
\!-\! 2 \eta_{\mu\nu}\Bigr) \; . \qquad \label{dydxdx'}
\end{eqnarray}
Here and subsequently
$\Delta x_{\mu} \equiv \eta_{\mu\nu} (x \!-\!x')^{\nu}$.

Acting covariant derivatives generates more basis tensors,
for example \cite{KW1},
\begin{eqnarray}
\frac{D^2 y(x;x')}{Dx^{\mu} Dx^{\nu}}
& = & H^2 (2 \!-\!y) \g_{\mu\nu}(x) \; , \\
\frac{D^2 y(x;x')}{Dx'^{\mu} Dx'^{\nu}}
& = & H^2 (2 \!-\!y)\g_{\mu\nu}(x') \; .
\end{eqnarray}
The contraction of any pair of the basis tensors also produces
more basis tensors \cite{KW1},
\begin{eqnarray}
\g^{\mu\nu}(x) \frac{\partial y}{\partial x^{\mu}} \frac{\partial y}{\partial x^{\nu}} & = & H^2 \Bigl(4 y - y^2\Bigr) =
\g^{\mu\nu}(x') \frac{\partial y}{
\partial x'^{\mu}} \frac{\partial y}{\partial x'^{\nu}} \; ,
\label{contraction1}\\
\g^{\mu\nu}(x) \frac{\partial y}{\partial x^{\nu}} \frac{\partial^2 y}{
\partial x^{\mu} \partial x'^{\sigma}} & = & H^2 (2-y) \frac{\partial y}{
\partial x'^{\sigma}} \; ,
\label{contraction2}\\
\g^{\rho\sigma}(x') \frac{\partial y}{\partial x'^{\sigma}}
\frac{\partial^2 y}{\partial x^{\mu} \partial x'^{\rho}} & = & H^2
(2-y)
\frac{\partial y}{\partial x^{\mu}} \; ,
\label{contraction3}\\
\g^{\mu\nu}(x) \frac{\partial^2 y}{\partial x^{\mu} \partial
x'^{\rho}} \frac{\partial^2 y}{\partial x^{\nu} \partial
x'^{\sigma}} & = & 4 H^4 \g_{\rho\sigma}(x') - H^2 \frac{\partial
y}{\partial x'^{\rho}}
\frac{\partial y}{\partial x'^{\sigma}} \; ,
\label{contraction4}\\
\g^{\rho\sigma}(x') \frac{\partial^2 y}{\partial x^{\mu}\partial
x'^{\rho}} \frac{\partial^2 y}{\partial x^{\nu} \partial
x'^{\sigma}} & = & 4 H^4 \g_{\mu\nu}(x) - H^2 \frac{\partial
y}{\partial x^{\mu}} \frac{\partial y}{
\partial x^{\nu}} \; .
\label{contraction5}
\end{eqnarray}

Our basis tensors are naturally covariant, but their indices can of
course be raised using the metric at the appropriate point. To save
space in writing this out we define the basis tensors with raised
indices as differentiation with respect to ``covariant''
coordinates,
\begin{eqnarray}
\frac{\partial y}{\partial x_{\mu}} & \equiv &
\overline{g}^{\mu\nu}(x)
\frac{\partial y}{\partial x^{\nu}} \; , \\
\frac{\partial y}{\partial x'_{\rho}} & \equiv &
\overline{g}^{\rho\sigma}(x') \frac{\partial y}{\partial x^{\prime
\sigma}} \; , \\
\frac{\partial^2 y}{\partial x_{\mu} \partial x'_{\rho}} & \equiv &
\overline{g}^{\mu\nu}(x) \overline{g}^{\rho\sigma}(x')
\frac{\partial^2 y}{\partial x^{\nu} \partial x^{\prime \sigma}} \;
.
\end{eqnarray}

\subsection{Scalar Propagator on de Sitter}

>From the MMC scalar Lagrangian (\ref{Lagrangian}) we see that the
propagator obeys
\begin{equation}
\partial_\mu\Bigl[\sqrt{-\g}~\g^{\mu\nu}\partial_\nu \Bigr]
i\triangle(x;x') = \sqrt{-\g}\square i\triangle(x;x') = i\delta^D(x-x')
\end{equation}
Although this equation is de Sitter invariant, there is no de Sitter
invariant solution for the propagator \cite{AF}, hence some of the
symmetries (\ref{sym1}-\ref{sym4}) must be broken. We choose to
preserve the homogeneity and isotropy of cosmology --- relations
(\ref{sym1}-\ref{sym2}) --- which corresponds to what is known as
the ``E3'' vacuum \cite{BA}. It can be realized in terms of plane
wave mode sums by making the spatial manifold $T^{D-1}$, rather than
$R^{D-1}$, with coordinate radius $H^{-1}$ in each direction, and
then using the integral approximation with the lower limit cut off
at $k=H$ \cite{TW2}. The final result consists of a de Sitter
invariant function of $y(x;x')$ plus a de Sitter breaking part which
depends upon the scale factors at the two points \cite{OW},
\begin{equation}
i\triangle(x;x') = A\Bigl( y(x;x')\Bigr) + k \ln(a a') \; .
\label{DeltaA}
\end{equation}
Here the constant $k$ is given as,
\begin{equation}
k \equiv \frac{H^{D-2}}{(4\pi)^{\frac{D}2}} \, \frac{\Gamma(D \!-\! 1)}{
\Gamma(\frac{D}2)} \; ,
\end{equation}
and the function $A(y)$ has the expansion,
\begin{eqnarray}
\lefteqn{A(y) \equiv \frac{H^{D-2}}{(4\pi)^{\frac{D}2}} \Biggl\{
\frac{\Gamma(\frac{D}2)}{\frac{D}2 \!-\! 1}
\Bigl(\frac{4}{y}\Bigr)^{ \frac{D}2 -1} \!+\!
\frac{\Gamma(\frac{D}2 \!+\! 1)}{\frac{D}2 \!-\! 2}
\Bigl(\frac{4}{y} \Bigr)^{\frac{D}2-2} \!-\! \pi
\cot\Bigl(\frac{\pi D}2\Bigr)
\frac{\Gamma(D \!-\! 1)}{\Gamma(\frac{D}2)} } \nonumber \\
& & \hspace{.5cm} + \sum_{n=1}^{\infty} \Biggl[\frac1{n}
\frac{\Gamma(n \!+\! D \!-\! 1)}{\Gamma(n \!+\! \frac{D}2)}
\Bigl(\frac{y}4 \Bigr)^n \!\!\!\! - \frac1{n \!-\! \frac{D}2 \!+\!
2} \frac{\Gamma(n \!+\!  \frac{D}2 \!+\! 1)}{ \Gamma(n \!+\! 2)}
\Bigl(\frac{y}4 \Bigr)^{n - \frac{D}2 +2} \Biggr] \Biggr\} . \qquad
\label{A(y)}
\end{eqnarray}
The infinite series terms of $A(y)$ vanish for $D=4$, so they only need to be
retained when multiplying a potentially divergent quantity, and even then
one only needs to include a handful of them. This makes loop computations
manageable.

We note that the MMC scalar propagator (\ref{DeltaA}) has a de
Sitter breaking term, $k\ln(aa')$. However, the one loop scalar
contribution to the graviton self-energy only involves the terms
like $\partial_{\alpha} \partial'_{\beta} i\triangle(x;x')$, which
are de Sitter invariant,
\begin{equation}
\partial_{\alpha} \partial'_{\beta} i\Delta(x;x') =
\frac{\partial}{\partial x^{\alpha}} \Biggl\{ A'(y) \frac{\partial
y}{\partial x^{\prime \beta}} + H a' \delta^0_{\beta} \Biggr\}
= A''(y) \frac{\partial y}{\partial x^{\alpha}} \frac{\partial
y}{\partial x^{\prime \beta}} + A'(y) \frac{\partial^2 y}{\partial
x^{\alpha} \partial x^{\prime \beta}} \; . \label{double}
\end{equation}
Another useful relation follows from the propagator equation,
\begin{equation}
(4 y \!-\! y^2) A''(y) + D (2 \!-\! y) A"(y) = (D \!-\! 1) k \; .
\label{proprel}
\end{equation}

\section{One Loop Graviton Self-energy}

In this section we calculate the first two, primitive, diagrams of
Figure~1. It turns out that the contribution from the 4-point vertex
(the middle diagram) vanishes in $D=4$ dimensions. The contribution
from two 3-point vertices (the leftmost diagram) is nonzero. For
noncoincident points it gives a relatively simple form which agrees
with the flat space limit \cite{FW} and with the de Sitter stress
tensor correlator recently derived by Perez-Nadal, Roura and
Verdaguer \cite{PNRV}.

\subsection{Contribution from 4-Point Vertices}

The 4-point contribution from the middle diagram of Figure~1 takes
the form,
\begin{equation}
-i\Bigl[\mbox{}^{\mu\nu}\Sigma^{\rho\sigma}\Bigr]_{\mbox{\tiny
4pt}}\!\!\!\!\!(x;x') \equiv \frac12 \sum_{I=1}^{4}
F_{I}^{\mu\nu\rho\sigma \alpha\beta}(x) \times
\partial_{\alpha}\partial'_{\beta}i\triangle(x;x')\times
\delta^D(x \!-\!x') \; . \label{middia}
\end{equation}
Recall that the four 4-point vertices
$F_I^{\mu\nu\rho\sigma\alpha\beta}(x)$ are given in
Table~\ref{4vertices}. Owing to the delta function, we need the
coincidence limit of the doubly differentiated propagator
(\ref{double}). The coincidence limits of the various tensor factors
follow from setting $a' = a$, $\Delta x^{\mu} = 0$ and $y = 0$ in
relations (\ref{dydx}-\ref{dydxdx'}),
\begin{eqnarray}
\lim_{x' \rightarrow x} \frac{\partial y(x;x')}{\partial x^{\mu}} &
= & 0 = \lim_{x' \rightarrow x} \frac{\partial y(x;x')}{\partial
x'^{\nu}} \; , \\
\lim_{x' \rightarrow x} \frac{\partial^2 y(x;x')}{\partial x^{\mu}
\partial x'^{\nu}} & = & = -2 H^2 \overline{g}_{\mu\nu}\; .
\end{eqnarray}
Hence the coincidence limit of the doubly differentiated propagator
can be expressed in terms of $A'(y)$ evaluated at $y=0$,
\begin{equation}
\lim_{x' \rightarrow x}
\partial_{\alpha} \partial'_{\beta} i \triangle(x;x') = A''(0) \times 0
+ A'(0) \times \Bigl[-2H^2g_{\mu\nu}\Bigl] \; .
\end{equation}

>From the definition (\ref{A(y)}) of $A(y)$, we see that $A'(y)$ is,
\begin{eqnarray}
\lefteqn{A'(y) = \frac14 \frac{H^{D-2}}{(4\pi)^{\frac{D}2}} \Biggl\{
-\Gamma(\frac{D}2)\Bigl(\frac{4}{y}\Bigr)^{ \frac{D}2 } \!-\!
\Gamma(\frac{D}2 \!+\! 1)\Bigl(\frac{4}{y} \Bigr)^{\frac{D}2-1}   }
\nonumber \\
& & \hspace{2cm} \!+\! \sum_{n=1} \Biggl[ \frac{\Gamma(n \!+\! D
\!-\! 1)}{\Gamma(n \!+\! \frac{D}2)} \Bigl(\frac{y}4\Bigr)^{n-1} -
\frac{\Gamma(n \!+\! \frac{D}2 \!-\! 1)}{\Gamma(n \!+\! 2)}
\Bigl(\frac{y}4\Bigr)^{n-\frac{D}2 + 1} \Biggr]\Biggr\} .
\label{A'(y)}
\end{eqnarray}
Now we recall that, in dimensional regularization, any $D$-dependent
power of zero vanishes. Therefore, only the $n=1$ term of the
infinite series in (\ref{A'(y)}) contributes to the coincidence
limit,
\begin{equation}
A'(0) = \frac14 \frac{H^{D-2}}{(4\pi)^{\frac{D}{2}}}
\frac{\Gamma(D)}{\Gamma(\frac{D}{2}+1)} \;,
\end{equation}
and we have,
\begin{eqnarray}
\lim_{x' \rightarrow x} \partial_{\alpha} \partial'_{\beta} i
\triangle(x;x') = -\frac12 \frac{H^{D}}{(4\pi)^{\frac{D}{2}}}
\frac{\Gamma(D)}{\Gamma(\frac{D}{2}+1)} \, \g_{\alpha\beta} \;.
\label{coinc}
\end{eqnarray}

Substituting (\ref{coinc}), and the 4-point vertices from
Table~\ref{4vertices}, into expression (\ref{middia}) gives,
\begin{eqnarray}
\lefteqn{-i\Bigl[\mbox{}^{\mu\nu}\Sigma^{\rho\sigma}\Bigr]_{\mbox{\tiny
4pt}}\!\!\!\!\!(x;x') } \nonumber \\
& & = -\frac12 \frac{H^{D}}{(4\pi)^{\frac{D}{2}}}
\frac{\Gamma(D)}{\Gamma(\frac{D}2 \!+\! 1)}
\overline{g}_{\alpha\beta} \times i \kappa^2 \sqrt{-\overline{g}} \,
\bigg\{-\frac14 \overline{g}^{\mu\nu} \overline{g}^{\rho\sigma}
\overline{g}^{\alpha\beta} + \frac12 \overline{g}^{\mu(\rho}
\overline{g}^{\sigma)\nu} \overline{g}^{\alpha\beta} \qquad
\nonumber \\
& & \hspace{.5cm} + \frac12 \Bigl[\overline{g}^{\mu(\alpha}
\overline{g}^{\beta)\nu} \overline{g}^{\rho\sigma} +
\overline{g}^{\mu\nu} \overline{g}^{\rho(\alpha}
\overline{g}^{\beta)\sigma} \Bigr] -2 \overline{g}^{\alpha(\mu}
\overline{g}^{\nu)(\rho} \overline{g}^{\sigma)\beta} \bigg\}
\delta^D(x \!-\! x') \; , \qquad \\
& & = \Bigl(\frac{D\!-\!4}4\Bigr) \frac{i \kappa^2
H^{D}}{(4\pi)^{\frac{D}{2}}} \frac{\Gamma(D)}{\Gamma(\frac{D}2 \!+\!
1)} \sqrt{-\overline{g}} \, \bigg\{ \frac12 \overline{g}^{\mu\nu}
\overline{g}^{\rho\sigma} - \overline{g}^{\mu(\rho}
\overline{g}^{\sigma)\nu} \bigg\}\delta^D(x \!-\! x') \; . \qquad
\label{fin4con}
\end{eqnarray}
Because the Gamma functions are finite for $D = 4$ dimensions so we
can dispense with dimensional regularization and set $D=4$. At that
point the net contribution (\ref{fin4con}) vanishes.

\subsection{Contribution from 3-Point Vertices}

The contribution from the leftmost diagram of Figure~1 takes the
form,
\begin{eqnarray}
\label{3-point}
\lefteqn{-i\Bigl[\mbox{}^{\mu\nu}\Sigma^{\rho\sigma}\Bigr]_{\mbox{\tiny
3pt}}\!\!\!\!\!(x;x') } \nonumber \\
& & = \frac{1}{2}\sum_{I=1}^{2}T_{I}^{\mu\nu\alpha\beta}(x) \sum_{J=1}^{2}
T_{J}^{\rho\sigma\gamma\delta}(x') \times \partial_{\alpha}
\partial'_{\gamma}i\triangle(x;x') \times
\partial_{\beta}\partial'_{\delta}i\triangle(x;x')  \; . \qquad
\end{eqnarray}
Recall from section 2 that any de Sitter invariant bitensor can be
expressed as a linear combination of functions of $y(x;x')$ times
the five basis tensors,
\begin{eqnarray}
\lefteqn{-i\Bigl[\mbox{}^{\mu\nu}\Sigma^{\rho\sigma}\Bigr]_{\mbox{\tiny
3pt}}\!\!\!\!\!(x;x') = \sqrt{-\g}\sqrt{-\g'}\Biggl\{
\frac{\partial^2 y}{\partial x_{\mu}\partial x'_{(\rho}}
\frac{\partial^2 y}{\partial x'_{\sigma)}\partial x_{\nu}} \times
\alpha(y) }
\; \nonumber \\
& & +
\frac{\partial y}{\partial x_{(\mu}}\frac{\partial^2 y}{\partial
x_{\nu)}\partial x'_{(\rho}}\frac{\partial y}{\partial x'_{\sigma)}}
\times \beta(y) +
\frac{\partial y}{\partial x_{\mu}}\frac{\partial y}{\partial
x_{\nu}} \frac{\partial y}{\partial x'_{\rho}}\frac{\partial
y}{\partial x'_{\sigma}} \times \gamma(y) \nonumber \\
& & +
\g^{\mu\nu}\g'^{\rho\sigma}H^4 \times \delta(y) +
\Bigl[ \g^{\mu\nu}\frac{\partial y}{\partial
x'_{\rho}}\frac{\partial y}{\partial x'_{\sigma}} + \frac{\partial
y}{\partial x_{\mu}}\frac{\partial y}{\partial x_{\nu}}\g'^{\rho
\sigma}\Bigr]H^2 \times \epsilon(y) \; \Bigg\} \; . \qquad
\label{3-point-A}
\end{eqnarray}
By substituting our result (\ref{double}) for the mixed second
derivative of the scalar propagator, along with the vertices from
Table~\ref{3vertices}, and then making use of the contraction
identities (\ref{contraction1}-\ref{contraction5}), it is
straightforward to obtain expressions for the five coefficient
functions,
\begin{eqnarray}
\alpha(y) & = & -\frac12 \kappa^2 (A')^2 \; , \label{alp1} \\
\beta(y) & = & - \kappa^2 A' A'' \; , \label{bet1} \\
\gamma(y) & = & -\frac12 \kappa^2 (A'')^2 \; , \label{gam1} \\
\delta(y) & = & -\frac18 \kappa^2 \biggl\{ (A'')^2(4y-y^2)^2 +
2A'A''(2-y)(4y-y^2) \nonumber \\
& & \hspace{4cm} + (A')^2  \Bigl[4(D\!-\!4) \!-\! (4y\!-\!y^2) \Bigr]
\biggr\} \; , \label{del1} \\
\epsilon(y) & = & \frac14 \kappa^2 \Bigl[ (4y-y^2) (A'')^2 + 2 (2 \!-\!y)
A'A'' - (A')^2 \Bigr] \; . \label{eps1}
\end{eqnarray}

Expressions (\ref{alp1}-\ref{eps1}) for the coefficient functions
have the advantage of being exact for any dimension $D$, but the
disadvantages of being neither very explicit nor very simple
functions of $y(x;x')$. We can obtain expressions which are both
simple and explicit, and totally adequate for use in the $D=4$
effective field equations, by noting that each pair of terms in the
infinite series part of $A(y)$ (\ref{A(y)}) vanishes for $D=4$
spacetime dimensions. Therefore, it is only neceesary to retain
those parts of the infinite series which can potentially multiply
potential a divergence. For our computation that turns out to mean
only the $n=1$ terms, and we can write the two derivatives as,
\begin{eqnarray}
A' & = & \frac{\Gamma(\frac{D}2) H^{D-2}}{4 (4\pi)^{\frac{D}2}}
\Biggl\{ -\Bigl(\frac4{y}\Bigr)^{\frac{D}2} - \frac{D}2
\Bigl(\frac4{y}\Bigr)^{\frac{D}2 -1} - \frac12 \frac{D}2
\Bigl(\frac{D}2 \!+\! 1\Bigr) \Bigl(\frac4{y}\Bigr)^{\frac{D}2 -2}
\nonumber \\
& & \hspace{5cm} + \frac{\Gamma(D)}{\Gamma(\frac{D}2)
\Gamma(\frac{D}2 \!+\! 1)} + \Bigl({\rm Irrelevant}\Bigr) \Biggr\}
\; , \qquad \\
A'' & = & \frac{\Gamma(\frac{D}2) H^{D-2}}{16 (4\pi)^{\frac{D}2}}
\Biggl\{ \frac{D}2 \Bigl(\frac4{y}\Bigr)^{\frac{D}2+1} +
\Bigl(\frac{D}2 \!-\! 1\Bigr) \frac{D}2
\Bigl(\frac4{y}\Bigr)^{\frac{D}2} \nonumber \\
& & \hspace{2.8cm} + \frac12 \Bigl(\frac{D}2 \!-\! 2) \frac{D}2
\Bigl(\frac{D}2 \!+\! 1\Bigr) \Bigl(\frac4{y}\Bigr)^{\frac{D}2 -1} +
\Bigl({\rm Irrelevant}\Bigr) \Biggr\} \; . \qquad
\end{eqnarray}
Substituting these expansions in (\ref{alp1}-\ref{eps1}) gives,
\begin{eqnarray}
\alpha &\!\!\!\!\!=\!\!\!\!\!& \frac{K}{2^5} \Biggl\{
-\Bigl(\frac{4}{y}\Bigr)^{D} \!-\! D\Bigl(\frac{4}{y}\Bigr)^{D-1}
\!-\! \frac{D (D\!+\!1)}{2} \Bigl(\frac{4}{y}\Bigr)^{D-2} \nonumber
\\
& & \hspace{4cm} + \frac{2 \Gamma(D)}{\Gamma(\frac{D}2)
\Gamma(\frac{D}2 \!+\! 1)} \, \Bigl(\frac{4}{y}\Bigr)^{\frac{D}2}
+ \Bigl({\rm Irrelevant}\Bigr) \Biggl\} \; , \qquad \label{alp2} \\
\beta &\!\!\!\!\!=\!\!\!\!\!& \frac{K}{2^7} \Biggl\{ D
\Bigl(\frac{4}{y}\Bigr)^{D+1} \!+\! (D \!-\! 1) D
\Bigl(\frac{4}{y}\Bigr)^{D} \!+\! \frac12 (D\!-\!2) D (D\!+\!1)
\Bigl(\frac{4}{y}\Bigr)^{D\!-\!1} \nonumber \\
& & \hspace{4cm} - \frac{D \Gamma(D)}{\Gamma(\frac{D}2)
\Gamma(\frac{D}2 \!+\! 1)} \, \Bigl(\frac{4}{y}\Bigr)^{\frac{D}2 +1}
+ \Bigl({\rm Irrelevant}\Bigr) \Biggl\} \; , \qquad \label{bet2} \\
\gamma &\!\!\!\!\!=\!\!\!\!\!& \frac{K}{2^{11}} \Biggl\{ -D^2
\Bigl(\frac{4}{y}\Bigr)^{D+2} \!-\!(D\!-\!2) D^2 \Bigl(\frac{4}{y}\Bigr)^{D+1}
\nonumber \\
& & \hspace{4cm} - \frac12 (D^2 \!-\! 3D\! - \!2) D^2
\Bigl(\frac{4}{y}\Bigr)^{D} \!+\! \Bigl({\rm Irrelevant}\Bigr)
\Biggl\} \; , \qquad \label{gam2} \\
\delta &\!\!\!\!\!=\!\!\!\!\!& \frac{K}{2^5}\Biggl\{- (D^2 \!-\! D \!-\!4)
\Bigl(\frac{4}{y}\Bigr)^{D} \!\!\! -(D^3 \!-\! 5 D^2 \!+\! 4 D \!-\!4 )
\Bigl(\frac{4}{y}\Bigr)^{D-1} \!\!\! - \frac12 \Bigl(D^4 \!-\! 8 D^3
\nonumber \\
& & \hspace{.5cm} +\! 19 D^2 \!-\! 28 D \!+\! 8\Bigr) \Bigl(\frac{4}{y}
\Bigr)^{D-2} \!\!\!- \frac{8 \Gamma(D)}{\Gamma(\frac{D}2)
\Gamma(\frac{D}2 \!+\! 1)} \Bigl(\frac{4}{y}\Bigr)^{\frac{D}2} \!\!\!+
({\rm Irrelevant)} \Biggl\} \; , \qquad \label{del2} \\
\epsilon &\!\!\!\!\!=\!\!\!\!\!& \frac{K}{2^8} \Biggl\{ (D\!-\!2) D
\Bigl(\frac{4}{y}\Bigr)^{D+1} \!\!\!+ (D^3 \!-\! 5 D^2 \!+\! 6 D \!-\!4 )
\Bigl(\frac{4}{y}\Bigr)^{D} \!\!\! + \frac12 D \Bigl(D^3 \!-\! 7 D^2
\nonumber \\
& & \hspace{1.4cm} +\! 12 D \!-\!12\Bigr) \Bigl(\frac{4}{y}\Bigr)^{D-1}
\!\!\!+ \frac{D \Gamma(D)}{\Gamma(\frac{D}2) \Gamma(\frac{D}2 \!+\! 1)}
\Bigl(\frac{4}{y}\Bigr)^{\frac{D}2 + 1} \!\!\!+ ({\rm Irrelevant}) \Biggl\}
\; . \qquad \label{eps2}
\end{eqnarray}
where the constant $K$ is,
\begin{equation}
K \equiv \frac{\kappa^2 H^{2D-4} \Gamma^2(\frac{D}{2})}{(4\pi)^D} \; .
\label{Kdef}
\end{equation}

\subsection{Correspondence with Flat Space}

An important and illuminating correspondence limit comes from taking
the Hubble constant to zero, with the conformal time going to minus
infinity so as to keep the physical time $t$ fixed,
\begin{equation}
\eta = -\frac1{H} \, e^{-H t} = -\frac1{H} + t + O(H) \; .
\label{ftime}
\end{equation}
When this is done the background geometry degenerates to flat space
and we should recover well-known results \cite{EMGR}. We will also
see in the next section that the flat space limit provides crucial
guidance in how to reorganize the de Sitter result for
renormalization.

Although each independent conformal time diverges under
(\ref{ftime}), the conformal coordinate separation just goes to the
usual temporal separation of flat space,
\begin{equation}
\Delta x^0 \longrightarrow t - t' \; .
\end{equation}
All scale factors approach unity, and the de Sitter length function
goes to $H^2$ times the invariant interval of flat space,
\begin{equation}
y(x;x') \longrightarrow H^2 \Delta x^2 \; . \label{flaty}
\end{equation}
In the flat space limit the leading behaviors of the various basis
tensors are,
\begin{equation}
\frac{\partial y}{\partial x_{\mu}} \longrightarrow 2 H^2 \Delta
x^{\mu} \;\; , \;\; \frac{\partial y}{\partial x'_{\nu}}
\longrightarrow -2 H^2 \Delta x^{\nu} \;\; , \;\; \frac{\partial
y^2}{\partial x_{\mu} \partial x'_{\nu}} \longrightarrow -2 H^2
\eta^{\mu\nu} \; . \label{flattens}
\end{equation}
And the leading behaviors for derivatives of the function $A(y)$
are,
\begin{eqnarray}
H^2 A'(y) & \longrightarrow & -\frac1{4 \pi^{\frac{D}2}}
\frac{\Gamma(\frac{D}2)}{(\Delta x^2)^{\frac{D}2}} \equiv -\frac1{4
\pi^{\frac{D}2}} \frac{\Gamma(\frac{D}2)}{\Delta x^D}
\; , \label{fA'} \\
H^4 A''(y) & \longrightarrow & \frac1{4 \pi^{\frac{D}2}}
\frac{\Gamma(\frac{D}2 \!+\! 1)}{(\Delta x^2)^{\frac{D}2 +1}} \equiv
\frac1{4 \pi^{\frac{D}2}} \frac{\Gamma(\frac{D}2 \!+\! 1)}{\Delta
x^{D+2}} \label{fA''} \; .
\end{eqnarray}

The 4-point contribution (\ref{fin4con}) to the graviton self-energy
vanishes in the flat space limit, even for $D\neq 4$. We can take
the flat space limit of the 3-point contribution (\ref{3-point-A})
in two steps. First, substitute the leading behaviors (\ref{flaty})
for $y(x;x')$ and (\ref{flattens}) for the basis tensors. Then use
expressions (\ref{fA'}-\ref{fA''}) on the derivatives of $A(y)$. The
result is,
\begin{eqnarray}
\lefteqn{-i \Bigl[\mbox{}^{\mu\nu}
\Sigma^{\rho\sigma}\Bigr]_{\mbox{\tiny flat}}\!\!\!\!\!(x;x') =
\lim_{H \rightarrow 0} \kappa^2 \Biggl\{ 4 H^4 \eta^{\mu (\rho}
\eta^{\sigma) \nu} \times -\frac12 (A')^2 } \nonumber \\
& & \hspace{0cm} + 8 H^6 \Delta x^{(\mu} \eta^{\nu) (\rho} \Delta
x^{\sigma)} \times -A' A'' + 16 H^8 \Delta x^{\mu} \Delta x^{\nu}
\Delta x^{\rho} \Delta x^{\sigma} \times -\frac12 (A'')^2 \nonumber \\
& & \hspace{0cm} + H^4 \eta^{\mu\nu} \eta^{\rho\sigma} \times
-\frac18 \Bigl[16 H^4 \Delta x^4 (A'')^2 \!+ \! 16 H^2 \Delta x^2 A'
A'' \!+\! 4(D\!-\!4) (A')^2\Bigr] \nonumber \\
& & \hspace{.5cm} + 4 H^6 \Bigl[\eta^{\mu\nu} \Delta x^{\rho}
\Delta x^{\sigma} \!+\! \Delta x^{\mu} \Delta x^{\nu}
\eta^{\rho\sigma}\Bigr] \times \frac14 \Bigl[4 H^2 \Delta x^2
(A'')^2 \!+\! 4 A' A''\Bigr] \Biggr\} \; ,
\qquad \\
& & = \frac{\kappa^2 \Gamma^2(\frac{D}2)}{16 \pi^D} \Biggl\{
\eta^{\mu(\rho} \eta^{\sigma)\nu} \times \Bigl[-\frac{2}{\Delta
x^{2D}}\Bigr] + \Delta x^{(\mu} \eta^{\nu)(\rho} \Delta x^{\sigma)}
\times \Bigl[ \frac{4D}{\Delta x^{2D+2}} \Bigr] \nonumber \\
& & \hspace{1cm} + \Delta x^{\mu} \Delta x^{\nu} \Delta x^{\rho}
\Delta x^{\sigma} \times \Bigl[- \frac{2D^2}{\Delta x^{2D+4}}\Bigr]
+ \eta^{\mu\nu} \eta^{\rho\sigma} \times \Bigl[-\frac12 \frac{(D^2
\!-\! D\!-\! 4)}{\Delta x^{2D}} \Bigr] \nonumber \\
& & \hspace{2cm} + \Bigl[\eta^{\mu\nu} \Delta x^{\rho} \Delta
x^{\sigma} \!+\! \Delta x^{\mu} \Delta x^{\nu} \eta^{\rho\sigma}
\Bigr] \times \Bigl[\frac{D(D \!-\! 2)}{\Delta x^{2D+2}} \Bigr]
\Biggr\} \;. \label{flat}
\end{eqnarray}
Our result (\ref{flat}) agrees with equation (26) of \cite{FW}.

\subsection{Correspondence with Stress Tensor Correlators}

Although the flat space limit (\ref{flat}) will prove a useful guide
when we renormalize in the next section, it does not check the
purely de Sitter parts of (\ref{3-point-A}). A true de Sitter check
is provided by the stress tensor correlators recently derived by
Perez-Nadal, Roura and Verdaguer \cite{PNRV}. To exploit their
result we first elucidate the relation between the graviton 2-point
1PI function and correlators of the stress tensor. Then we convert
their notation to ours.

The Heisenberg equation for the metric field operator coupled to a
matter stress tensor $T^{\mu\nu}$ is,
\begin{equation}
R^{\mu\nu} - \frac12 g^{\mu\nu} R + \frac12 (D\!-\!2) (D\!-\!1) H^2
g^{\mu\nu} = \frac12 \kappa^2 T^{\mu\nu} \; . \label{Einstein}
\end{equation}
Perturbation theory is implemented by expressing the full metric
$g_{\mu\nu} = \overline{g}_{\mu\nu} + \kappa h_{\mu\nu}$ as the sum
of a vacuum solution $\overline{g}_{\mu\nu}$ plus $\kappa$ times the
graviton field $h_{\mu\nu}$. Expanding the left hand side of
(\ref{Einstein}) in powers of the graviton field gives,
\begin{equation}
R^{\mu\nu} - \frac12 g^{\mu\nu} R + \frac12 (D\!-\!2) (D\!-\!1) H^2
g^{\mu\nu} = \kappa \mathcal{D}^{\mu\nu\rho\sigma} h_{\rho\sigma} -
\frac12 \kappa^2 \Delta \mathcal{T}^{\mu\nu} \; ,
\end{equation}
where the nonlinear terms comprise the graviton pseudo-stress tensor
$\Delta \mathcal{T}^{\mu\nu}$. The Lichnerowicz operator of the linear
term is,
\begin{eqnarray}
\lefteqn{ \mathcal{D}^{\mu\nu\rho\sigma} \equiv D^{(\rho}
\overline{g}^{\sigma) (\mu} D^{\nu)} -\frac12
\Bigl[\overline{g}^{\rho\sigma} D^{\mu\nu} \!+\!
\overline{g}^{\mu\nu} D^{\rho} D^{\sigma}\Bigr] } \nonumber \\
& & \hspace{1.5cm} + \frac12 \Bigl[ \overline{g}^{\mu\nu}
\overline{g}^{\rho\sigma} \!-\! \overline{g}^{\mu (\rho}
\overline{g}^{\sigma) \nu} \Bigr] D^2 + (D\!-\!1) \Bigl[\frac12
\overline{g}^{\mu\nu} \overline{g}^{\rho\sigma} \!-\!
\overline{g}^{\mu (\rho} \overline{g}^{\sigma) \nu}\Bigr] H^2 \; ,
\qquad \label{Lich}
\end{eqnarray}
where $D^{\mu}$ is the covariant derivative operator in the background
geometry. Substituting these expansions in (\ref{Einstein}) and
rearranging gives,
\begin{equation}
\mathcal{D}^{\mu\nu\rho\sigma} h_{\rho\sigma} = \frac12 \kappa
\Bigl(T^{\mu\nu} + \Delta \mathcal{T}^{\mu\nu}\Bigr) \equiv \frac12
\kappa \mathcal{T}^{\mu\nu} \; .
\end{equation}

We are computing the 1PI graviton 2-point function, which can be
obtained from the full graviton 2-point function by eliminating the
one particle reducible parts and amputating the external leg
propagators. At the one loop order we are working, the one particle
reducible part drops out if one computes the correlator of the field
minus its expectation value,
\begin{eqnarray}
\delta h_{\mu\nu}(x) & \equiv & h_{\mu\nu}(x) - \Bigl\langle \Omega
\Bigl\vert h_{\mu\nu}(x) \Bigr\vert \Omega \Bigr\rangle \; , \\
\delta \mathcal{T}^{\mu\nu}(x) & \equiv & \mathcal{T}^{\mu\nu}(x) -
\Bigl\langle \Omega \Bigl\vert \mathcal{T}^{\mu\nu}(x) \Bigr\vert
\Omega \Bigr\rangle \; .
\end{eqnarray}
To amputate, recall that the graviton propagator obeys,
\begin{equation}
\sqrt{-\overline{g}(x)} \mathcal{D}^{\mu\nu\alpha\beta} i\Bigl[
\mbox{}_{\alpha\beta} \Delta_{\rho\sigma}\Bigr](x;x') =
\delta^{\mu}_{(\rho} \delta^{\nu}_{\sigma)} i\delta^D(x \!-\! x') +
\Bigl({\rm Gauge\ Terms}\Bigr) \; ,
\end{equation}
where ``Gauge Terms'' refers to the extra pieces needed to complete
the projection operator onto whatever gauge condition is employed.
(For example, the projection operator for de Donder grauge is given
in equation (120) of \cite{MTW}.) This means that external leg
propagators are amputated by $-i \sqrt{-\overline{g}}$ times the
Lichnerowicz operator. Hence the desired relation between the
2-point graviton 1PI function and a 2-point correlator of the stress
tensor is,
\begin{eqnarray}
\lefteqn{-i\Bigl[\mbox{}^{\mu\nu} \Delta^{\rho\sigma}\Bigr](x;x') }
\nonumber \\
& & \hspace{.1cm} = \Bigl\langle \Omega \Bigl\vert \Bigl(-i
\sqrt{-\overline{g}} \mathcal{D}^{\mu\nu\alpha\beta} \delta
h_{\alpha\beta}(x)\Bigr) \Bigl(-i \sqrt{-\overline{g}}
\mathcal{D}^{\rho\sigma\gamma\delta} \delta h_{\gamma\delta}(x')
\Bigr) \Bigr\vert \Omega \Bigr\rangle +
O(\kappa^4) \; , \qquad \\
& & \hspace{.1cm} =  -\frac14 \kappa^2 \sqrt{-\overline{g}(x)} \,
\sqrt{-\overline{g}(x')} \Bigl\langle \Omega \Bigl\vert \delta
\mathcal{T}^{\mu\nu}(x) \delta \mathcal{T}^{\rho\sigma}(x')
\Bigr\vert \Omega \Bigr\rangle + O(\kappa^4) \; . \label{desired}
\end{eqnarray}
The expectation value on the right hand side of (\ref{desired}) is
the stress tensor correlator $F^{\mu\nu\rho\sigma}$ of Perez-Nadal,
Roura and Verdaguer \cite{PNRV}.

Perez-Nadal, Roura and Verdaguer actually derived
$F^{\mu\nu\rho\sigma}$ for a scalar with arbitrary mass, but we can
compare our result (\ref{3-point-A}) for the massless case with
their equation (28) \cite{PNRV}
\begin{eqnarray}
 \lefteqn{F_{\mu\nu\rho\sigma} =
 P(\mu)n_{\mu}n_{\nu}n_{\rho}n_{\sigma} + Q(\mu)(n_{\mu}n_{\nu}\g_{\rho\sigma} + n_{\rho}n_{\sigma}\g_{\mu\nu} )}
\nonumber \\
&& \hspace{1cm}
 +R(\mu)(n_{\mu}n_{\rho}\g_{\nu\sigma} + n_{\nu}n_{\sigma}\g_{\mu\rho} + n_{\mu}n_{\sigma}\g_{\nu\rho} 
 + n_{\nu}n_{\rho}\g_{\mu\sigma})
\nonumber \\
&& \hspace{1cm}
 +S(\mu)(\g_{\mu\rho}\g_{\nu\sigma} + \g_{\nu\rho}\g_{\mu\sigma}) + T(\mu)\g_{\mu\nu}\g_{\rho\sigma}\;.
\label{Roura_correlator}
\end{eqnarray}
Note that here they expressed the stress tensor correlator in terms of 
five basis tensors which are different from ours given in equation 
(\ref{3-point-A}). Each of these five bitensors are formed as a linear 
combination of products of the de Sitter invariant bitensors, $n_{a}, 
n_{a'}, \g_{ab}, \g_{a'b'}$ and $\g_{ab'}$. The variable $\mu$ and 
bitensors are defined as \cite{PNRV}:
\begin{itemize}
\item{$\mu(x,x')$: the distance along the shortest geodesic joining 
$x$ and $x'$, also called the geodesic distance; }
\item{$n_{a}$ and $n_{a'}$: the unit vectors tangent to the geodesic 
at the points $x$ and $x'$ respectively, pointing outward from it; }
\item{$\g_{ab'}$: the parallel propagator which parallel-transports a vector 
from $x$ to $x'$ along the geodesic; }
\item{$\g_{ab}$ and $\g_{a'b'}$: the metric tensors at the points, 
at the points $x$ and $x'$ respectively.}
\end{itemize}
The distance $\mu(x,x')$ (in our notation $\mu(x,x') = H\ell(x;x')$ 
which is given in section 2) corresponds to our de Sitter invariant 
function $y(x,x')$ with the relation,
\begin{equation}
  \cos(\mu) \equiv Z = 1 - \frac{y}{2}\;.
\end{equation}


In comparing their results with
ours it is also useful to note the relations between their basis
tensors and ours,
\begin{eqnarray}
n_a & = & \frac1{H \sqrt{y (4 \!-\! y)}} \,
\frac{\partial y}{\partial x^a} \; , \\
n_{b'} & = & \frac1{H \sqrt{y (4 \!-\! y)}} \,
\frac{\partial y}{\partial x^{\prime b'}} \; , \\
\g_{ab'} & = & -\frac1{2 H^2} \biggl\{ \frac{\partial^2 y}{\partial
x^a \partial x^{\prime b'}} + \frac1{4 \!-\! y} \, \frac{\partial
y}{\partial x^a} \frac{\partial y}{\partial x^{\prime b'}} \biggr\}
\; . \label{partran}
\end{eqnarray}
Thus the five basis tensors given in (\ref{Roura_correlator}) are converted into our basis tensors as,
\begin{eqnarray}
 n_{a}n_{b}n_{c'}n_{d'} 
&\!\!\!=\!\!\!& \frac{1}{H^4(4y -y^2)^2}\frac{\partial y}{\partial x^a}\frac{\partial y}{\partial x^b}
\frac{\partial y}{\partial {x'}^{c'}}\frac{\partial y}{\partial {x'}^{d'}} \;,
\nonumber \\
 n_{a}n_{b}\g_{c'd'} + n_{c'}n_{d'}\g_{ab} 
&\!\!\!=\!\!\!& \frac{1}{H^2(4y -y^2)}\bigg[
\g_{ab}\frac{\partial y}{\partial {x'}^{c'}}\frac{\partial y}{\partial {x'}^{d'}} 
+ \frac{\partial y}{\partial x^a}\frac{\partial y}{\partial x^b}\g_{c'd'}\bigg]\;,
\nonumber \\
 4n_{(a}\g_{b)(c'}n_{d')}
&\!\!\!=\!\!\!& -\frac{2}{H^4(4y -y^2)}\frac{\partial y}{\partial x^{(a}}
\frac{\partial^2 y}{\partial x^{b)}\partial {x'}^{(c'}}
\frac{\partial y}{\partial {x'}^{d')}} \nonumber \\
&& -\frac{2}{H^4(4y -y^2)(4-y)}
\frac{\partial y}{\partial x^a}\frac{\partial y}{\partial x^b}
\frac{\partial y}{\partial {x'}^{c'}}\frac{\partial y}{\partial {x'}^{d'}}\;,
\nonumber \\
2\g_{a(c'}\g_{d')b} &\!\!\!=\!\!\!& 
\frac{1}{2H^4}\frac{\partial^2 y}{\partial x^{a}\partial {x'}^{(c'}}
\frac{\partial^2 y}{\partial x^{d')}\partial {x'}^{b}}
\nonumber \\
&& + \frac{1}{H^4(4-y)}\frac{\partial y}{\partial x^{(a}}
\frac{\partial^2 y}{\partial x^{b)}\partial {x'}^{(c'}}
\frac{\partial y}{\partial {x'}^{d')}}
\nonumber \\
&& + \frac{1}{2H^4}\frac{1}{(4-y)^2}
\frac{\partial y}{\partial x^a}\frac{\partial y}{\partial x^b}
\frac{\partial y}{\partial {x'}^{c'}}\frac{\partial y}{\partial {x'}^{d'}}\;,  
\nonumber \\
\g_{ab}\g_{c'd'} &\!\!\!=\!\!\!& \g_{ab}\g_{c'd'}\;.
\label{convert_basis}
\end{eqnarray}
(Note that we have restored the factor of $H$ which
Perez-Nadal, Roura and Veraguer set to unity.) 

For a massless, minimally coupled scalar field, the 
$\mu$-dependent coefficients are \cite{PNRV},
\begin{eqnarray}
 P &\!\!\!=\!\!\!& 2G_1^2 \;, \nonumber \\
 Q &\!\!\!=\!\!\!& -G_1^2 + 2G_1G_2 \;, \nonumber \\
 R &\!\!\!=\!\!\!& G_1G_2 \;, \nonumber \\
 S &\!\!\!=\!\!\!& G_2^2\;, \nonumber \\
 T &\!\!\!=\!\!\!& \frac{1}{2}G_1^2 -G_1G_2 + \frac{D-4}{2}G_2^2\;.
\label{PQRST}
\end{eqnarray}
Here the $G_1$ and $G_2$ are defined as 
\begin{eqnarray}
 G_1(\mu) &\!\!\!=\!\!\!& G''(\mu) - G'(\mu)\csc(\mu) \;, \nonumber \\
 G_2(\mu) &\!\!\!=\!\!\!& -G'(\mu)\csc(\mu) \;,
\label{G_1G_2}
\end{eqnarray}
where prime stands for derivative with respect to $\mu$.

The comparison can be completed by noting that the Wightman function $G(\mu)$ becomes almost the same as our $A(y)$ for the case of MMC scalar. 
In the massless limit, their propagator has the formal expansion,
\begin{eqnarray}
G(\mu) = \frac{H^{D-2}}{(4\pi)^{D/2}}\sum_{n=0}^{\infty}
\frac{\Gamma(D-1 + n)\Gamma(n)}{\Gamma(\frac{D}{2} + n)}
\frac{1}{n!}\Bigl(\frac{1+Z}{2}\Bigr)^n \;.
\end{eqnarray}
(Note that we have restored the factor of $H^{D-2}$ which
Perez-Nadal, Roura and Veraguer set to unity.)
Recalling the hypergeometric function,
\begin{eqnarray}
{}_{2}F_{1}\Bigl(\alpha, \beta; \gamma; z\Bigr) =
\sum_{n=0}^{\infty} \frac{\Gamma(\alpha \!+\!n)}{\Gamma(\alpha)}
\frac{\Gamma(\beta \!+\! n)}{\Gamma(\beta)}
\frac{\Gamma(\gamma)}{\Gamma(\gamma \!+\! n)} \frac{z^n}{n!} \; ,
\label{hyper_series}
\end{eqnarray}
we see that $G(Z)$ can be written as,
\begin{eqnarray}
G(y) = \frac{H^{D-2}}{(4\pi)^{D/2}} \frac{\Gamma(D\!-\!1)
\Gamma(0)}{\Gamma(\frac{D}2)} \, \mbox{}{}_2F_1\Bigl(D\!-\!1, 0;
\frac{D}2; 1 \!-\! \frac{y}{4}\Bigr) \; .
\end{eqnarray}
Now we use one of the transformation formulae for hypergeometric
functions (See for example, 9.131 of \cite{GR}) to expand $G^+$ in
powers of $y/4$:
\begin{eqnarray}
\lefteqn{G(y)= \frac{H^{D-2}}{(4\pi)^{\frac{D}2}} \Biggl\{
\frac{\Gamma(\frac{D}2)}{\frac{D}2 \!-\! 1}
\Bigl(\frac{4}{y}\Bigr)^{ \frac{D}2 -1} \!+\! \frac{\Gamma(\frac{D}2
\!+\! 1)}{\frac{D}2 \!-\! 2} \Bigl(\frac{4}{y} \Bigr)^{\frac{D}2-2}
\!-\! \Gamma(0)
\frac{\Gamma(D \!-\! 1)}{\Gamma(\frac{D}2)} } \nonumber \\
& & \hspace{.5cm} + \sum_{n=1}^{\infty} \Biggl[\frac1{n}
\frac{\Gamma(n \!+\! D \!-\! 1)}{\Gamma(n \!+\! \frac{D}2)}
\Bigl(\frac{y}4 \Bigr)^n \!\!\!\! - \frac1{n \!-\! \frac{D}2 \!+\!
2} \frac{\Gamma(n \!+\!  \frac{D}2 \!+\! 1)}{ \Gamma(n \!+\! 2)}
\Bigl(\frac{y}4 \Bigr)^{n - \frac{D}2 +2} \Biggr] \Biggr\} . \qquad
\end{eqnarray}
So we see that $G(y)$ is the same as the function $A(y)$ except for
the replacement,
\begin{eqnarray}
\Gamma(0) \frac{\Gamma(D \!-\! 1)}{\Gamma(\frac{D}2)}
\longrightarrow \pi \cot\Bigl(\frac{\pi D}2\Bigr)
 \frac{\Gamma(D \!-\! 1)}{\Gamma(\frac{D}2)}\;.
\end{eqnarray}
This makes no difference because $G(y)$ only enters the stress tensor
correlator (\ref{Roura_correlator}) differentiated (See equations (\ref{PQRST}-\ref{G_1G_2})). Thus for comparison, we replace the derivatives of $G$ by the ones of $A$:
\begin{eqnarray}
  \frac{\partial G}{\partial \mu} 
&\!\!\!=\!\!\!& \sqrt{4y-y^2}G' \equiv \sqrt{4y-y^2}A'\;,
\nonumber \\
 \frac{\partial^2 G}{\partial \mu^2}
&\!\!\!=\!\!\!& (4y-y^2)G''
+ (2-y)G'
\equiv (4y-y^2)A''
+ (2-y)A'\;.
\end{eqnarray}
Here the prime stand for derivative with respect to $y$.
Then the coefficients $P, Q, R, S$ and $T$ given in equation (\ref{PQRST}) are written in terms of $y$ as
\begin{eqnarray}
 P &\!=\!& 
2(4y-y^2)^2(A'')^2 -4y(4y-y^2)A''A' + 2y^2(A')^2\;,
\nonumber \\
 Q &\!=\!&
-(4y-y^2)^2(A'')^2 - 2(2-y)(4y-y^2)A''A' +(4y-y^2)(A')^2\;.
\nonumber \\
 R &\!=\!&
-2(4y-y^2)A''A'+ 2y(A')^2\;, \nonumber \\
 S &\!=\!&
4(A')^2\;, \nonumber \\
 T  &\!=\!& 
\frac{1}{2}\Bigl[(4y-y^2)^2(A'')^2 + 2(2-y)(4y-y^2)A''A' 
\nonumber \\
&&  \hspace{5cm} + \{4(D-4)-(4y-y^2)\}(A')^2\Bigr]\;.
\label{PQRST_y}
\end{eqnarray}
With this equation (\ref{PQRST_y}) and the conversion of basis given in equation (\ref{convert_basis}) we can arrange $F_{\mu\nu\rho\sigma}$ for the MMC scalar in terms of our basis tensors, 
\begin{eqnarray}
\lefteqn{F_{\mu\nu\rho\sigma} = - \frac{4}{\kappa^2}\Biggl\{
\frac{\partial^2 y}{\partial x^{\mu}\partial x'^{(\rho}}
\frac{\partial^2 y}{\partial x'^{\sigma)}\partial x^{\nu}} \times
\alpha(y) }
\; \nonumber \\
& & +
\frac{\partial y}{\partial x^{(\mu}}\frac{\partial^2 y}{\partial
x^{\nu)}\partial x'^{(\rho}}\frac{\partial y}{\partial x'^{\sigma)}}
\times \beta(y) +
\frac{\partial y}{\partial x^{\mu}}\frac{\partial y}{\partial
x^{\nu}} \frac{\partial y}{\partial x'^{\rho}}\frac{\partial
y}{\partial x'^{\sigma}} \times \gamma(y) \nonumber \\
& & +
\g^{\mu\nu}\g'^{\rho\sigma}H^4 \times \delta(y) +
\Bigl[ \g^{\mu\nu}\frac{\partial y}{\partial
x'^{\rho}}\frac{\partial y}{\partial x'^{\sigma}} + \frac{\partial
y}{\partial x^{\mu}}\frac{\partial y}{\partial x^{\nu}}\g'^{\rho
\sigma}\Bigr]H^2 \times \epsilon(y) \; \Bigg\} \; . \qquad
\\
& & = - \frac{4}{\kappa^2} \times \frac{1}{\sqrt{-\overline{g}(x)}
\, \sqrt{-\overline{g}(x')}} \times -i
\Bigl[\mbox{}_{\mu\nu}\Sigma_{\rho\sigma}\Bigr]_{\mbox{\tiny
3pt}}\!\!\!\!\!(x;x') \; .
\end{eqnarray}


\section{Renormalization}

Our result (\ref{3-point-A}) is valid as long as $x^{\prime \mu}
\neq x^{\mu}$, either with the exact coefficient functions
(\ref{alp1}-\ref{eps1}) or with the relevant expansions
(\ref{alp2}-\ref{eps2}) for $D=4$. However, it is not immediately
usable in the quantum-corrected, linearized Einstein equations
because they involve an integration over $x^{\prime \mu}$,
\begin{equation}
\sqrt{-\overline{g}} \, \mathcal{D}^{\mu\nu\rho\sigma} h_{\rho\sigma}(x)
- \int \!\! d^4x' \Bigl[ \mbox{}^{\mu\nu} \Sigma^{\rho\sigma}\Bigr]_{
\mbox{\tiny ren}}\!\!\!\!\!(x;x') h_{\rho\sigma}(x') = \frac12 \kappa
\sqrt{-\overline{g}} \, T^{\mu\nu}_{\mbox{\tiny lin}}(x) \; . \label{QMEin}
\end{equation}
To obtain a usable form we must express (\ref{3-point-A}) as a
product of up to six differential operators acting upon a function
of $y(x;x')$ which is integrable in $D=4$ spacetime dimensions. The
derivatives with respect to $x^{\mu}$ can be pulled outside the integral,
and those with respect to $x^{\prime \mu}$ can be partially integrated
to act back on the $h_{\rho\sigma}(x')$,\footnote{The resulting surface
terms can be absorbed by correcting the initial state \cite{KOW}.}
leaving an expression for which the $D=4$ limit could be taken were it
not for some factors of $1/(D-4)$. At this stage one adds zero in the
form of identities such as,
\begin{equation}
\Biggl[\square - \frac{D}2 \Bigl( \frac{D}2 \!-\! 1\Bigr) H^2 \Biggr]
\Bigl(\frac4{y}\Bigr)^{\frac{D}2 - 1} - \frac{(4 \pi)^{\frac{D}2} \, i
\delta^D(x \!-\! x')}{\Gamma(\frac{D}2 \!-\!1) H^{D-2}
\sqrt{-\overline{g}}} = 0 \; . \label{ZERO}
\end{equation}
We combine (\ref{ZERO}) with terms which arise from extracting
derivatives to segregate the divergences on local, delta function
terms, for example,
\begin{eqnarray}
\lefteqn{ \frac1{D \!-\!4} \Biggl[\square - \frac{D}2 \Bigl(\frac{D}2
\!-\!1 \Bigr) H^2\Biggr] \Bigl(\frac{4}{y}\Bigr)^{D-3} } \nonumber \\
& & \hspace{-.5cm} = \Biggl[\square \!-\! \frac{D}2 \Bigl(\frac{D}2
\!-\!1 \Bigr) H^2\Biggr] \Biggl\{\frac{(\frac{4}{y})^{D-3} \! -
(\frac{4}{y})^{\frac{D}2 -1}}{D \!-\! 4}\Biggr\} + \frac{(4 \pi)^{\frac{D}2}
\, i \delta^D(x \!-\! x')/\!\sqrt{-\overline{g}}}{(D \!-\!4)
\Gamma(\frac{D}2 \!-\!1) H^{D-2}} \; , \qquad \\
& & \hspace{-.5cm} = -\frac12 \Bigl[ \square \!-\! 2 H^2\Bigr]
\Biggl\{\frac{4}{y} \, \ln\Bigl(\frac{y}4\Bigr) \Biggr\} +
O(D \!-\!4) + \frac{(4 \pi)^{\frac{D}2} \, i \delta^D(x \!-\! x')/\!
\sqrt{-\overline{g}}}{(D \!-\!4) \Gamma(\frac{D}2 \!-\!1) H^{D-2}}
\; . \qquad \label{limit}
\end{eqnarray}
Renormalization consists of subtracting off the divergent delta
functions with counterterms. In subsection 4.1 we exhibit the one
loop counterterms for quantum gravity. We review how to renormalize
the flat space limit (\ref{flat}) in subsection 4.2. That suggests a
convenient way of organizing the tensor algebra into two transverse,
4th order differential operators, one with spin zero and the other
with spin two. In subsection 4.3 we implement this for de Sitter. The
spin zero part is renormalized in subsection 4.4, and the spin two
part in subsection 4.5.

\subsection{One Loop Counterterms}

Gravity + Scalar is not renormalizable in $D = 4$ dimensions \cite{HV}.
However, the theorem of Bogoliubov, Parasiuk, Hepp and Zimmerman (BPHZ)
shows us how to construct local counterterms which absorb the
ultraviolet divergences of any quantum field theory to any fixed order
in the loop expansion \cite{BPHZ}. For quantum gravity at one loop
order the necessary counterterms can be taken to be the squares of
the Ricci scalar and the Weyl tensor \cite{HV}. The problem of quantum
gravity is that the Weyl counterterm would destabilize the universe if
it were regarded as a fundamental, nonperturbative interaction \cite{RPW}.
We shall therefore consider it only perturbatively, in the sense of
effective field theory, as a proxy for the yet unknown ultraviolet
completion of quantum gravity. The quantum effects we seek to study
derive from infrared virtual scalars with wavelengths on the order
of the Hubble radius, and they will manifest as nonlocal and ultraviolet
finite contributions to the graviton self-energy which are not affected
by how nature resolves the ultraviolet problem of quantum gravity.

Because the background Ricci scalar is nonzero it is useful to
reorganize $R^2$ into a part which is quadratic in the graviton field,
\begin{equation}
R^2 = \Bigl[ R - D (D \!-\! 1) H^2\Bigr]^2 + 2 D (D \!-\! 1) H^2 R
- D^2 (D \!-\! 1)^2 H^4 \; .
\end{equation}
So we will employ four counterterms,
\begin{eqnarray}
\Delta \mathcal{L}_1 & \equiv & c_1 \Bigl[ R - D (D \!-\! 1) H^2\Bigr]^2
\sqrt{-g} \; , \label{LC1} \\
\Delta \mathcal{L}_2 & \equiv & c_2 C^{\alpha\beta\gamma\delta}
C_{\alpha\beta\gamma\delta} \sqrt{-g} \; , \label{LC2} \\
\Delta \mathcal{L}_3 & \equiv & c_3 H^2 \Bigl[ R - (D \!-\! 1)
(D \!-\! 2) H^2\Bigr] \sqrt{-g} \; , \label{LC3} \\
\Delta \mathcal{L}_4 & \equiv & c_4 H^4 \sqrt{-g} \; . \label{LC4}
\end{eqnarray}
Of course the divergences can really be eliminated with just
$\Delta \mathcal{L}_2$ and the particular linear combination of
$\Delta \mathcal{L}_1$, $\Delta \mathcal{L}_3$ and $\Delta \mathcal{L}_4$
which is proportional to just $R^2 \sqrt{-g}$. It must therefore be
that two linear combinations of the coefficients are finite,
\begin{eqnarray}
\lim_{D \rightarrow 4} \Bigl[-2 D (D \!-\! 1) c_1 + c_3\Bigr] & = &
{\rm Finite} \; , \label{cor1} \\
\lim_{D \rightarrow 4} \Bigl[D^2 (D \!-\! 1)^2 c_1 - (D \!-\! 1)
(D \!-\! 2) c_3 + c_4\Bigr] & = & {\rm Finite} \; . \label{cor2}
\end{eqnarray}
And the divergent parts of $c_1$ and $c_2$ must agree with the values
obtained long ago by `t Hooft and Veltman \cite{HV}.

At this point we digress to define two 2nd order differential
operators of great importance to our subsequent analysis. They come
from expanding the scalar and Weyl curvatures around de Sitter background,
\begin{eqnarray}
R - D (D \!-\! 1) H^2 & \equiv & \mathcal{P}^{\mu\nu} \kappa h_{\mu\nu}
+ O(\kappa^2 h^2) \; , \label{Rexp} \\
C_{\alpha\beta\gamma\delta}  & \equiv & \mathcal{P}^{\mu\nu}_{\alpha\beta
\gamma\delta} \kappa h_{\mu\nu} + O(\kappa^2 h^2) \; . \label{Cexp}
\end{eqnarray}
>From (\ref{Rexp}) we have,
\begin{equation}
\mathcal{P}^{\mu\nu} = D^{\mu} D^{\nu} - \overline{g}^{\mu\nu}
\Bigl[ D^2 + (D \!-\! 1) H^2\Bigr] \; , \label{P0}
\end{equation}
where $D^{\mu}$ is the covariant derivative operator in de Sitter
background. The more difficult expansion of the Weyl tensor gives,
\begin{eqnarray}
\lefteqn{\mathcal{P}^{\mu\nu}_{\alpha\beta\gamma\delta} =
\mathcal{D}^{\mu\nu}_{\alpha\beta\gamma\delta} + \frac1{D \!-\!2}
\Bigl[ \overline{g}_{\alpha\delta} \mathcal{D}^{\mu\nu}_{\beta\gamma}
\!-\! \overline{g}_{\beta\delta} \mathcal{D}^{\mu\nu}_{\alpha\gamma}
\!-\! \overline{g}_{\alpha\gamma} \mathcal{D}^{\mu\nu}_{\beta\delta}
\!+\! \overline{g}_{\beta\gamma} \mathcal{D}^{\mu\nu}_{\alpha\delta}
\Bigr] } \nonumber \\
& & \hspace{5cm} + \frac1{(D \!-\! 1) (D \!-\! 2)}
\Bigl[\overline{g}_{\alpha\gamma} \overline{g}_{\beta\delta} \!-\!
\overline{g}_{\alpha\delta} \overline{g}_{\beta\gamma} \Bigr]
\mathcal{D}^{\mu\nu} \; , \qquad \label{P2}
\end{eqnarray}
where we define,
\begin{eqnarray}
\mathcal{D}^{\mu\nu}_{\alpha\beta\gamma\delta} & \equiv & \frac12 \Bigl[
\delta^{(\mu}_{\alpha} \delta^{\nu)}_{\delta} D_{\gamma} D_{\beta}
\!-\! \delta^{(\mu}_{\beta} \delta^{\nu)}_{\delta} D_{\gamma} D_{\alpha}
\!-\! \delta^{(\mu}_{\alpha} \delta^{\nu)}_{\gamma} D_{\delta} D_{\beta}
\!+\! \delta^{(\mu}_{\beta} \delta^{\nu)}_{\gamma} D_{\delta} D_{\alpha}
\Bigr] \; , \qquad \\
\mathcal{D}^{\mu\nu}_{\beta\delta} & \equiv & \overline{g}^{\alpha\gamma}
\mathcal{D}^{\mu\nu}_{\alpha\beta\gamma\delta} = \frac12 \Bigl[
\delta^{(\mu}_{\delta} D^{\nu)} D_{\beta} \!-\! \delta^{(\mu}_{\beta}
\delta^{\nu)}_{\delta} D^2 \!-\! \overline{g}^{\mu\nu} D_{\delta}
D_{\beta} \!+\! \delta^{(\mu}_{\beta} D_{\delta} D^{\nu)}
\Bigr] \; , \qquad \\
\mathcal{D}^{\mu\nu} & \equiv & \overline{g}^{\alpha\gamma}
\overline{g}^{\beta\delta} \mathcal{D}^{\mu\nu}_{\alpha\beta\gamma\delta}
= D^{(\mu} D^{\nu)} - \overline{g}^{\mu\nu} D^2 \; .
\end{eqnarray}

One obtains the counterterm vertices by functionally differentiating
$i$ times each counterterm action twice, and then setting the graviton
field to zero. They are,
\begin{eqnarray}
\frac{i \delta \Delta S_1}{\delta h_{\mu\nu}(x) \delta h_{\rho\sigma}(x')}
\Biggl\vert_{h = 0} & \!\!\!=\!\!\! & 2 c_1 \kappa^2 \sqrt{-\overline{g}} \,
\mathcal{P}^{\mu\nu} \mathcal{P}^{\rho\sigma} i\delta^D(x \!-\! x')
\; , \label{C1} \\
\frac{i \delta \Delta S_2}{\delta h_{\mu\nu}(x) \delta h_{\rho\sigma}(x')}
\Biggl\vert_{h = 0} &\!\!\! =\!\!\! & 2 c_2 \kappa^2 \sqrt{-\overline{g}}
\, \overline{g}^{\alpha\kappa} \overline{g}^{\beta\lambda}
\overline{g}^{\gamma\theta} \overline{g}^{\delta\phi}
\mathcal{P}^{\mu\nu}_{\alpha\beta\gamma\delta}
\mathcal{P}^{\rho\sigma}_{\kappa\lambda\theta\phi}
i\delta^D(x \!-\! x') \; , \qquad \label{C2} \\
\frac{i \delta \Delta S_3}{\delta h_{\mu\nu}(x) \delta h_{\rho\sigma}(x')}
\Biggl\vert_{h = 0} & \!\!\!=\!\!\! & -c_3 \kappa^2 H^2 \sqrt{-\overline{g}}
\, \mathcal{D}^{\mu\nu\rho\sigma} i \delta^D(x \!-\! x') \; , \label{C3} \\
\frac{i \delta \Delta S_4}{\delta h_{\mu\nu}(x) \delta h_{\rho\sigma}(x')}
\Biggl\vert_{h = 0} & \!\!\!=\!\!\! & c_4 \kappa^2 H^4 \sqrt{-\overline{g}}
\, \Bigl[\frac14 \overline{g}^{\mu\nu} \overline{g}^{\rho\sigma} \!-\!
\frac12 \overline{g}^{\mu (\rho} \overline{g}^{\sigma) \nu} \Bigr]
i \delta^D(x \!-\! x') \; . \qquad \label{C4}
\end{eqnarray}
Recall that the Lichnerowicz operator in expression (\ref{C3}) was
defined in expression (\ref{Lich}). Also note the flat space
limits,
\begin{eqnarray}
\frac{i \delta \Delta S_1}{\delta h_{\mu\nu}(x) \delta h_{\rho\sigma}(x')}
\Biggl\vert_{h = 0} & \!\!\!\longrightarrow\!\!\! & 2 c_1 \kappa^2
\Pi^{\mu\nu} \Pi^{\rho\sigma} i\delta^D(x \!-\! x') \; , \label{C1flat} \\
\frac{i \delta \Delta S_2}{\delta h_{\mu\nu}(x) \delta h_{\rho\sigma}(x')}
\Biggl\vert_{h = 0} &\!\!\! \longrightarrow \!\!\! & 2 c_2 \kappa^2
\Bigl(\frac{D \!-\!3}{D \!-\! 2}\Bigr) \Bigl[\Pi^{\mu (\rho} \Pi^{\sigma)\nu}
\!-\! \frac{\Pi^{\mu\nu} \Pi^{\rho\sigma}}{D \!-\! 1} \Bigr] i
\delta^D(x \!-\! x') \; , \qquad \label{C2flat} \\
\frac{i \delta \Delta S_3}{\delta h_{\mu\nu}(x) \delta h_{\rho\sigma}(x')}
\Biggl\vert_{h = 0} & \!\!\! \longrightarrow \!\!\! & 0 \; , \label{C3flat} \\
\frac{i \delta \Delta S_4}{\delta h_{\mu\nu}(x) \delta h_{\rho\sigma}(x')}
\Biggl\vert_{h = 0} & \!\!\! \longrightarrow \!\!\! & 0 \; , \label{C4flat}
\end{eqnarray}
where we define,
\begin{equation}
\Pi^{\mu\nu} \equiv \partial^{\mu} \partial^{\nu} - \eta^{\mu\nu}
\partial^2 \; . \label{Pi}
\end{equation}

\subsection{Renormalizing the Flat Space Result}

Renormalizing the flat space result (\ref{flat}) provides an excellent
guide for the vastly more complicated reduction required on de Sitter
background. We begin by extracting a 4th order differential operator
from each term using the identities,
\begin{eqnarray}
\frac1{\Delta x^{2D}} & = & \frac{\partial^4}{4 (D \!-\!2)^2 (D \!-\! 1)
D} \, \frac1{\Delta x^{2D -4}} \; , \\
\frac{\Delta x^{\mu} \Delta x^{\nu}}{\Delta x^{2D+2}} & = &
\frac1{8 (D \!-\! 2)^2 (D \!-\!1) D} \Biggl\{ \partial^{\mu} \partial^{\nu}
\partial^2 + \frac{\eta^{\mu\nu} \partial^4}{D} \Biggr\} \,
\frac1{\Delta x^{2D-4}} \; , \qquad \\
\frac{\Delta x^{\mu} \Delta x^{\nu} \Delta x^{\rho}
\Delta x^{\sigma}}{\Delta x^{2D+4}} & = & \frac1{16 (D \!-\! 2) (D \!-\!1) D
(D \!+\!1)} \Biggl\{\partial^{\mu} \partial^{\nu} \partial^{\rho}
\partial^{\sigma} \nonumber \\
& & \hspace{-1cm} + \frac{6}{D \!-\! 2} \, \eta^{(\mu\nu}
\partial^{\rho} \partial^{\sigma)} \partial^2
+ \frac{3}{(D \!-\! 2) D}  \, \eta^{(\mu\nu} \eta^{\rho\sigma)} \partial^4
\Biggr\} \, \frac1{\Delta x^{2D-4}} \; . \qquad
\end{eqnarray}
Substituting these relations into (\ref{flat}), and then organizing the
various derivatives into factors of the transverse operator $\Pi^{\mu\nu}$
of expression (\ref{Pi}), gives a manifestly transverse form,
\begin{eqnarray}
\lefteqn{-i \Bigl[ \mbox{}^{\mu\nu} \Sigma^{\rho\sigma}\Bigr]_{\mbox{\tiny
flat}}\!\!\!\!(x;x') } \nonumber \\
& & \hspace{.9cm} = \frac{\kappa^2 \Gamma^2(\frac{D}2)}{16 \pi^D} \Biggl\{-
\frac{ \Pi^{\mu\nu} \Pi^{\rho\sigma}}{8 (D \!-\!1)^2} -
\frac{[ \Pi^{\mu (\rho} \Pi^{\sigma) \nu} \!-\! \frac1{D-1} \Pi^{\mu\nu}
\Pi^{\rho\sigma} ]}{4 (D \!-\!2)^2 (D \!-\!1) (D \!+\! 1)} \Biggr\} \,
\frac1{\Delta x^{2D - 4}} \; . \qquad \label{transflat}
\end{eqnarray}

Let us pause at this point to note that we could have guessed most of
the form of expression (\ref{transflat}). Gauge invariance implies
transversality. We also have Poincar\'e invariance, symmetry under
interchange the interchanges $\mu \leftrightarrow \nu$ and $\rho
\leftrightarrow \sigma$, and symmetry under interchange of the primed
and unprimed coordinates and indices. All this implies the form,
\begin{equation}
-i \Bigl[ \mbox{}^{\mu\nu} \Sigma^{\rho\sigma}\Bigr]_{\mbox{\tiny
flat}}\!\!\!\!(x;x') = \Pi^{\mu\nu} \Pi^{\rho\sigma} F_1(\Delta x^2)
+ \Bigl[ \Pi^{\mu (\rho} \Pi^{\sigma) \nu} \!-\! \frac{ \Pi^{\mu\nu}
\Pi^{\rho \sigma}}{D \!-\! 1} \Bigr] F_2(\Delta x^2) \; . \label{flatans}
\end{equation}
Taking the trace of this and our result (\ref{flat}) against
$\eta_{\mu\nu} \eta_{\rho\sigma}$ gives an equation for the spin zero
structure function $F_1(\Delta x^2)$,
\begin{equation}
\eta_{\mu\nu} \eta_{\rho\sigma} \times -i \Bigl[ \mbox{}^{\mu\nu}
\Sigma^{\rho\sigma}\Bigr]_{\mbox{\tiny flat}} =
(D \!-\! 1)^2 \partial^4 F_1(\Delta x^2) = \frac{\kappa^2
\Gamma^2(\frac{D}2)}{16 \pi^D} \times -\frac{(D \!-\! 2)^2 (D \!-\!1) D}{2
\Delta x^{2D}} \; .
\end{equation}
Of course the solution is just what we found in (\ref{transflat})
by direct computation,
\begin{equation}
F_1(\Delta x^2) = \frac{\kappa^2 \Gamma^2(\frac{D}2)}{16 \pi^D} \times
-\frac1{8 (D \!-\! 1)^2} \, \Bigl(\frac1{\Delta x^2} \Bigr)^{D-2} \; .
\end{equation}

Determining the spin two structure function $F_2(\Delta x^2)$ is done
by first acting the derivatives on the spin zero structure function,
\begin{eqnarray}
\lefteqn{\Pi^{\mu\nu} \Pi^{\rho\sigma} F_1 = \eta^{\mu (\rho}
\eta^{\sigma) \nu} \times 8 F_1'' + \Delta x^{(\mu} \eta^{\nu) (\rho}
\Delta x^{\sigma)} \times 32 F_1''' + \Delta x^{\mu} \Delta x^{\nu}
\Delta x^{\rho} \Delta x^{\sigma} } \nonumber \\
& & \hspace{.5cm} \times 16 F_1'''' + \eta^{\mu\nu} \eta^{\rho\sigma}
\times \Bigl[ 4(D^2 \!-\! 3) F_1'' + 16 (D \!+\! 1) \Delta x^2 F_1'''
+ 16 \Delta x^4 F_1''''\Bigr] \nonumber \\
& & \hspace{1cm} + \Bigl[\eta^{\mu\nu} \Delta x^{\rho} \Delta x^{\sigma}
\!+\! \Delta x^{\mu} \Delta x^{\nu} \eta^{\rho\sigma}\Bigr] \times
\Bigl[-8(D \!+\! 3) F_1''' - 16 \Delta x^2 F_1''''\Bigr] \; . \qquad
\end{eqnarray}
We subtract these from each tensor factor in (\ref{flat}) and then
act the spintwo operator $[\Pi^{\mu (\rho} \Pi^{\sigma) \nu} - \frac1{D-1}
\Pi^{\mu\nu} \Pi^{\rho\sigma}]$ on $F_2(\Delta x^2)$ to read off an
equation for  each of the five tensor factors,
\begin{eqnarray}
& & \hspace{-.7cm} \eta^{\mu (\rho} \eta^{\sigma) \nu} \Rightarrow
\frac{4 (D \!-\!2) D (D \!+\! 1)}{D \!-\! 1} \, F_2'' + 16 (D \!+\! 1)
\Delta x^2 F_2''' + 16 \Delta x^4 F_2'''' \nonumber \\
& & \hspace{6.5cm} = \frac{\kappa^2 \Gamma^2(\frac{D}2)}{16 \pi^D} \,
\Biggl\{ -\frac{D}{D \!-\! 1} \frac1{\Delta x^{2D}} \Biggr\} \; , \qquad
\label{alpflat2} \\
& & \hspace{-.7cm} \Delta x^{(\mu} \eta^{\nu) (\rho} \Delta x^{\sigma)}
\Rightarrow -\frac{16 D (D \!+\! 1)}{D \!-\! 1} \, F_2''' - 32 \Delta x^2
F_2'''' \nonumber \\
& & \hspace{6.5cm} = \frac{\kappa^2 \Gamma^2(\frac{D}2)}{16 \pi^D} \, \Biggl\{
\frac{4 D}{D \!-\! 1} \frac1{\Delta x^{2D}} \Biggr\} \; , \qquad
\label{betflat2} \\
& & \hspace{-.7cm} \Delta x^{\mu} \Delta x^{\nu} \Delta x^{\rho}
\Delta x^{\sigma} \Rightarrow 16 \Bigl(\frac{D \!-\! 2}{D \!-\! 1} \Bigr)
F_2'''' = \frac{\kappa^2 \Gamma^2(\frac{D}2)}{16 \pi^D} \, \Biggl\{-
\frac{4 D}{D \!-\! 1} \frac1{\Delta x^{2D}} \Biggr\} \; , \qquad
\label{gamflat2} \\
& & \hspace{-.7cm} \eta^{\mu \nu} \eta^{\rho \sigma} \Rightarrow -
\frac{4}{D \!-\! 1} \Bigl[(D \!-\!2) (D \!+\! 1) F_2'' + 4 (D \!+\! 1)
\Delta x^2 F_2''' + 4 \Delta x^4 F_2'''' \Bigr]\nonumber \\
& & \hspace{6.5cm} = \frac{\kappa^2 \Gamma^2(\frac{D}2)}{16 \pi^D} \,
\Biggl\{ \frac1{D \!-\! 1} \frac1{\Delta x^{2D}} \Biggr\} \; , \qquad
\label{delflat2} \\
& & \hspace{-.7cm} \Bigl[ \eta^{\mu\nu} \Delta x^{\rho} \Delta x^{\sigma}
\!+\! \Delta x^{\mu} \Delta x^{\nu} \eta^{\rho\sigma}\Bigr] \Rightarrow
\frac{16}{D \!-\! 1} \Bigl[ (D \!+\! 1) F_2''' + \Delta x^2 F_2''''\Bigr]
= 0 \; . \qquad \label{epsflat2}
\end{eqnarray}
Each of these equations has the same solution, which of course agrees
with (\ref{transflat}),
\begin{equation}
F_2(\Delta x^2) = \frac{\kappa^2 \Gamma^2(\frac{D}2)}{16 \pi^D} \times
-\frac1{4 (D \!-\! 2)^2 (D \!-\! 1) (D \!+\! 1)} \, \Bigl(\frac1{\Delta x^2}
\Bigr)^{D-2} \; .
\end{equation}
We note for future reference that a particular linear combination of
the five relations (\ref{alpflat2}-\ref{epsflat2}) gives a second order
equation for $F_2(\Delta x^2)$,
\begin{equation}
(\ref{delflat2}) + \Delta x^2 (\ref{epsflat2}) = -\frac4{D \!-\! 1} \,
(D \!-\! 2) (D \!+\! 1) F_2'' = \frac{\kappa^2 \Gamma^2(\frac{D}2)}{16 \pi^D}
\, \Biggl\{ \frac1{D \!-\! 1} \frac1{\Delta x^{2D}} \Biggr\} \; .
\label{2ndorder}
\end{equation}

Even after extracting the 4th order differential operators from the
integration of (\ref{QMEin}), the factor of $1/\Delta x^{2D-4}$ is
logarithmically divergent. We must therefore extract one more d'Alembertian,
\begin{equation}
\Bigl( \frac1{\Delta x^2} \Bigr)^{D-2} = \frac{\partial^2}{2 (D \!-\! 3)
(D \!-\! 4)} \Bigl( \frac1{\Delta x^2}\Bigr)^{D-3} \; . \label{finder}
\end{equation}
After this final derivative is extracted the integrand converges, however,
we still cannot take the $D = 4$ limit owing to the factor of $1/(D-4)$.
The solution is to add zero in the form of the identity,
\begin{equation}
\partial^2 \Bigl( \frac1{\Delta x^2}\Bigr)^{\frac{D}2-1} -
\frac{4 \pi^{\frac{D}2} \, i \delta^D(x \!-\! x')}{\Gamma(\frac{D}2
\!-\! 1)} = 0 \; .
\end{equation}
To make this dimensionally consistent with (\ref{finder}) we must
multiply by the dimensional regualrization mass scale $\mu$ raised to
the $(D-4)$ power,
\begin{eqnarray}
\lefteqn{\Bigl( \frac1{\Delta x^2} \Bigr)^{D-2} } \nonumber \\
& & \hspace{-.3cm} = \frac{\partial^2}{2 (D \!-\! 3) (D \!-\! 4)} \Biggl\{
\frac1{\Delta x^{2D-6}} - \frac{\mu^{D-4}}{\Delta x^{D-2}} \Biggr\} +
\frac{4 \pi^{\frac{D}2} \mu^{D-4} i \delta^D(x \!-\! x')}{2 (D \!-\! 3)
(D \!-\! 4) \Gamma(\frac{D}2 \!-\! 1)} \; , \qquad \\
& & = -\frac14 \partial^2 \Biggl\{ \frac{\ln(\mu^2 \Delta x^2)}{\Delta x^2}
+ O(D \!-\! 4)\Biggr\} + \frac{4 \pi^{\frac{D}2} \mu^{D-4} i
\delta^D(x \!-\! x')}{2 (D \!-\! 3) (D \!-\! 4) \Gamma(\frac{D}2 \!-\! 1)}
\; . \qquad
\end{eqnarray}

The divergences have now been segregated on delta function terms
which can be removed with local counterterms. From expressions
(\ref{C1flat}-\ref{C4flat}) we see that the counterterms make the
following contribution to the graviton self-energy,
\begin{eqnarray}
\lefteqn{-i \Bigl[\mbox{}^{\mu\nu} \Delta \Sigma^{\rho\sigma}\Bigr]_{
\mbox{\tiny flat}}\!\!\!\!\!(x;x') = \Pi^{\mu\nu} \Pi^{\rho\sigma}
\Biggl\{2 c_1 \kappa^2 i \delta^D(x \!-\! x') \Biggr\} } \nonumber \\
& & \hspace{2.5cm} + \Bigl[ \Pi^{\mu (\rho} \Pi^{\sigma) \nu} -
\frac{\Pi^{\mu\nu} \Pi^{\rho\sigma}}{D \!-\! 1} \Bigr] \Biggl\{2
\Bigl(\frac{D \!-\! 3}{D \!-\! 2}\Bigr) c_2 \kappa^2 i \delta^D(x \!-\! x')
\Biggr\} \; . \qquad
\end{eqnarray}
The delta function terms will be entirely absorbed by choosing the
constants $c_1$ and $c_2$ as,
\begin{eqnarray}
c_1 & = & \frac{\mu^{D-4} \Gamma(\frac{D}2)}{2^8 \pi^{\frac{D}2}} \,
\frac{(D \!-\! 2)}{(D \!-\! 1)^2 (D \!-\! 3) (D \!-\!4)} \; , \\
c_2 & = & \frac{\mu^{D-4} \Gamma(\frac{D}2)}{2^8 \pi^{\frac{D}2}} \,
\frac{2}{(D \!+\! 1) (D \!-\! 1) (D \!-\! 3)^2 (D \!-\!4)} \; .
\end{eqnarray}
Of course the divergent parts agree with the results obtained long ago
by `t Hooft and Veltman \cite{HV}, with the arbitrary finite parts
represented by $\mu$. The fully renormalized graviton self-energy (for
flat space background) is,
\begin{eqnarray}
\lefteqn{-i \Bigl[\mbox{}^{\mu\nu} \Sigma^{\rho\sigma}\Bigr]_{\mbox{\tiny
ren} \atop \mbox{\tiny flat}} = \lim_{D \rightarrow 4}
\Biggl\{ -i \Bigl[\mbox{}^{\mu\nu} \Sigma^{\rho\sigma}\Bigr]_{\mbox{\tiny
flat}}\!\!\!\!\!(x;x') -i \Bigl[\mbox{}^{\mu\nu} \Delta \Sigma^{\rho\sigma}
\Bigr]_{\mbox{\tiny flat}}\!\!\!\!\!(x;x') \Biggr\} \; , } \\
& & \hspace{1cm} = \Pi^{\mu\nu} \Pi^{\rho\sigma} \partial^2 \Biggl\{
\frac{\kappa^2}{2^9 3^2 \pi^4} \, \frac{\ln(\mu^2 \Delta x^2)}{\Delta x^2}
\Biggr\} \nonumber \\
& & \hspace{2.5cm} + \Bigl[\Pi^{\mu (\rho} \Pi^{\sigma) \nu} - \frac13
\Pi^{\mu\nu} \Pi^{\rho\sigma} \Bigr] \partial^2 \Biggl\{
\frac{\kappa^2}{2^{10} 3^1 5^1 \pi^4} \, \frac{\ln(\mu^2 \Delta x^2)}{
\Delta x^2} \Biggr\} \; . \qquad
\end{eqnarray}

\subsection{The de Sitter Structure Functions}

We must now extend the flat space ansatz (\ref{flatans}) to de Sitter and
determine the resulting structure functions by comparison with the explicit
result (\ref{3-point-A}) of section 3. As before, gauge invariance implies
transversality, which suggests that we make use of the differential
operators $\mathcal{P}^{\mu\nu}$ and $\mathcal{P}^{\mu\nu}_{\alpha\beta
\gamma\delta}$ which were defined in expressions (\ref{P0}) and (\ref{P2}),
respectively. In place of Poincar\'e invariance we now have de Sitter
invariance. We also have symmetry under the interchanges $\mu \leftrightarrow
\nu$ and $\rho \leftrightarrow \sigma$, and under interchange of the primed
and unprimed coordinates and indices. A simple generalization is,
\begin{eqnarray}
\lefteqn{ -i \Bigl[\mbox{}^{\mu\nu} \Sigma^{\rho\sigma}\Bigr](x;x') =
\sqrt{-\overline{g}(x)} \, \mathcal{P}^{\mu\nu}(x)
\sqrt{-\overline{g}(x')} \, \mathcal{P}^{\rho\sigma}(x') \Bigl\{
\mathcal{F}_1(y)\Bigr\} } \nonumber \\
& & \hspace{-.5cm} + \sqrt{-\overline{g}(x)} \,
\mathcal{P}^{\mu\nu}_{\alpha\beta \gamma\delta}(x)
\sqrt{-\overline{g}(x')} \, \mathcal{P}^{\rho\sigma}_{\kappa\lambda
\theta\phi}(x') \Biggl\{\mathcal{T}^{\alpha\kappa} \mathcal{T}^{\beta\lambda}
\mathcal{T}^{\gamma\theta} \mathcal{T}^{\delta\phi}
\Bigl( \frac{D \!-\! 2}{D \!-\! 3}\Bigr) \mathcal{F}_2(y) \Biggr\} ,
\qquad \label{ansatz}
\end{eqnarray}
where the bitensor $\mathcal{T}^{\alpha\kappa}$ is,\footnote{One could
actually employ any bitensor --- for example, the parallel transport
matrix (\ref{partran}) --- which reduces to $\eta^{\alpha\kappa}$ in the
flat space limit. Different choices for $\mathcal{T}^{\alpha\kappa}(x;x')$
make corresponding changes in the subdominant parts of the spin two
structure function $\mathcal{F}_2(y)$. We have not troubled to determine
the ``simplest'' choice.}
\begin{equation}
\mathcal{T}^{\alpha\kappa}(x;x') \equiv -\frac1{2 H^2} \,
\frac{\partial^2 y(x;x')}{\partial x_{\alpha} \partial x'_{\kappa}}
\; . \label{Tak}
\end{equation} As in flat space, the second term is
traceless.

Note the flat space limits of the bitensor and the two structure
functions,
\begin{equation}
\lim_{H \rightarrow 0} \mathcal{T}^{\alpha\kappa} = \eta^{\kappa\lambda}
\;\; , \;\; \lim_{H \rightarrow 0} \mathcal{F}_1(y) = F_1(\Delta x^2)
\;\; , \;\; \lim_{H \rightarrow 0} \mathcal{F}_2(y) = F_2(\Delta x^2) \; .
\end{equation}
These limits mean one can immediately read off the most singular parts of the
expansions for each structure function from the corresponding flat space
result,
\begin{eqnarray}
\mathcal{F}_1(y) & = & \frac{\kappa^2 H^{2D-4} \Gamma^2(\frac{D}2)}{(4\pi)^{D}}
\Biggl\{\frac{-1}{8 (D \!-\!1)^2} \Bigl(\frac{4}{y}\Bigr)^{D-2} + \dots
\Biggr\} , \\
\mathcal{F}_2(y) & = & \frac{\kappa^2 H^{2D-4} \Gamma^2(\frac{D}2)}{(4\pi)^{D}}
\Biggl\{\frac{-1}{4 (D \!-\! 3) (D \!-\! 2) (D \!-\!1) (D \!+\! 1)}
\Bigl(\frac{4}{y}\Bigr)^{D-2} \!\!\!\!\!\!+ \dots \Biggr\} . \qquad
\end{eqnarray}
The interesting de Sitter physics we seek to elucidate derives from the
subdominant terms.

Just as for the flat space limit, we can obtain an equation for the
spin zero structure function by tracing (\ref{ansatz}) and then comparing
with the trace of the explicit computation (\ref{3-point-A}). Tracing the
ansatz gives,
\begin{equation}
\frac{\overline{g}_{\mu\nu}(x)}{\sqrt{-\overline{g}(x)}} \times
\frac{\overline{g}_{\rho\sigma}(x')}{\sqrt{-\overline{g}(x')}} \times
-i \Bigl[\mbox{}^{\mu\nu} \Sigma^{\rho\sigma}\Bigr](x;x') = (D \!-\! 1)^2
\Bigl[\square \!+\! D H^2\Bigr] \Bigl[\square' \!+\! D H^2\Bigr]
\mathcal{F}_1(y) \; . \label{righthand}
\end{equation}
Tracing the explicit result (\ref{3-point-A}), substituting
(\ref{alp1}-\ref{eps1}), and then making use of (\ref{proprel}) gives,
\begin{eqnarray}
\lefteqn{\frac{\overline{g}_{\mu\nu}(x)}{\sqrt{-\overline{g}(x)}} \times
\frac{\overline{g}_{\rho\sigma}(x')}{\sqrt{-\overline{g}(x')}} \times
-i \Bigl[\mbox{}^{\mu\nu} \Sigma^{\rho\sigma}\Bigr]_{\mbox{\tiny 3pt}}\!\!\!\!
(x;x') = H^4 \Biggl\{\Bigl[4D \!-\! (4y \!-\! y^2)\Bigr] \alpha } \nonumber \\
& & \hspace{2cm} + (2 \!-\!y) (4y \!-\! y^2) \beta + (4y \!-\! y^2)^2 \gamma +
D^2 \delta + 2 D (4y \!-\! y^2) \epsilon \Biggr\} , \qquad \\
& & = \frac18 (D \!-\! 2)^2 \kappa^2 H^4 \Biggl\{
\Bigl[(4y \!-\! y^2) - 4 D\Bigr] (A')^2 \nonumber \\
& & \hspace{4cm} - 2 (2 \!-\! y) (4y \!-\! y^2) A' A'' - (4y \!-\! y^2)^2
(A'')^2 \Biggr\} , \qquad \\
& & = -\frac18 (D \!-\! 1)^2 (D \!-\! 2)^2 \kappa^2 H^4
\Biggl\{ \frac4{D \!-\! 1} \, (A')^2 + \Bigl[(2 \!-\! y) A' - k\Bigr]^2
\Biggr\} \; . \qquad \label{lefthand}
\end{eqnarray}
Now note that the primed and unprimed scalar d'Alembertian's agree
when acting on any function of only $y(x;x')$. Equating (\ref{righthand})
and (\ref{lefthand}) and expanding implies,
\begin{eqnarray}
\lefteqn{\Bigl[ \frac{\square}{H^2} \!+\! D\Bigr]^2 \mathcal{F}_1(y) =
-\frac18 (D \!-\!2)^2 \kappa^2 \Biggl\{ \frac{4}{D \!-\! 1} \, (A')^2
+ \Bigl[ (2 \!-\! y) A' - k\Bigr]^2 \Biggr\} \; . } \\
& & = -\frac{K}{32} \frac{(D \!-\! 2)^2}{(D \!-\! 1)} \Biggl\{ D
\Bigl(\frac{4}{y}\Bigr)^D + (D \!-\! 2)^2 \Bigl(\frac{4}{y}\Bigr)^{D-1}
\nonumber \\
& & \hspace{3cm}
+ \frac12 (D^3 \!-\! 7 D^2 \!+\! 16 D \!-\! 8) \Bigl(\frac{4}{y}\Bigr)^{D-2}
+ \Bigl({\rm Irrelevant}\Bigr) \Biggr\} \; , \qquad \label{F1eqn}
\end{eqnarray}
where the constant $K$ was defined in (\ref{Kdef}) and ``Irrelevant''
means terms which are both integrable at coincidence, and which vanish in
$D = 4$ dimensions.

Let us first note that we can find a Green's function for the differential
operator $[\square/H^2 + D]$. To see this, act the operator on some function
$f(y)$ which is free of the unique power $y^{\frac{D}2 -1}$ which produces
a delta function,
\begin{equation}
\Bigl[ \frac{\square}{H^2} \!+\! D\Bigr] f(y) = (4 y \!-\! y^2) f'' +
D (2 \!-\! y) f' + D f \; .
\end{equation}
Now note that $f_1(y) = (2 - y)$ is a homogeneous solution, which
means we can factor to obtain a first order equation (and hence solvable)
for the second solution,
\begin{equation}
f_1(y) = (2 \!-\! y) \Longrightarrow f_2(y) \equiv f_1(y) g(y)
\quad {\rm with} \quad g'(y) = \frac1{(4y \!-\! y^2)^{\frac{D}2}
f_1^2(y)} \; .
\end{equation}
With the two, linearly independent solutions one can construct a
Green's function,
\begin{equation}
G_1(y;y') = \theta((y \!-\! y') \Bigl[f_2(y) f_1(y') \!-\! f_1(y) f_2(y')\Bigr]
(4y' \!-\! y^{\prime 2})^{\frac{D}2 - 1} \; . \label{Gfunc}
\end{equation}
Hence we can solve (\ref{F1eqn}) to obtain on integral epxression
for the spin zero structure function,
\begin{equation}
\mathcal{F}_1(y) = \Biggl[ \frac1{\frac{\square}{H^2} \!+\! D} \Biggr]^2
\Biggr\{ {\rm Right\ hand\ side\ of} \; (\ref{F1eqn})\Biggr\} \; .
\label{formal}
\end{equation}

Although we will eventually make use of the Green's function (\ref{Gfunc}),
it is best to delay this until the point at which one can set $D = 4$. For
the more singular terms the best strategy is to exploit the fact that the
``source'' terms on the right hand side of (\ref{F1eqn}) upon which we wish
to act the inverse of $[\square/H^2 + D]^2$ are just powers of $y$. Consider
acting the operator upon a power $p -2 \neq \frac{D}-1$ or $\frac{D}2 - 2$
(those powers produce delta functions),
\begin{eqnarray}
\lefteqn{\Bigl[ \frac{\square}{H^2} \!+\! D\Bigr]^2 \Bigl(\frac{4}{y}
\Bigr)^{p-2} = (p \!-\! 2)(p \!-\! 1) ( p \!-\! 1 \!-\! \frac{D}2)
(p \!-\! \frac{D}2) \Bigl(\frac{4}{y}\Bigr)^p + (p \!-\! 2) (p \!-\! 1 \!-\!
\frac{D}2) } \nonumber \\
& & \hspace{1.7cm} \times \Bigl[D (2p \!-\! 1) \!-\! 2 (p \!-\!1)^2\Bigr]
\Bigl(\frac{4}{y}\Bigr)^{p-1} \!\!\! + (p \!-\! 1)^2 (D \!-\! p \!+\! 2)^2
\Bigl(\frac{4}{y}\Bigr)^{p-2} \!\!\! . \qquad
\end{eqnarray}
We can therefore develop a recursive procedure for reducing the power
of the source,
\begin{eqnarray}
\lefteqn{\Biggl[ \frac1{\frac{\square}{H^2} \!+\! D} \Biggr]^2
\Bigl( \frac{4}{y} \Bigr)^p = \frac1{(p \!-\! 2) (p \!-\! 1) (p \!-\! 1
\!-\! \frac{D}2) (p \!-\! \frac{D}2)} \Bigl( \frac{4}{y}\Bigr)^{p-2}
- \Biggl[ \frac1{\frac{\square}{H^2} \!+\! D} \Biggr]^2 } \nonumber \\
& & \hspace{.2cm} \times \Biggl\{ \frac{[D (2 p \!-\! 1) \!-\! 2 (p \!-\!
1)^2]}{(p \!-\! 1) p \!-\! \frac{D}2)} \Bigl( \frac{4}{y}\Bigr)^{p-1} \!\!\!+
\frac{(p \!-\! 1) (D \!+\! 2 \!-\! p)^2}{(p \!-\! 2) (p \!-\! 1 \!-\!
\frac{D}2) (p \!-\! \frac{D}2)} \Bigl(\frac{4}{y}\Bigr)^{p-2} \Biggr\} .\qquad
\end{eqnarray}
The strategy is to apply this until the source is integrable, at
which point the dimension can be set to $D = 4$ (unless there are factors
of $1/(D-4)$) and the $D=4$ Green's function can be used to obtain the
full solution for $\mathcal{F}_1(y)$.

It is useful to examine the sorts of terms generated when this recursive
procedure is applied to the source terms on the right hand side of
(\ref{F1eqn}). The most singular term introduces no factors of $1/(D-4)$,
nor does it produce remainder terms different from those in the original
source term (\ref{F1eqn}),
\begin{eqnarray}
\lefteqn{ \Biggl[ \frac1{\frac{\square}{H^2} \!+\! D} \Biggr]^2
\Bigl(\frac{4}{y}\Bigr)^D = \frac4{(D \!-\!2) D (D \!-\! 2) (D \!-\! 1)}
\Bigl(\frac{4}{y}\Bigr)^{D-2} } \nonumber \\
& & \hspace{1.4cm} -\Biggl[ \frac1{\frac{\square}{H^2} \!+\! D} \Biggr]^2
\Biggl\{ \frac{2 (3 D \!-\! 2)}{D (D \!-\! 1)} \Bigl(\frac{4}{y}\Bigr)^{D-1}
\!\!\! + \frac{16 (D \!-\! 1)}{(D \!-\! 2) D (D \!-\! 2)} \Bigl(\frac{4}{y}
\Bigr)^{D-2} \Biggr\} . \qquad
\end{eqnarray}
Neither statement is true for the remaining two source terms,
\begin{eqnarray}
\lefteqn{ \Biggl[ \frac1{\frac{\square}{H^2} \!+\! D} \Biggr]^2
\Bigl(\frac{4}{y}\Bigr)^{D-1} = \frac4{(D \!-\!4) (D \!-\! 2) (D \!-\! 3)
(D \!-\! 2)} \Bigl(\frac{4}{y}\Bigr)^{D-3} } \nonumber \\
& & \hspace{-.3cm} -\Biggl[ \frac1{\frac{\square}{H^2} \!+\! D} \Biggr]^2
\Biggl\{ \frac{2 (5 D \!-\! 8)}{(D \!-\! 2) (D \!-\! 2)}
\Bigl(\frac{4}{y}\Bigr)^{D-2} \!\!\! + \frac{36 (D \!-\! 2)}{(D \!-\! 4)
(D \!-\! 2) (D \!-\! 3)} \Bigl(\frac{4}{y} \Bigr)^{D-3} \Biggr\} , \qquad \\
\lefteqn{ \Biggl[ \frac1{\frac{\square}{H^2} \!+\! D} \Biggr]^2
\Bigl(\frac{4}{y}\Bigr)^{D-2} = \frac4{(D \!-\!6) (D \!-\! 4) (D \!-\! 4)
(D \!-\! 3)} \Bigl(\frac{4}{y}\Bigr)^{D-4} } \nonumber \\
& & \hspace{-.3cm} -\Biggl[ \frac1{\frac{\square}{H^2} \!+\! D} \Biggr]^2
\Biggl\{ \frac{2 (7 D \!-\! 18)}{(D \!-\! 4) (D \!-\! 3)}
\Bigl(\frac{4}{y}\Bigr)^{D-3} \!\!\! + \frac{64 (D \!-\! 3)}{(D \!-\! 6)
(D \!-\! 4) (D \!-\! 4)} \Bigl(\frac{4}{y} \Bigr)^{D-4} \Biggr\} . \qquad
\end{eqnarray}
These relations allow the the spin zero structure function to be
expressed as a ``quotient'' and a ``remainder'' of the form,
\begin{eqnarray}
\mathcal{F}_1(y) & = & \mathcal{Q}_1(y) + \Bigl[ \frac1{\frac{\square}{H^2}
\!+\! D} \Bigr]^2 \mathcal{R}_1(y) \; , \label{F1form} \\
\mathcal{Q}_1(y) & = & - K \Biggl\{ f_{1a} \Bigl(\frac{4}{y}\Bigr)^{D-2}
+ \frac{f_{1b}}{D \!-\! 4} \, \Bigl(\frac{4}{y}\Bigr)^{D-3}
+ \frac{f_{1c}}{(D \!-\! 4)^2} \Bigl(\frac{4}{y}\Bigr)^{D-4} \Biggr\}
\; , \qquad \label{Q1form} \\
\mathcal{R}_1(y) & = & -K \Biggl\{ \frac{f_{1d}}{D \!-\! 4} \, \Bigl(
\frac{4}{y}\Bigr)^{D-3} + \frac{f_{1e}}{(D \!-\! 4)^2} \Bigl(\frac{4}{y}
\Bigr)^{D-4} + \Bigl({\rm Irrelevant}\Bigr) \Biggr\} \; , \qquad \label{R1form}
\end{eqnarray}
where the coefficients are,
\begin{eqnarray}
f_{1a} & = & \frac1{8 (D \!-\! 1)^2} \; , \label{f1a} \\
f_{1b} & = & \frac{D (D^2 \!-\! 5 D \!+\!2)}{ 8 (D\!-\!3) (D\!-\!1)^2} \; ,
\label{f1b} \\
f_{1c} & = & \frac{D^2 (D^4 \!-\! 12 D^3 \!+\! 39 D^2 \!-\!16 D \!-\! 36)}{
16 (D \!-\! 6) (D \!-\!3) (D\!-\!1)^2} \; , \label{f1c} \\
f_{1d} & = & -\frac83 + \frac{79}9 (D \!-\! 4) + O\Bigl( (D \!-\! 4)^2\Bigr)
\; , \qquad \label{f1d} \\
f_{1e} & = & \frac{32}3 - \frac{64}9 (D \!-\! 4) - \frac{274}9 (D \!-\! 4)^2
+ O\Bigl( (D \!-\! 4)^3\Bigr) \; . \label{f1e} \qquad
\end{eqnarray}

Although the powers $y^{D-3}$ and $y^{D-4}$ in the remainder term of
(\ref{F1form}) are integrable, the factors of $1/(D - 4)$ they carry
preclude us setting $D=4$ and then obtaining an explicit form using
the $D=4$ Green's function. In the next subsection we will see how to
add zero so as to localize the divergences, and then absorb them into
counterterms. For now, let us assume $\mathcal{F}_1(y)$ has been
derived and explain the procedure for computing the spin two structure
function $\mathcal{F}_2(y)$.

The spin zero part of the graviton self-energy can be expressed as a
sum of the five de Sitter invariant bitensors times functions of $y$,
\begin{eqnarray}
\lefteqn{ \mathcal{P}^{\mu\nu}(x) \!\times\! \mathcal{P}^{\rho\sigma}(x')
\!\times\! \mathcal{F}_1(y) =
\frac{\partial^2 y}{\partial x_{\mu}\partial x'_{(\rho}}
\frac{\partial^2 y}{\partial x'_{\sigma)} \partial x_{\nu}}
\!\times\! \alpha_1(y) + \frac{\partial y}{\partial x_{(\mu}}
\frac{\partial^2 y}{\partial x_{\nu)}\partial x'_{(\rho}}
\frac{\partial y}{\partial x'_{\sigma)}} } \nonumber \\
& & \hspace{1cm} \times \beta_1(y) + \frac{\partial y}{\partial x_{\mu}}
\frac{\partial y}{\partial x_{\nu}} \frac{\partial y}{\partial x'_{\rho}}
\frac{\partial y}{\partial x'_{\sigma}} \!\times\! \gamma_1(y) + H^4
\overline{g}^{\mu\nu}(x) \overline{g}^{\rho\sigma}(x') \!\times\!
\delta_1(y) \nonumber \\
& & \hspace{3.5cm}
+ H^2 \Bigl[ \overline{g}^{\mu\nu}(x) \frac{\partial y}{\partial x'_{\rho}}
\frac{\partial y}{\partial x'_{\sigma}} \!+\! \frac{\partial y}{\partial
x_{\mu}} \frac{\partial y}{\partial x_{\nu}} \overline{g}^{\rho \sigma}(x')
\Bigl] \times \epsilon_1(y) \; , \qquad \label{3-point-A-F1}
\end{eqnarray}
Here the spin zero coefficient functions are,
\begin{eqnarray}
\alpha_1 &\!\!\!=\!\!\!&  2 \mathcal{F}_1'' \;, \\
\beta_1 &\!\!\!=\!\!\!& 4 \mathcal{F}_1''' \;, \\
\gamma_1 &\!\!\!=\!\!\!&  \mathcal{F}_1'''' \;, \\
\delta_1 &\!\!\!=\!\!\!& (4y\!-\!y^2)^2 \mathcal{F}_1'''' + 2 (D\!+\!1) (
2\!-\!y) (4y\!-\!y^2) \mathcal{F}_1''' - 4 (4y\!-\!y^2) \mathcal{F}_1''
\nonumber \\
&& \hspace{1.5cm} + (D^2\!-\!3) (2\!-\!y)^2 \mathcal{F}_1'' + (D\!-\!1)^2
(2\!-\!y) \mathcal{F}_1' + (D\!-\!1)^2 \mathcal{F}_1 \; , \qquad \\
\epsilon_1 &\!\!\!=\!\!\!& -(4y\!-\!y^2) \mathcal{F}_1'''' - (D\!+\!3)
(2\!-\!y) \mathcal{F}_1''' + (D\!+\!1) \mathcal{F}_1'' \; .
\end{eqnarray}
Of course the spin two contribution can be reduced to the same form,
\begin{eqnarray}
\lefteqn{ \mathcal{P}^{\mu\nu}_{\alpha\beta \gamma\delta}(x) \times
\mathcal{P}^{\rho\sigma}_{\kappa\lambda\theta\phi}(x') \times
\Biggl\{\mathcal{T}^{\alpha\kappa} \mathcal{T}^{\beta\lambda}
\mathcal{T}^{\gamma\theta} \mathcal{T}^{\delta\phi} \Bigl(
\frac{D \!-\! 2}{D \!-\! 3}\Bigr) \mathcal{F}_2(y) \Biggr\} } \nonumber \\
& & \hspace{1cm} = \frac{\partial^2 y}{\partial x_{\mu}\partial x'_{(\rho}}
\frac{\partial^2 y}{\partial x'_{\sigma)} \partial x_{\nu}}
\!\times\! \alpha_2(y) + \frac{\partial y}{\partial x_{(\mu}}
\frac{\partial^2 y}{\partial x_{\nu)}\partial x'_{(\rho}}
\frac{\partial y}{\partial x'_{\sigma)}} \times \beta_2(y) \nonumber \\
& & \hspace{2cm} + \frac{\partial y}{\partial x_{\mu}}
\frac{\partial y}{\partial x_{\nu}} \frac{\partial y}{\partial x'_{\rho}}
\frac{\partial y}{\partial x'_{\sigma}} \!\times\! \gamma_2(y) + H^4
\overline{g}^{\mu\nu}(x) \overline{g}^{\rho\sigma}(x') \!\times\!
\delta_2(y) \nonumber \\
& & \hspace{3.5cm}
+ H^2 \Bigl[ \overline{g}^{\mu\nu}(x) \frac{\partial y}{\partial x'_{\rho}}
\frac{\partial y}{\partial x'_{\sigma}} \!+\! \frac{\partial y}{\partial
x_{\mu}} \frac{\partial y}{\partial x_{\nu}} \overline{g}^{\rho \sigma}(x')
\Bigl] \times \epsilon_2(y) \; , \qquad \label{3-point-A-F2}
\end{eqnarray}
Determining the coefficient functions is an extremely tedious exercise
that was done by computer. The results for each coefficient function are
expressed as an expansion in powers of derivatives of the spin two
structure function, for example,
\begin{equation}
\alpha_2 = \sum_{k=0}^4 \alpha_{2k} \frac{d^k \mathcal{F}_2}{dy^k} \; .
\end{equation}
The various coefficients, which are functions of $D$ and $y$, are reported
in Tables~\ref{coF2}-\ref{coF2pppp}.

\begin{table}[ht]

\vbox{\tabskip=0pt \offinterlineskip
\def\tablerule{\noalign{\hrule}}
\halign to390pt {\strut#& \vrule#\tabskip=1em plus2em& \hfil#\hfil&
\vrule#& \hfil#\hfil&  \hfil#\hfil& & \vrule#\tabskip=0pt\cr
\tablerule \omit&height4pt&\omit&&\omit&&\cr
\omit&height2pt&\omit&&\omit&&\cr &&\omit\hidewidth $ \!\!\!\! $ \hidewidth &&
Coefficient of $F_2$
&&\cr
\omit&height4pt&\omit&&\omit&&\cr \tablerule
\omit&height2pt&\omit&&\omit&&\cr && $\alpha_{20}$ &&
$ -(D\!-\!3) D^2 (D\!+\!1)^2  \Bigl[-4 (D\!-\!2) + (D\!-\!1) (4 y \!-\! y^2)\Bigr] $
&&\cr
\omit&height2pt&\omit&&\omit&&\cr \tablerule
\omit&height2pt&\omit&&\omit&&\cr && $\beta_{20}$ &&
$ 2 (D\!-\!3) (D\!-\!1) D^2 (D\!+\!1)^2 (2 \!-\!y)$
&&\cr
\omit&height2pt&\omit&&\omit&&\cr \tablerule
\omit&height2pt&\omit&&\omit&&\cr && $\gamma_{20}$ &&
$ (D\!-\!3) (D\!-\!1) D^2 (D\!+\!1)^2$
&&\cr
\omit&height2pt&\omit&&\omit&&\cr \tablerule
\omit&height2pt&\omit&&\omit&&\cr && $\delta_{20}$ &&
$ 4 (D\!-\!3) D (D\!+\!1)^2 \Bigl[-4 (D\!-\!2) + D (4 y \!-\! y^2)\Bigr]$
&&\cr
\omit&height2pt&\omit&&\omit&&\cr \tablerule
\omit&height2pt&\omit&&\omit&&\cr && $\epsilon_{20}$ &&
$ -4 (D\!-\!3) D^2 (D\!+\!1)^2$
&& \cr
\omit&height2pt&\omit&&\omit&&\cr \tablerule}}

\caption{Coefficient of $F_2$: each term is multiplied
by $\frac{1}{16(D-2)(D-1)}$ }

\label{coF2}

\end{table}

\begin{table}[ht]

\vbox{\tabskip=0pt \offinterlineskip
\def\tablerule{\noalign{\hrule}}
\halign to390pt {\strut#& \vrule#\tabskip=1em plus2em& \hfil#\hfil&
\vrule#& \hfil#\hfil&  \hfil#\hfil& & \vrule#\tabskip=0pt\cr
\tablerule \omit&height4pt&\omit&&\omit&&\cr
\omit&height2pt&\omit&&\omit&&\cr &&\omit\hidewidth $ \!\!\!\! $ \hidewidth &&
Coefficient of $F'_2$
&&\cr
\omit&height4pt&\omit&&\omit&&\cr \tablerule
\omit&height2pt&\omit&&\omit&&\cr && $\!\!\alpha_{21}\!\!$ &&
$4 (D \!-\!3) (D\!+\!1)^2 (2 \!-\!y) \Bigl[-2 (D\!-\!2 ) D
   + (D\!-\!1) (D\!+\!1) (4 y \!-\! y^2)\Bigr] $
&&\cr
\omit&height2pt&\omit&&\omit&&\cr \tablerule
\omit&height2pt&\omit&&\omit&&\cr && $\!\!\beta_{21}\!\!$ &&
$8 (D \!-\!3) (D\!+\!1)^2  \Bigl[-3 D^2 + (D\!-\!1) (D\!+\!1) (4 y \!-\! y^2)\Bigr] $
&&\cr
\omit&height2pt&\omit&&\omit&&\cr \tablerule
\omit&height2pt&\omit&&\omit&&\cr && $\!\!\gamma_{21}\!\!$ &&
$-4 (D \!-\!3) (D\!-\!1) (D\!+\!1)^3 (2 \!-\!y) $
&&\cr
\omit&height2pt&\omit&&\omit&&\cr \tablerule
\omit&height2pt&\omit&&\omit&&\cr && $\!\!\delta_{21}\!\!$ &&
$-16 (D \!-\!3) (D\!+\!1)^2 (2 \!-\!y) \Bigl[-2 (D\!-\!2) + (D\!+\!1) (4 y \!-\! y^2)\Bigr] $
&&\cr
\omit&height2pt&\omit&&\omit&&\cr \tablerule
\omit&height2pt&\omit&&\omit&&\cr && $\!\!\epsilon_{21}\!\!$ &&
$16 (D \!-\!3) (D\!+\!1)^3 (2 \!-\!y) $
&& \cr
\omit&height2pt&\omit&&\omit&&\cr \tablerule}}

\caption{Coefficient of $F'_2$: each term is multiplied
by $\frac{1}{16(D-2)(D-1)}$}

\label{coF2p}

\end{table}

\begin{table}[ht]

\vbox{\tabskip=0pt \offinterlineskip
\def\tablerule{\noalign{\hrule}}
\halign to390pt {\strut#& \vrule#\tabskip=1em plus2em& \hfil#\hfil&
\vrule#& \hfil#\hfil&  \hfil#\hfil& & \vrule#\tabskip=0pt\cr
\tablerule \omit&height4pt&\omit&&\omit&&\cr
\omit&height2pt&\omit&&\omit&&\cr &&\omit\hidewidth $\!\!\!\! $ \hidewidth &&
Coefficient of $F''_2$
&&\cr
\omit&height4pt&\omit&&\omit&&\cr \tablerule
\omit&height2pt&\omit&&\omit&&\cr && $\!\!\alpha_{22}\!\!$ &&
$2\Bigl[8 (D\!-\!2 )^2 D (D \!+\! 1 ) -
   4 (D\!+\!1 ) (3 D^3\!-\!8 D^2\!-\!6 D\!+\!12) (4 y \!-\! y^2) $
&&\cr
\omit&height2pt&\omit&&\omit&&\cr &&  &&
$+ (D \!-\!3 ) (D\!-\!1 ) (3 D^2\!+\!9 D\!+\!7 ) (4 y - y^2)^2\Bigr] $
&&\cr
\omit&height2pt&\omit&&\omit&&\cr \tablerule
\omit&height2pt&\omit&&\omit&&\cr && $\!\!\beta_{22}\!\!$ &&
$-4 (2 \!-\! y)  \Bigl[-2 D (D\!+\!1) (3 D^2\!-\!5 D\!-\!10) $
&&\cr
\omit&height2pt&\omit&&\omit&&\cr &&  &&
$ + (D \!-\!3 ) (D\!-\!1) (3 D^2\!+\!9 D\!+\!7 ) (4 y \!-\! y^2) \Bigr] $
&&\cr
\omit&height2pt&\omit&&\omit&&\cr \tablerule
\omit&height2pt&\omit&&\omit&&\cr && $\!\!\gamma_{22}\!\!$ &&
$-2 \Bigl[-12 (D^4\!-\!D^3\!-\!7 D^2\!+\!D\!+\!10) $
&&\cr
\omit&height2pt&\omit&&\omit&&\cr &&  &&
$+ (D \!-\!3 ) (D\!-\!1) (3 D^2\!+\!9 D\!+\!72 ) (4 y \!-\! y^2)\Bigr] $
&&\cr
\omit&height2pt&\omit&&\omit&&\cr \tablerule
\omit&height2pt&\omit&&\omit&&\cr && $\!\!\delta_{22}\!\!$ &&
$-8 \Bigl[8 (D\!-\!2 )^2 (D\!+\!1 ) - 2 (D \!+\!1 ) (6 D^2\!-\!11
D\!-\!18) (4 y \!-\! y^2)$
&&\cr
\omit&height2pt&\omit&&\omit&&\cr &&  &&
$+ (D\!-\!3 ) (3 D^2\!+\!9 D\!+\!7 ) (4 y \!-\! y^2)^2\Bigr] $
&&\cr
\omit&height2pt&\omit&&\omit&&\cr \tablerule
\omit&height2pt&\omit&&\omit&&\cr && $\!\!\epsilon_{22}\!\!$ &&
$\!8 \Bigl[-2 (D\!+\!1 ) (5 D^2\!-\!6 D\!-\!24 )
+ (D\!-\!3 ) (3 D^2\!+\!9 D\!+\!7 ) (4 y \!-\!y^2)\Bigr] \!$
&& \cr
\omit&height2pt&\omit&&\omit&&\cr \tablerule}}

\caption{Coefficient of $F''_2$: each term is multiplied
by $\frac{1}{16(D-2)(D-1)}$}

\label{coF2pp}

\end{table}

\begin{table}[ht]

\vbox{\tabskip=0pt \offinterlineskip
\def\tablerule{\noalign{\hrule}}
\halign to390pt {\strut#& \vrule#\tabskip=1em plus2em& \hfil#\hfil&
\vrule#& \hfil#\hfil&  \hfil#\hfil& & \vrule#\tabskip=0pt\cr
\tablerule \omit&height4pt&\omit&&\omit&&\cr
\omit&height2pt&\omit&&\omit&&\cr &&\omit\hidewidth  \hidewidth &&
Coefficient of $F'''_2$
&&\cr
\omit&height4pt&\omit&&\omit&&\cr \tablerule
\omit&height2pt&\omit&&\omit&&\cr && $\!\!\!\alpha_{23}\!\!\!$ &&
$ -4 (D \!-\!1 ) (2\! -\! y) (4 y\! -\! y^2) \Bigl[-2 (D\!-\!2 ) (D\!+\!1 )$
&&\cr
\omit&height2pt&\omit&&\omit&&\cr &&  &&
$+ (D\!-\!3 ) (D\!+\!2 ) (4 y \!-\! y^2)\Bigr]\! $
&&\cr
\omit&height2pt&\omit&&\omit&&\cr \tablerule
\omit&height2pt&\omit&&\omit&&\cr && $\!\!\!\beta_{23}\!\!\!$ &&
$-8 \Bigl[4 (D\!-\!2 ) D (D\!+\!1) - (5 D^3\!-\!8 D^2\!-\!23 D\!+\!22)
(4 y \!-\! y^2)$ &&\cr
\omit&height2pt&\omit&&\omit&&\cr &&  &&
$ + (D\!-\!3) (D\!-\!1 ) (D\!+\!2) (4 y \!-\! y^2)^2\Bigr]\! $
&&\cr
\omit&height2pt&\omit&&\omit&&\cr \tablerule
\omit&height2pt&\omit&&\omit&&\cr && $\!\!\!\gamma_{23}\!\!\!$ &&
$\!4 (2 \!-\! y) \Bigl[-4 (D \!-\! 2) (D^2 \!-\! 5)
 + (D\!-\!3 ) (D\!-\!1 ) (D\! +\! 2 ) (4 y \!-\! y^2)\Bigr]\! $
&&\cr
\omit&height2pt&\omit&&\omit&&\cr \tablerule
\omit&height2pt&\omit&&\omit&&\cr && $\!\!\!\delta_{23} \!\!\!$ &&
$\!16 (2 \!-\! y) (4 y \!-\! y^2) \Bigl[-2 (D\!-\!2 ) (D\!+\!1 )
+ (D\!-\!3 ) (D\!+\!2 ) (4 y \!-\! y^2)\Bigr] $
&&\cr
\omit&height2pt&\omit&&\omit&&\cr \tablerule
\omit&height2pt&\omit&&\omit&&\cr && $\!\!\!\epsilon_{23}\!\!\!$ &&
$-16 (2 \!-\! y) \Bigl[-2 (D\!-\!2 ) ( D\!+\!1) + (D\!-\!3 ) (D\!+\!2 )
(4 y \!-\! y^2)\Bigr]\! $
&& \cr
\omit&height2pt&\omit&&\omit&&\cr \tablerule}}

\caption{Coefficient of $F'''_2$: each term is multiplied
by $\frac{1}{16(D-2)(D-1)}$}

\label{coF2ppp}

\end{table}

\begin{table}[ht]

\vbox{\tabskip=0pt \offinterlineskip
\def\tablerule{\noalign{\hrule}}
\halign to390pt {\strut#& \vrule#\tabskip=1em plus2em& \hfil#\hfil&
\vrule#& \hfil#\hfil&  \hfil#\hfil& & \vrule#\tabskip=0pt\cr
\tablerule \omit&height4pt&\omit&&\omit&&\cr
\omit&height2pt&\omit&&\omit&&\cr &&\omit\hidewidth  \hidewidth &&
Coefficient of $F''''_2$
&&\cr
\omit&height4pt&\omit&&\omit&&\cr \tablerule
\omit&height2pt&\omit&&\omit&&\cr && $\!\!\alpha_{24}\!\!$ &&
$-(D\!-\!1) (4 y \!-\! y^2)^2 \Bigl[-4 (D\!-\!2) + (D\!-\!3)
(4 y \!-\! y^2)\Bigr] $
&&\cr
\omit&height2pt&\omit&&\omit&&\cr \tablerule
\omit&height2pt&\omit&&\omit&&\cr && $\!\!\beta_{24}\!\!$ &&
$2 (D\!-\!1 ) (2 \!-\! y) (4 y \!-\! y^2) \Bigl[-4 (D\!-\!2 ) + (D\!-\!3)
(4 y \!-\! y^2)\Bigr] $
&&\cr
\omit&height2pt&\omit&&\omit&&\cr \tablerule
\omit&height2pt&\omit&&\omit&&\cr && $\!\!\gamma_{24}\!\!$ &&
$\Bigl[4 (D\!-\!2 ) - (D\!-\!3 ) (4 y \!-\! y^2)\Bigr]
\Bigl[4 (D\!-\!2 ) - (D\!-\!1 ) (4 y \!-\! y^2)\Bigr] $
&&\cr
\omit&height2pt&\omit&&\omit&&\cr \tablerule
\omit&height2pt&\omit&&\omit&&\cr && $\!\!\delta_{24}\!\!$ &&
$4 (4 y \!-\! y^2)^2 \Bigl[-4 (D\!-\!2) + (D\!-\!3) (4 y \!-\! y^2)\Bigr] $
&&\cr
\omit&height2pt&\omit&&\omit&&\cr \tablerule
\omit&height2pt&\omit&&\omit&&\cr && $\!\!\epsilon_{24}\!\!$ &&
$-4 (4 y \!-\! y^2) \Bigl[-4 (D\!-\!2) + (D\!-\!3) (4 y \!-\! y^2)\Bigr] $
&& \cr
\omit&height2pt&\omit&&\omit&&\cr \tablerule}}

\caption{Coefficient of $F''''_2$: each term is multiplied
by $\frac{1}{16(D-2)(D-1)}$}

\label{coF2pppp}

\end{table}

Now recall the second order equation (\ref{2ndorder}) we were able to
find for the flat space structure function $F_2(\Delta x^2)$ by adding
$\delta$ and $\Delta x^2 \epsilon$. After long contemplation of the
bewildering data in Tables~\ref{coF2}-\ref{coF2pppp} it becomes
apparent that a similar second order equation for $\mathcal{F}_2(y)$
derives from the combination,
\begin{eqnarray}
\lefteqn{ \delta_2(y) + (4 y \!-\! y^2) \epsilon_2(y) = \Bigl[\delta(y) \!-\!
\delta_1(y)\Bigr] + (4 y \!-\! y^2) \Bigl[\epsilon(y) \!-\! \epsilon_1(y)
\Bigr] \; , } \\
& & = -\Bigl(\frac{D \!+\! 1}{D \!-\! 1}\Bigr) \Biggl\{
(D \!-\! 2)  \mathcal{F}_2'' - (D \!-\! 3) \Biggl[ (4 y \!-\! y^2)
\mathcal{F}_2'' \nonumber \\
& & \hspace{5.5cm} + 2 (D \!+\! 1) (2 \!-\! y) \mathcal{F}_2' - D (D \!+\! 1)
\mathcal{F}_2\Biggr\} . \qquad
\end{eqnarray}
Hence we can express the equation for $\mathcal{F}_2(y)$ as,
\begin{equation}
\mathcal{D} \mathcal{F}_2 = -\Bigl(\frac{D \!- \! 1}{D \!+\! 1}\Bigr)
\Biggl\{ \Bigl[\delta(y) \!-\! \delta_1(y)\Bigr] + (4 y \!-\! y^2)
\Bigl[\epsilon(y) \!-\! \epsilon_1(y) \Bigr] \Biggr\} , \label{F2eqn}
\end{equation}
where the second order operator $\mathcal{D}$ is,
\begin{eqnarray}
\lefteqn{\mathcal{D} \equiv 4 (D \!-\! 2) \Bigl(\frac{d}{dy}\Bigr)^2 }
\nonumber \\
& & - (D \!-\! 3) \Biggl[ (4 y \!-\! y^2) \Bigl(\frac{d}{dy}
\Bigr)^2 + 2 (D \!+\! 1) (2 \!-\! y) \frac{d}{d y} - D (D \!+\! 1)\Biggr]
\; , \qquad \label{1stform} \\
& & \hspace{-.5cm} = 4 \Bigl(\frac{d}{dy}\Bigr)^2 + (D \!-\! 3) \Biggl[
(2 \!-\! y)^2 \Bigl(\frac{d}{dy}\Bigr)^2 - 2 (D \!+\! 1) (2 \!-\! y)
\frac{d}{d y} + D (D \!+\! 1)\Biggr] \; . \qquad \label{2ndform}
\end{eqnarray}
The source term on the right hand side of (\ref{F2eqn}) has the form,
\begin{eqnarray}
\lefteqn{ -\Bigl(\frac{D \!- \! 1}{D \!+\! 1}\Bigr) \Biggl\{ \Bigl[\delta(y)
\!-\! \delta_1(y)\Bigr] + (4 y \!-\! y^2) \Bigl[\epsilon(y) \!-\!
\epsilon_1(y) \Bigr] \Biggr\} } \nonumber \\
& & \hspace{.5cm} = K \Biggl\{ s_a \Bigl(\frac{4}{y}\Bigr)^D + \frac{s_b}{D
\!-\! 4} \Bigl(\frac{4}{y}\Bigr)^{D-1} + \frac{s_c}{D \!-\! 4} \Bigl(
\frac{4}{y} \Bigr)^{D-2} + s_{c'} \Bigl(\frac{4}{y}\Bigr)^{\frac{D}2}
\nonumber \\
& & \hspace{1.5cm} + \frac{s_d}{D \!-\! 4} \Bigl(\frac{4}{y} \Bigr)^{D-3}
+ \frac{s_e}{(D \!-\! 4)^2} \Bigl(\frac{4}{y} \Bigr)^{D-4} + \Bigl({\rm
Irrelevant}\Bigr) \Biggr\} + \mathcal{R} \; , \qquad \label{F2source}
\end{eqnarray}
where the remainder term $\mathcal{R}$ derives from the remainder
$\mathcal{R}_1$ of $\mathcal{F}_1$,
\begin{eqnarray}
\lefteqn{ \mathcal{R} = \Bigl(\frac{D \!-\! 1}{D \!+\! 1}\Bigr) \Biggl\{
(D \!-\! 1) (2 \!-\! y) (4 y \!-\! y^2) \Bigl(\frac{\partial}{\partial y}
\Bigr)^3 - D (D \!-\! 1) (4 y \!-\! y^2) \Bigl( \frac{\partial}{\partial y}
\Bigr)^2 } \nonumber \\
& & \hspace{0cm} + 4 (D^2 \!-\! 3) \Bigl( \frac{\partial}{\partial y}\Bigr)^2
+ (D \!-\! 1)^2 (2 \!-\! y) \Bigl(\frac{\partial}{\partial y} \Bigr) +
(D \!-\! 1)^2 \Biggr\} \Biggl[ \frac1{\frac{\square}{H^2} \!+\! D} \Biggr]^2
\, \mathcal{R}_1 \; . \qquad
\end{eqnarray}
The coefficients in (\ref{F2source}) are,
\begin{eqnarray}
s_a & = & -\frac1{16 (D \!+\! 1)} \; , \\
s_b & = & -\frac{(D \!-\! 2) D}{16 (D \!-\!1)} \; , \\
s_c & = & - \frac{(D \!-\! 4) (D \!-\! 2) D (D \!+\! 3)}{32 (D\!-\!6)
(D \!-\! 1)} \; , \\
s_{c'} & = & - \frac{(D \!-\!4) (D \!-\! 1) \Gamma(D)}{16 (D \!+\!1)
\Gamma(\frac{D}2) \Gamma(\frac{D}2 \!+\! 1)} \; , \\
s_d & = & -\frac75 + \frac{263}{100} (D \!-\! 4) + O\Bigl( (D \!-\! 4)^2\Bigr)
\; , \\
s_e & = & \frac{18}5 - \frac{18}{25} (D \!-\! 4) - \frac{11331}{1000}
(D \!-\! 4)^2 + O\Bigl( (D \!-\! 4)^3\Bigr) \; .
\end{eqnarray}

Just as for the differential operator $(\frac{\square}{H^2} + D)$, it
is straightforward to construct a Green's function to invert $\mathcal{D}$.
The first step is to change variables in the second form (\ref{2ndform}),
\begin{equation}
w \equiv \sqrt{ \frac{D \!-\! 3}4 } \, (2 \!-\! y) \Longrightarrow
\mathcal{D} = (D \!-\! 3) \Bigl[(1 \!+\! w^2) \Bigl(\frac{d}{d w}\Bigr)^2 +
2 (D \!+\! 1) w \frac{d}{d w} + D (D \!+\! 1)\Bigr] \; .
\end{equation}
The homogeneous equation $\mathcal{D} f(w) = 0$ gives rise to a simple,
2-term recursion relation which generates even and odd solutions. These
series solutions can be expressed as hypergeometric functions that reduce
to elementary functions for $D = 4$,
\begin{eqnarray}
f_e(w) & = & \mbox{}_2 F_1\Bigl(\frac{D}2, \frac{D \!+\! 1}2 ; \frac12 ;
w^2\Bigr) \longrightarrow \frac{(1 \!-\! 6 w^2 \!+\! w^4)}{(1 \!+\! w^2)^4}
\; , \\
f_o(w) & = & w \times \mbox{}_2 F_1\Bigl(\frac{D \!+\! 1}2, \frac{D \!+\! 2}2 ;
\frac32 ; w^2\Bigr) \longrightarrow \frac{(w \!-\! w^3)}{(1 \!+\! w^2)^4} \; .
\end{eqnarray}
Because we again have both homogeneous solutions it is simple to
write down a Green's function,
\begin{equation}
G_2(w;w') = \frac{ \theta(w \!-\! w')}{D \!-\! 3} \Bigl[ f_o(w) f_e(w')
\!-\! f_e(w) f_o(w')\Bigr] (1 \!+\! w^{\prime 2})^D \; . \label{G2}
\end{equation}

As was the case for it spin zero cousin (\ref{Gfunc}), the spin two
Green's function (\ref{G2}) is not simple to use for arbitrary $D$. We
therefore adopt the same strategy we used for $\mathcal{F}_1$, of
recursively extracting powers until the remainder is integrable and
the $D = 4$ forms can be employed. Acting $\mathcal{D}$ on a power gives,
\begin{eqnarray}
\lefteqn{\mathcal{D} \Bigl(\frac{4}{y}\Bigr)^{p-2} = \frac14 (D \!-\! 2)
(p \!-\! 2) (p \!-\! 1) \Bigl(\frac{4}{y}\Bigr)^p } \nonumber \\
& & \hspace{-.5cm} + (D \!-\! 3) (p \!-\! 2) (D \!+\! 2 \!-\! p)
\Bigl(\frac{4}{y}\Bigr)^{p-1} \!\!\! + (D \!-\! 3) (D \!+\! 2 \!-\! p)
(D \!+\! 3 \!-\! p) \Bigl(\frac{4}{y}\Bigr)^{p-2} \!\! . \qquad
\end{eqnarray}
Hence we conclude,
\begin{eqnarray}
\lefteqn{\frac1{\mathcal{D}} \Bigl(\frac{4}{y}\Bigr)^p = \frac{4}{(D \!-\!2)
(p \!-\! 2)(p \!-\! 1)} \Bigl(\frac{4}{y}\Bigr)^{p-2} } \nonumber \\
& & \hspace{-.3cm} - \frac4{\mathcal{D}} \Biggl\{ \frac{(D \!-\! 3)
(D \!+\! 2 \!-\! p)}{(D \!-\! 2) (p \!-\! 1)} \Bigl(\frac{4}{y}\Bigr)^{p-1}
\!\!\!+ \frac{(D \!-\! 3) (D \!+\! 2 \!-\! p) (D \!+\! 3 \!-\! p)}{(D \!-\! 2)
(p \!-\! 2) (p \!-\! 1)} \Bigl(\frac{4}{y}\Bigr)^{p-2} \Biggr\} . \qquad
\label{recurs2}
\end{eqnarray}

For the four powers of relevance expression (\ref{recurs2}) gives,
\begin{eqnarray}
\lefteqn{\frac1{\mathcal{D}} \Bigl(\frac{4}{y}\Bigr)^D = \frac{4}{(D \!-\!2)^2
(D \!-\! 1)} \Bigl(\frac{4}{y}\Bigr)^{D-2} } \nonumber \\
& & \hspace{2cm} - \frac1{\mathcal{D}} \Biggl\{ \frac{8 (D \!-\! 3)
}{(D \!-\! 2) (D \!-\! 1)} \Bigl(\frac{4}{y}\Bigr)^{D-1} +
\frac{24 (D \!-\! 3)}{(D \!-\! 2)^2 (D \!-\! 1)} \Bigl(\frac{4}{y}\Bigr)^{D-2}
\Biggr\} , \qquad \\
\lefteqn{\frac1{\mathcal{D}} \Bigl(\frac{4}{y}\Bigr)^{D-1} =
\frac{4}{(D \!-\!3) (D \!-\! 2)^2} \Bigl(\frac{4}{y}\Bigr)^{D-3} } \nonumber \\
& & \hspace{3cm} - \frac1{\mathcal{D}} \Biggl\{ \frac{12 (D \!-\! 3)
}{(D \!-\! 2)^2} \Bigl(\frac{4}{y}\Bigr)^{D-2} +
\frac{48}{(D \!-\! 2)^2 } \Bigl(\frac{4}{y}\Bigr)^{D-3} \Biggr\} , \qquad \\
\lefteqn{\frac1{\mathcal{D}} \Bigl(\frac{4}{y}\Bigr)^{D-2} =
\frac{4}{(D \!-\! 4) (D \!-\!3) (D \!-\! 2)} \Bigl(\frac{4}{y}\Bigr)^{D-4} }
\nonumber \\
& & \hspace{3cm} - \frac1{\mathcal{D}} \Biggl\{ \frac{16}{(D \!-\! 2)}
\Bigl(\frac{4}{y}\Bigr)^{D-3} + \frac{80}{(D \!-\! 4) (D \!-\! 2)}
\Bigl(\frac{4}{y}\Bigr)^{D-4} \Biggr\} , \qquad \\
\lefteqn{\frac1{\mathcal{D}} \Bigl(\frac{4}{y}\Bigr)^{\frac{D}2} =
\frac{16}{(D \!-\! 4) (D \!-\! 2)^2} \Bigl(\frac{4}{y}\Bigr)^{\frac{D}2-2} }
\nonumber \\
& & \hspace{1cm} - \frac4{\mathcal{D}} \Biggl\{ \frac{(D \!-\! 3)
(D \!+\! 4)}{(D \!-\! 2)^2} \Bigl(\frac{4}{y}\Bigr)^{\frac{D}2 - 1} \!+
\frac{(D \!-\! 3) (D \!+\! 4) (D \!+\! 6)}{(D \!-\! 4) (D \!-\! 2)^2}
\Bigl(\frac{4}{y}\Bigr)^{\frac{D}2 - 2} \Biggr\} . \qquad \label{D/2}
\end{eqnarray}
These relations allow the spin two structure function to be
expressed as a ``quotient'' and ``remainder'' of the form,
\begin{eqnarray}
\mathcal{F}_2 &\!\!\! =\!\!\! & \mathcal{Q}_2(y) + \frac1{\mathcal{D}}
\mathcal{R}_2(y) \; , \label{F2form} \\
\mathcal{Q}_2 &\!\!\! =\!\!\! & - K \Biggl\{ f_{2a} \Bigl(\frac{4}{y}
\Bigr)^{D-2} \!\!\!\!\! + \frac{f_{2b}}{D \!-\! 4} \, \Bigl(\frac{4}{y}
\Bigr)^{D-3} \!\!\!\!\!+ \frac{f_{2c}}{(D \!-\! 4)^2} \Bigl(\frac{4}{y}
\Bigr)^{D-4} \!\!\!\!\!+ \frac{f_{2c'}}{D \!-\! 4} \Bigl(\frac{4}{y}
\Bigr)^{\frac{D}2-2} \Biggr\} , \qquad \label{Q2} \\
\mathcal{R}_2 &\!\!\! =\!\!\! & -K \Biggl\{ \frac{f_{2d}}{D \!-\! 4} \,
\Bigl(\frac{4}{y}\Bigr)^{D-3} + \frac{f_{2e}}{(D \!-\! 4)^2} \Bigl(\frac{4}{y}
\Bigr)^{D-4} + \Bigl({\rm Irrelevant}\Bigr) \Biggr\} + \mathcal{R} \; ,
\qquad \label{R2}
\end{eqnarray}
where the coefficients are,
\begin{eqnarray}
\lefteqn{f_{2a} = \frac{1}{4 (D-2)^2 (D-1) (D+1)} \; , } \label{f2a} \\
\lefteqn{f_{2b} = \frac{D^4 \!-\! 3 D^3 \!-\! 8 D^2\!+\! 60 D\!-\! 96}
{4 (D \!-\! 3) (D\!-\!2)^3 (D\!-\!1) (D\!+\!1)} \; , } \label{f2b} \\
\lefteqn{f_{2c} = } \nonumber \\
& & \hspace{-.7cm} \frac{D^8 \!\!\!-\! 8 D^7 \!\!\!-\! 13 D^6 \!\!+\! 348 D^5
\!\!\!-\! 1136 D^4 \!\!\!-\! 2^{10} D^3 \!\!+\! 15056 D^2 \!\!\!-\! 38208 D
\!\!+\! 34560}{8 (D\!-\!6) (D\!-\!3) (D\!-\!2)^4 (D\!-\!1) (D\!+\!1)} ,
\label{f2c} \\
\lefteqn{f_{2c'} = \frac{(D \!-\! 4) (D \!-\! 1) \Gamma(D)}{(D \!-\! 2)^2 
(D \!+\! 1) \Gamma(\frac{D}2) \Gamma(\frac{D}2 \!+\! 1)} \; , } \label{f2c'} \\
\lefteqn{f_{2d} = \frac{17}5 + \frac{161}{300} (D \!-\! 4) +
O\Bigl( (D \!-\! 4)^2\Bigr) \; , } \nonumber \label{f2d} \\
\lefteqn{f_{2e} = \frac{82}5 + \frac{243}{25} (D \!-\! 4) + \frac{13343}{3000}
(D \!-\! 4)^2 + O\Bigl( (D \!-\! 4)^3\Bigr) \; . } \label{f2e}
\end{eqnarray}

\subsection{Renormalizing the Spin Zero Structure Function}

Recall the form (\ref{F1form}) we obtained for the spin zero struncture
function from taking the trace of the graviton self-energy,
\begin{equation}
\mathcal{F}_1(y) = \mathcal{Q}_1(y) + \Bigl[ \frac1{\frac{\square}{H^2}
\!+\! D} \Bigr]^2 \mathcal{R}_1(y) \; .
\end{equation}
Recall also that the quotient $\mathcal{Q}_1(y)$ and the remainder
$\mathcal{R}_1(y)$ are given in relations (\ref{Q1form}-\ref{f1e}).
>From these expressions we perceive three sorts of ultraviolet divergences:
\begin{itemize}
\item{The factor of $(\frac{4}{y})^{D-2}$ in $\mathcal{Q}_1$, which has a
finite coefficient but is still not integrable in $D=4$ dimensions;}
\item{The factors of $\frac1{D - 4} (\frac{4}{y})^{D-3}$ in $\mathcal{Q}_1$
and $\mathcal{R}_1$ which are integrable in $D = 4$ dimensions but have
divergent coefficients that preclude taking the unregulated limits; and}
\item{The factors of $(\frac1{D - 4})^2 (\frac{4}{y})^{D-4}$ in $\mathcal{Q}_1$
and $\mathcal{R}_1$ which are integrable in $D = 4$ dimensions but have even
more divergent coefficients.}
\end{itemize}
In this subsection we will explain how to localize all three divergences
onto delta function terms which can be absorbed by the counterterms
(\ref{C1}), (\ref{C3}) and (\ref{C4}). We will also take the unregulated
limits of the remaining, finite parts, and use the $D=4$ Green's function
(\ref{Gfunc}) to obtain an explicit result for the renormalized structure
function.

In dealing with the factor of $(\frac{4}{y})^{D-2}$ in $\mathcal{Q}_1$,
the first step is to extract a d'Alembertian,
\begin{equation}
\Bigl(\frac{4}{y}\Bigr)^{D-2} = \frac{2}{(D \!-\! 4) (D \!-\! 3)}
\Biggl[ \frac{\square}{H^2} \Bigl(\frac{4}{y}\Bigr)^{D-3} - 2 (D \!-\! 3)
\Bigl(\frac{4}{y}\Bigr)^{D-3} \Biggr] \; . \label{finalext}
\end{equation}
The resulting factors of $(\frac{4}{y})^{D-3}$ are integrable in $D = 4$
dimensions, at which point we could take the unregulated limit except for
the factor of $1/(D - 4)$ in (\ref{finalext}). We can localize the divergence
on a delta function by adding zero in the form of the identity (\ref{ZERO}),
\begin{eqnarray}
\lefteqn{ \Bigl(\frac{4}{y}\Bigr)^{D-2} = \frac2{(D \!-\! 4) (D \!-\! 3)}
\Biggl\{ \frac{\square}{H^2} \Biggl[ \Bigl(\frac{4}{y}\Bigr)^{D-3} \!\!\!-
\Bigl(\frac{4}{y}\Bigr)^{\frac{D}2 -1} \Biggr] } \nonumber \\
& & \hspace{.5cm} -2 (D \!-\! 3) \Bigl(\frac{4}{y}\Bigr)^{D-3} +
\frac{D}2 \Bigl(\frac{D}2 \!-\! 1\Bigr) \Bigl(\frac{4}{y}\Bigr)^{\frac{D}2 -1}
+ \frac{(4\pi)^{\frac{D}2}}{\Gamma(\frac{D}2 \!-\! 1)}
\frac{i \delta^D(x \!-\! x')}{H^{D} \sqrt{-\overline{g}}} \Biggr\} \; , \\
& & \hspace{-.5cm} = - \Bigl[ \frac{\square}{H^2} \!-\! 2\Bigr] \Bigl\{
\frac{4}{y} \ln\Bigl(\frac{y}{4}\Bigr) \Bigr\} - \frac{4}{y} +
O(D \!-\! 4) + \frac{2 (4\pi)^{\frac{D}2} i \delta^D(x \!-\! x')/\!
\sqrt{-\overline{g}}}{(D \!-\! 4) (D \!-\! 3) \Gamma(\frac{D}2 \!-\! 1) H^D}
\; . \qquad \label{lead}
\end{eqnarray}

We turn now to the factors of $\frac1{D-4} (\frac{4}{y})^{D-3}$ and
$(\frac1{D-4})^2 (\frac{4}{y})^{D-4}$ in $\mathcal{Q}_1$ and $\mathcal{R}_1$.
The key relations for resolving these terms follow from (\ref{ZERO}),
\begin{eqnarray}
\lefteqn{\Bigl[ \frac{\square}{H^2} \!+\! D\Bigr]^2
\Bigl(\frac{4}{y}\Bigr)^{\frac{D}2 -1} = \frac1{16} D^2 (D \!+\! 2)^2
\Bigl( \frac{4}{y}\Bigr)^{\frac{D}2 -1} } \nonumber \\
& & \hspace{1cm} + \frac{(4\pi)^{\frac{D}2}}{
\Gamma(\frac{D}2 \!-\! 1) H^D \sqrt{-\overline{g}}} \Biggl[
\frac{\square}{H^2} \!+\! D \!+\! \frac14 D (D \!+\! 2)\Biggr]
i \delta^D(x \!-\! x') \; , \qquad \label{key1} \\
\lefteqn{\Bigl[ \frac{\square}{H^2} \!+\! D\Bigr]^2
\Bigl(\frac{4}{y}\Bigr)^{\frac{D}2 -2} = -\frac14 (D \!-\! 4)
(D^2 \!+\! 2 D \!-\! 4) \Bigl( \frac{4}{y}\Bigr)^{\frac{D}2 -1} }
\nonumber \\
& & \hspace{1cm} + \frac1{16} (D \!-\! 2)^2 (D \!+\! 4)^2
\Bigl(\frac{4}{y}\Bigr)^{\frac{D}2-2} - \frac{(D \!-\! 4)
(4\pi)^{\frac{D}2}}{2 \Gamma(\frac{D}2 \!-\! 1) H^D \sqrt{-\overline{g}}}
i \delta^D(x \!-\! x') \; , \qquad \label{key2} \\
\lefteqn{\Bigl[ \frac{\square}{H^2} \!+\! D\Bigr]^2 1 = D^2 \; . }
\end{eqnarray}
One add zero using these relations so as to resolve the problematic terms in
$\mathcal{Q}_1$, and the remainder automatically resolves the problematic
terms in $\mathcal{R}_1$,
\begin{eqnarray}
\lefteqn{\frac{f_{1b}}{D \!-\! 4} \Bigl(\frac{4}{y}\Bigr)^{D-3} \!\!\!\!\!+
\frac{f_{1c}}{(D \!-\! 4)^2} \Bigl(\frac{4}{y}\Bigr)^{D-4} \!\!\!\!\!+
\Biggl[ \frac1{\frac{\square}{H^2} \!+\! D} \Biggr]^2 \Biggl\{
\frac{f_{1d}}{D \!-\! 4} \Bigl( \frac{4}{y}\Bigr)^{D-3} \!\!\!\!\!+
\frac{f_{1e}}{(D \!-\! 4)^2} \Bigl( \frac{4}{y}\Bigr)^{D-4} \Biggr\} }
\nonumber \\
& & \hspace{-.7cm} = \frac{f_{1b}}{D \!-\! 4} \Biggl\{ \Bigl( \frac{4}{y}
\Bigr)^{D-3} \!\!\!\!\!- \Bigl(\frac{4}{y}\Bigr)^{\frac{D}2 -1} \Biggr\} +
\frac{f_{1c}}{(D \!-\! 4)^2} \Biggl\{ \Bigl( \frac{4}{y} \Bigr)^{D-4}
\!\!\!\!\!- 2 \Bigl(\frac{4}{y}\Bigr)^{\frac{D}2 -2} \!\!\!\!\!+ 1 \Biggr\}
\nonumber \\
& & \hspace{-.4cm} + \Biggl[ \frac1{\frac{\square}{H^2} \!+\! D} \Biggr]^2
\Biggl\{ \frac{f_{1d}}{D \!-\! 4} \Bigl(\frac{4}{y}\Bigr)^{D-3} \!\!\!+
\frac{[D^2 (D \!+\! 2)^2 f_{1b} \!-\! 8 (D^2 \!+\! 2 D \!-\! 4) f_{1c}]}{16
(D \!-\! 4)} \Bigl(\frac{4}{y} \Bigr)^{\frac{D}2 -1} \nonumber \\
& & \hspace{-.2cm} + \frac{f_{1e}}{(D \!-\! 4)^2} \Bigl(\frac{4}{y}\Bigr)^{D-4}
\!\!\!+ \frac{(D \!-\! 2)^2 (D \!+\! 4)^2 f_{1c}}{8 (D \!-\! 4)^2}
\Bigl(\frac{4}{y}\Bigr)^{\frac{D}2 -2} \!\!\!- \frac{D^2 f_{1c}}{(D \!-\! 4)^2}
\nonumber \\
& & \hspace{0cm} + \frac{(4\pi)^{\frac{D}2}/\sqrt{-\overline{g}}}{
\Gamma(\frac{D}2 \!-\! 1) H^D } \Biggl[ \frac{f_{1b}}{D \!-\! 4}
\Bigl[\frac{\square}{H^2} \!+\! D\Bigr] + \frac{D (D \!+\! 2) f_{1b} \!-\!
4 f_{1c}}{4 (D \!-\! 4)}  \Biggr] i \delta^D(x \!-\! x') \Biggr\} , \qquad \\
& & \hspace{-.7cm} = \frac1{18} \times \frac{4}{y} \ln\Bigl(\frac{y}{4}\Bigr)
\!-\! \frac16 \times \ln^2\Bigl(\frac{y}{4}\Bigr) + O(D \!-\!4) + \Biggl[
\frac1{\frac{\square}{H^2} \!+\! 4} \Biggr]^2 \Biggl\{ \frac43 \times
\frac{4}{y} \ln\Bigl(\frac{y}{4}\Bigr) \nonumber \\
& & \hspace{1cm} + \frac83 \times \frac{4}{y} \!+\! \frac83
\ln^2\Bigl(\frac{y}{4}\Bigr) \!-\! 8 \ln\Bigl(\frac{y}{4}\Bigr) \!+\!
\frac13 \Biggr\} + \Biggl[ \frac1{\frac{\square}{H^2} \!+\! D} \Biggr]^2
\Biggl\{ \frac{(4\pi)^{\frac{D}2}/\sqrt{-\overline{g}}}{
\Gamma(\frac{D}2 \!-\! 1) H^D } \nonumber \\
& & \hspace{2.5cm} \times \Biggl[ \frac{f_{1b}}{D \!-\! 4}
\Bigl[\frac{\square}{H^2} \!+\! D\Bigr] + \frac{D (D \!+\! 2) f_{1b} \!-\!
4 f_{1c}}{4 (D \!-\! 4)}  \Biggr] i \delta^D(x \!-\! x') \Biggr\} . \qquad
\label{sub}
\end{eqnarray}

Employing expressions (\ref{lead}) and (\ref{sub}) in (\ref{F1form})
allows us to separate the spin zero structure function into a finite part
and a divergent part,
\begin{equation}
\mathcal{F}_1 = \mathcal{F}_{1R} + O(D \!-\! 4) + \Delta \mathcal{F}_1 \; .
\end{equation}
The finite part consists of the renormalized spin zero structure function,
\begin{eqnarray}
\lefteqn{\mathcal{F}_{1R} = \frac{\kappa^2 H^4}{(4\pi)^4} \Biggl\{
\frac{\square}{H^2} \Biggl[ \frac1{72} \!\times\! \frac{4}{y} \ln\Bigl(
\frac{y}4\Bigr) \Biggr] \!-\! \frac1{12} \!\times\! \frac{4}{y} \ln\Bigl(
\frac{y}{4}\Bigr) \!+\! \frac1{72} \!\times\! \frac{4}{y} \!+\! \frac16
\ln^2\Bigl(\frac{y}4\Bigr) \Biggr\} } \nonumber \\
& & \hspace{0cm} + \frac{\kappa^2 H^4}{(4 \pi)^4}
\Biggl[ \frac1{\frac{\square}{H^2} \!+\! 4} \Biggr]^2 \Biggl\{ -\frac43
\!\times\! \frac{4}{y} \ln\Bigl(\frac{y}{4}\Bigr) \!-\! \frac83 \!\times\!
\frac{4}{y} \!-\! \frac{8}3 \ln^2\Bigl(\frac{y}{4} \Bigr) \!+\! 8
\ln\Bigl(\frac{y}{4}\Bigr) \!-\! \frac13\Biggr\} . \qquad \label{F1R}
\end{eqnarray}
The divergent part consists of $[\frac{\square}{H^2} + D]^{-2}$ acting
on a sum of three local terms,
\begin{eqnarray}
\lefteqn{ \Delta \mathcal{F}_1 = \frac{\kappa^2 H^{D-4}(\frac{D}2 \!-\! 1)
\Gamma(\frac{D}2)}{(4 \pi)^{\frac{D}2}} \Biggl[ \frac1{\frac{\square}{H^2}
\!+\! D}\Biggr]^2 \!\! \Biggl\{ \frac{-2 f_{1a}}{(D \!-\! 4)(D \!-\! 3)}
\Biggl[ \frac{\square}{H^2} \!+\! D \Biggr]^2 \frac{i \delta^D(x \!-\! x')}{
\sqrt{-\overline{g}}} } \nonumber \\
& & \hspace{.5cm} - \frac{f_{1b}}{D \!-\! 4} \Biggl[ \frac{\square}{H^2}
\!+\! D\Biggr] \frac{i \delta^D(x \!-\! x')}{\sqrt{-\overline{g}}}
- \Biggl[ \frac{D (D \!+\! 2) f_{1b} \!-\! 4 f_{1c}}{4 (D \!-\! 4)} \Biggr]
\frac{i \delta^D(x \!-\! x')}{\sqrt{-\overline{g}}} \Biggr\} . \qquad
\end{eqnarray}

Of course one cancels $\Delta \mathcal{F}_1$ with counterterms. From
expressions (\ref{C1}-\ref{C4}) we wee that the four counterterms contribute
to the graviton self-energy as,
\begin{eqnarray}
\lefteqn{-i \Bigl[\mbox{}^{\mu\nu} \Delta \Sigma^{\rho\sigma}\Bigr](x;x')
= \sqrt{-\overline{g}} \Biggl[ 2 c_1 \kappa^2  \mathcal{P}^{\mu\nu}
\mathcal{P}^{\rho\sigma} + 2 c_2 \kappa^2 \overline{g}^{\alpha\kappa}
\overline{g}^{\beta\lambda} \overline{g}^{\gamma\theta}
\overline{g}^{\delta\phi} \mathcal{P}^{\mu\nu}_{\alpha\beta\gamma\delta}
\mathcal{P}^{\rho\sigma}_{\kappa\lambda\theta\phi} } \nonumber \\
& & \hspace{.5cm} - c_3 \kappa^2 H^2 \mathcal{D}^{\mu\nu\rho\sigma}
+ c_4 \kappa^2 H^4 \sqrt{-\overline{g}} \, \Bigl[ \frac14
\overline{g}^{\mu\nu} \overline{g}^{\rho\sigma} \!-\! \frac12
\overline{g}^{\mu (\rho} \overline{g}^{\sigma) \nu} \Bigr] \Biggr]
i \delta^D(x \!-\! x') \; . \qquad
\end{eqnarray}
Tracing as we did in (\ref{righthand}) gives,
\begin{eqnarray}
\lefteqn{ \frac{\overline{g}_{\mu\nu}(x)}{\sqrt{-\overline{g}(x)}} \!\times\!
\frac{\overline{g}_{\rho\sigma}(x')}{\sqrt{-\overline{g}(x')}} \!\times\!
-i \Bigl[\mbox{}^{\mu\nu} \Delta \Sigma^{\rho\sigma}\Bigr](x;x')
= (D \!-\! 1)^2 H^4 \Biggl[ 2 c_1 \kappa^2 \Bigl[ \frac{\square}{H^2}
\!+\! D\Bigr]^2 } \nonumber \\
& & \hspace{1.7cm} + 0 \!-\! \frac12 \Bigl( \frac{D \!-\! 2}{D \!-\! 1} \Bigr)
c_3 \kappa^2 \Bigl[ \frac{\square}{H^2} \!+\! D\Bigr] +
\frac{D (D \!-\! 2)}{4 (D \!-\! 1)^2} c_4 \kappa^2 \Biggr]
\frac{i \delta^D(x \!-\! x')}{\sqrt{-\overline{g}}} \; . \qquad
\end{eqnarray}
We can entirely absorb $\Delta \mathcal{F}_1$ by making the choices,
\begin{eqnarray}
\lefteqn{c_1 = \frac{H^{D-4} (\frac{D}2 \!-\! 1) \Gamma(\frac{D}2)}{(\pi)^{
\frac{D}2}} \times \frac{f_{1a}}{(D \!-\! 4)(D\!-\! 3)} } \nonumber \\
& & \hspace{3.5cm} = \frac{H^{D-4} \Gamma(\frac{D}2)}{16 (4\pi)^{\frac{D}2}}
\times \frac{(D \!-\!2)}{(D \!-\! 4)(D \!-\! 3) (D\!-\! 1)^2} \; , \label{c1}\\
\lefteqn{c_3 = \frac{H^{D-4} (\frac{D}2 \!-\! 1) \Gamma(\frac{D}2)}{(\pi)^{
\frac{D}2}} \times -2 \Bigl(\frac{D \!-\! 1}{D \!-\! 2}\Bigr) \times
\frac{f_{1b}}{D \!-\! 4} } \nonumber \\
& & \hspace{3.5cm} = \frac{H^{D-4} \Gamma(\frac{D}2)}{16 (4\pi)^{\frac{D}2}}
\times -\frac{2 D (D^2 \!-\! 5 D \!+\! 2)}{(D \!-\! 4)(D \!-\! 3) (D\!-\! 1)}
\; , \label{c3} \\
\lefteqn{c_4 = \frac{H^{D-4} (\frac{D}2 \!-\! 1) \Gamma(\frac{D}2)}{(\pi)^{
\frac{D}2}} \times \frac{ 4 (D \!-\! 1)^2}{D (D \!-\! 2)} \times
\Biggl[\frac{D (D \!+\! 2) f_{1b} \!-\! 4 f_{1c}}{4 (D \!-\! 4)} \Biggr] }
\nonumber \\
& & \hspace{3.5cm} = \frac{H^{D-4} \Gamma(\frac{D}2)}{16 (4\pi)^{\frac{D}2}}
\times -\frac{D (D^3 \!-\! 11 D^2 \!+\! 24 D \!+\! 12)}{(D \!-\! 6)
(D \!-\! 3) (D\!-\! 2)} \; . \qquad \label{c4}
\end{eqnarray}
The linear combinations (\ref{cor1}) and (\ref{cor2}) are finite,
\begin{eqnarray}
\lefteqn{-2 (D \!-\! 1) D c_1 + c_3 = \frac{H^{D-4} \Gamma(\frac{D}2)}{16
(4\pi)^{\frac{D}2}} \times \frac{-2D^2}{(D \!-\!3) (D\!-\! 1)} \; , } \\
\lefteqn{(D \!-\! 1)^2 D^2 c_1 - (D \!-\! 2) (D\!-\!1) c_3 + c_4 } \nonumber \\
& & \hspace{4cm} = \frac{H^{D-4} \Gamma(\frac{D}2)}{16 (4\pi)^{\frac{D}2}}
\times \frac{D (D^3 \!-\! 6 D^2 \!+\! 8 D \!-\! 24)}{(D \!-\!6) (D\!-\! 3)}
\; . \qquad
\end{eqnarray}
Therefore neither the Newton constant nor the cosmological constant
requires a divergent renormalization, although we are free to continue
making the finite renormalizations of these constants which are implied
by equations (\ref{c1}-\ref{c4}).

It remains to act the $D=4$ Green's function (\ref{Gfunc}) twice on the
renormalized remainder term in expression (\ref{F1R}). The result is,
\begin{eqnarray}
\lefteqn{ \Biggl[ \frac1{\frac{\square}{H^2} \!+\! 4} \Biggr]^2
\Biggl\{ -\frac43 \!\times\! \frac{4}{y} \ln\Bigl(\frac{y}{4}\Bigr) \!-\!
\frac83 \!\times\! \frac{4}{y} \!-\! \frac{8}3 \ln^2\Bigl(\frac{y}{4} \Bigr)
\!+\! 8 \ln\Bigl(\frac{y}{4}\Bigr) \!-\! \frac13\Biggr\} } 
\nonumber \\
& & \hspace{-0.5cm} = -\frac13 \!\times\! \frac{y}{4} \ln^2\Bigl(\frac{y}{4}\Bigr)
\!+\! \frac13 \!\times\! \frac{y}{4} \ln\Bigl(\frac{y}{4}\Bigr) 
\nonumber \\
& & \hspace{0cm}
-\frac{7}{540}(12 \pi^2 +265) \!\times\! \frac{y}{4}
+\frac{84 \pi^2 - 131}{1080}
\nonumber \\
& & \hspace{0cm}
+ \frac19 \!\times\! \frac{y}{4}\ln(\frac{y}{4})
-\frac{1}{45} \ln(\frac{y}{4})
+\frac{1}{45} \!\times\! \frac4{4-y}\ln(\frac{y}{4})
\nonumber \\
& & \hspace{0cm}
-\frac1{30} (2 - y) \biggl[ 7\mbox{Li}_2(1-\frac{y}{4}) 
- 2 \mbox{Li}_2(\frac{y}{4}) +  5\ln(1 - \frac{y}{4})\ln(\frac{y}{4}) \biggr]
\nonumber \\
& & \hspace{0cm}
+\frac{43}{216} \!\!\times\!\! \frac4{4-y}
-\frac56 \!\times\!\frac{y}{4}\ln(1 - \frac{y}{4})
-\frac1{20}\ln(1 - \frac{y}{4})
+\frac{7}{90} \!\!\times\!\!\frac{4}{y}\ln(1 - \frac{y}{4})\;. \qquad 
\end{eqnarray}
Here ${\rm Li}_2(z)$ is the dilogarithm function,
\begin{equation}
{\rm Li}_2(z) \equiv -\int_0^z \!\! dt \, \frac{\ln(1 \!-\! t)}{t} =
\sum_{k=1}^{\infty} \frac{z^k}{k^2} \; . \label{dilog}
\end{equation}
Hence our final result for the renormalized spin zero structure function is,
\begin{eqnarray}
\lefteqn{\mathcal{F}_{1R} = \frac{\kappa^2 H^4}{(4\pi)^4} \Biggl\{
\frac{\square}{H^2} \Biggl[ \frac1{72} \!\times\! \frac{4}{y} \ln\Bigl(
\frac{y}4\Bigr) \Biggr] \!-\! \frac1{12} \!\times\! \frac{4}{y} \ln\Bigl(
\frac{y}{4}\Bigr) \!+\! \frac1{72} \!\times\! \frac{4}{y} \!+\! \frac16
\ln^2\Bigl(\frac{y}{4}\Bigr) } 
\nonumber \\
& & \hspace{1.5cm} 
+\frac{1}{45} \!\times\! \frac4{4-y}\ln(\frac{y}{4})
-\frac{1}{45} \ln(\frac{y}{4})
+\frac{43}{216} \!\!\times\!\! \frac4{4-y}
-\frac56 \!\times\!\frac{y}{4}\ln(1 - \frac{y}{4})
\nonumber \\
& & \hspace{1.5cm} 
+\frac{7}{90} \!\times\!\frac{4}{y}\ln(1 - \frac{y}{4})
-\frac1{20}\ln(1 - \frac{y}{4})
-\frac{7(12 \pi^2 +265)}{540} \!\times\! \frac{y}{4}
\nonumber \\
& & \hspace{1.5cm}
+\frac{84 \pi^2 - 131}{1080}
-\frac13 \!\times\! \frac{y}{4} \ln^2\Bigl(\frac{y}{4}\Bigr)
+ \frac49 \!\times\! \frac{y}{4} \ln\Bigl(\frac{y}{4}\Bigr) 
\nonumber \\
& & \hspace{1.5cm}
-\frac1{30} (2 - y) \biggl[ 7\mbox{Li}_2(1-\frac{y}{4}) 
- 2 \mbox{Li}_2(\frac{y}{4}) +  5\ln(1 - \frac{y}{4})\ln(\frac{y}{4}) \biggr]
\Biggr\} \;. \qquad 
\label{F1renprm}
\end{eqnarray}

\subsection{Renormalizing the Spin Two Structure Function}

Recall the form (\ref{F2form}) we obtained for the spin two structure
function,
\begin{equation}
\mathcal{F}_2(y) = \mathcal{Q}_2(y) + \frac1{\mathcal{D}} \mathcal{R}_2(y) \; ,
\end{equation}
where the second order differential operator $\mathcal{D}$ was defined in
(\ref{2ndform}). Recall also that the quotient $\mathcal{Q}_2(y)$ and the
remainder $\mathcal{R}_2(y)$ are given in relations (\ref{Q2}-\ref{f2e}).
These expression imply that $\mathcal{F}_2$ harbors the same sort of
ultraviolet divergences as $\mathcal{F}_1$:
\begin{itemize}
\item{The factor of $(\frac{4}{y})^{D-2}$ in $\mathcal{Q}_2$, which has a
finite coefficient but is still not integrable in $D=4$ dimensions;}
\item{The factors of $\frac1{D - 4} (\frac{4}{y})^{D-3}$ in $\mathcal{Q}_2$
and $\mathcal{R}_2$ which are integrable in $D = 4$ dimensions but have
divergent coefficients that preclude taking the unregulated limits; and}
\item{The factors of $(\frac1{D - 4})^2 (\frac{4}{y})^{D-4}$ in $\mathcal{Q}_2$
and $\mathcal{R}_2$ which are integrable in $D = 4$ dimensions but have even
more divergent coefficients.}
\end{itemize}

Only the leading divergence requires a new counterterm. It is handled by
first extracting another derivative and then adding zero in the form
(\ref{ZERO}), just as we did in equations (\ref{finalext}) and (\ref{lead}).
The final result is,
\begin{eqnarray}
\lefteqn{ -K f_{2a} \Bigl(\frac{4}{y}\Bigr)^{D-2} =
\frac{\kappa^2 H^4}{(4\pi)^4} \Biggl\{ \frac{\square}{H^2} \Biggl[
\frac1{240} \!\times\! \frac{4}{y} \ln\Bigl(\frac{y}{4}\Bigr)\Biggr]
\!-\! \frac1{120} \!\times\! \frac{4}{y} \ln\Bigl(\frac{y}{4}\Bigr)
\!+\! \frac1{240} \!\times\! \frac{4}{y} \Biggr\} } \nonumber \\
& & \hspace{0cm} +O(D \!-\! 4) -\frac{\kappa^2 H^{D-4} \Gamma(\frac{D}2)}{
16 (4\pi)^{\frac{D}2}} \times \frac{4 i \delta^D(x \!-\! x')/\!
\sqrt{-\overline{g}}}{(D\!-\!4) (D\!-\!3) (D\!-\!2) (D\!-\!1)(D\!+\!1)} 
\; . \qquad
\end{eqnarray}
Comparing expressions (\ref{C2}) and (\ref{ansatz}) implies that
the divergent part can be entirely absorbed by choosing the coefficient
$c_2$ of the Weyl counterterm (\ref{LC2}) to be,
\begin{equation}
c_2 = \frac{H^{D-4} \Gamma(\frac{D}2)}{16 (4\pi)^{\frac{D}2}} \times
\frac{2}{(D \!-\!4)(D \!-\!3)^2 (D\!-\!1) (D\!+\!1)} \; .
\end{equation}
Of course the divergent part agrees with \cite{HV}.

It turns out that the lower divergences of $\mathcal{F}_2$ are canceled
by the three factors we added to $\mathcal{Q}_1$ to cancel its lower
divergences,
\begin{equation}
\delta \mathcal{Q}_1 = K \Biggl\{ \frac{f_{1b}}{D \!-\! 4}
\Bigl(\frac{4}{y}\Bigr)^{\frac{D}2-1} + \frac{2 f_{1c}}{(D \!-\! 4)^2}
\Bigl(\frac{4}{y}\Bigr)^{\frac{D}2-2} - \frac{f_{1c}}{(D \!-\!4)^2} \Biggr\} .
\end{equation}
These changes in $\mathcal{Q}_1$ induce changes in the source term upon
which we act $\mathcal{D}^{-1}$ to get $\mathcal{F}_2$,
\begin{eqnarray}
\lefteqn{\delta S \equiv \Bigl(\frac{D \!-\! 1}{D \!+\! 1}\Bigr) \Biggl\{
(D \!-\! 1) (2 \!-\! y) (4 y \!-\! y^2) \delta \mathcal{Q}_1'''
- D (D \!-\! 1) (4 y \!-\! y^2) \delta \mathcal{Q}_1''} \nonumber \\
& & \hspace{2.5cm} + 4 (D^2 \!-\! 3) \delta \mathcal{Q}_1''
+ (D \!-\! 1)^2 (2 \!-\! y) \delta \mathcal{Q}_1' + (D \!-\! 1)^2
\delta \mathcal{Q}_1\Biggr\} \; , \qquad \\
& & \hspace{-.5cm}= K \Biggl\{ \frac{\delta s_{b}}{D \!-\!4}
\Bigl(\frac{4}{y}\Bigr)^{\frac{D}2+1} \!\!\!\!\!\!\!\!\!+\!
\frac{\delta s_{c}}{D \!-\!4} \Bigl(\frac{4}{y}\Bigr)^{\frac{D}2} \!\!\!\!+\!
\frac{\delta s_{d}}{D \!-\!4} \Bigl(\frac{4}{y}\Bigr)^{\frac{D}2-1}
\!\!\!\!\!\!\!\!\!+\! \frac{\delta s_{e}}{(D \!-\!4)^2}
\Bigl(\frac{4}{y}\Bigr)^{\frac{D}2-2} \!\!\!\!\!\!\!\!\!+\!
\frac{\delta s_{e'}}{(D \!-\!4)^2} \Biggr\} . \qquad
\end{eqnarray}
Here the coefficients are,
\begin{eqnarray}
\delta s_b & = & -\frac1{16} (D\!-\!2) (D\!-\!1) D f_{1b} \; , \\
\delta s_c & = & \frac{(D \!-\!2)(D \!-\!1)}{16 (D \!+\!1)}
\Bigl[-(D \!-\!1) (D^2 \!-\!2D \!-\! 4) f_{1b} + 2 (D \!-\!3) f_{1c}
\Bigr] \; , \qquad \\
\delta s_d & = & \frac{(D \!-\!1)^2}{8 (D \!+\!1)} \Bigl[D^3 f_{1b} -
(D^2 \!+\! 2 D \!-\! 4) f_{1c} \Bigr] \; , \\
\delta s_{e} & = & \frac{(D-2)^2 (D\!-\!1)^2 (D \!+\! 2)}{4 (D \!+\! 1)} \,
f_{1c} \; , \\
\delta s_{e'} & = & -\frac{(D \!-\! 1)^3}{(D \!+\! 1)} \, f_{1c} \; .
\end{eqnarray}

To infer the corresponding changes in the spin two quotient and remainder
we need to invert $\mathcal{D}$ on $(\frac{4}{y})^{\frac{D}2+1}$,
$(\frac{4}{y})^{\frac{D}2}$ and $1$. The second one was given in (\ref{D/2}).
>From expression (\ref{recurs2}) we find,
\begin{eqnarray}
\lefteqn{\frac1{\mathcal{D}} \Bigl(\frac{4}{y}\Bigr)^{\frac{D}2+1} =
\frac{16}{(D \!-\! 2)^2 D} \Bigl(\frac{4}{y}\Bigr)^{\frac{D}2-1} }
\nonumber \\
& & \hspace{.9cm} - \frac4{\mathcal{D}} \Biggl\{ \frac{(D \!-\! 3)
(D \!+\! 2)}{(D \!-\! 2) D} \Bigl(\frac{4}{y}\Bigr)^{\frac{D}2} \!+
\frac{(D \!-\! 3) (D \!+\! 2) (D \!+\! 4)}{(D \!-\! 2)^2 D}
\Bigl(\frac{4}{y}\Bigr)^{\frac{D}2 - 1} \Biggr\} , \qquad \\
\lefteqn{\frac1{\mathcal{D}} \Bigl(1\Bigr) = \frac1{(D \!-\! 3) D (D \!+\! 1)}
\; . }
\end{eqnarray}
Although we want to move all the $(\frac{4}{y})^{\frac{D}2 +1}$ and 
$(\frac{4}{y})^{\frac{D}2}$ terms from the remainder to the quotient,
we must allow for an arbitrary amount $\delta f_{2c'}$ of the $1$ term. 
Hence the changes in the quotient and the remainder take the form,
\begin{eqnarray}
\delta \mathcal{Q}_2 & = & K \Biggl\{ \frac{\delta f_{2b}}{D \!-\! 4}
\Bigl(\frac{4}{y}\Bigr)^{\frac{D}2-1} + \frac{\delta f_{2c}}{(D \!-\! 4)^2}
\Bigl(\frac{4}{y}\Bigr)^{\frac{D}2-2} + \frac{\delta f_{2c'}}{ (D \!-\! 4)^2}
\Biggr\} , \\
\delta \mathcal{R}_2 & = & K \Biggl\{ \frac{\delta f_{2d}}{D \!-\! 4}
\Bigl(\frac{4}{y}\Bigr)^{\frac{D}2-1} + \frac{\delta f_{2e}}{(D \!-\! 4)^2}
\Bigl(\frac{4}{y}\Bigr)^{\frac{D}2-2} + \frac{\delta f_{e'}}{(D \!-\! 4)^2}
\Biggr\} .
\end{eqnarray}
The various coefficients are,
\begin{eqnarray}
\delta f_{2b} & = & \frac{16}{(D \!-\! 2)^2 D} \times \delta s_{b} \; , \\
\delta f_{2c} & = & -\frac{64 (D \!-\! 3) (D \!+\! 2)}{(D \!-\! 2)^3 D}
\times \delta s_{b} + \frac{16}{(D \!-\! 2)^2} \times \delta s_{c}
\; , \qquad \\
\delta f_{2d} & = &  \frac{4 (D \!-\!3) (D \!+\! 2)(D \!+\! 4)
(3 D \!-\! 10)}{(D \!-\!2)^3 D} \times \delta s_{b} \nonumber \\
& & \hspace{4cm} - \frac{4 (D \!-\! 3) (D \!+\! 4)}{(D \!-\! 2)^2}
\times \delta s_{c} + \delta s_{d} \; , \qquad \\
\delta f_{2e} & = & \frac{16 (D \!-\!3)^2 (D \!+\! 2)(D \!+\! 4)(D \!+\! 6)}{
(D \!-\! 2)^3 D} \times \delta s_{b} \nonumber \\
& & \hspace{4cm} - \frac{4 (D \!-\! 3)(D \!+\! 4) (D \!+\! 6)}{(D \!-\!2)^2}
\times \delta s_{c} + \delta s_{e} \; , \qquad \\
\delta f_{2e'} & = & -(D \!-\!3)D(D \!+\! 1) \delta f_{2c'} +
\delta s_{e'} \; .
\end{eqnarray}

It is possible to make the combination $\mathcal{Q}_2 + \delta 
\mathcal{Q}_2$ possess a finite unregulated limit by choosing,
\begin{equation}
\delta f_{2c'} = 1 -\frac{271}{60} \, (D \!-\!4) + \frac{11057}{3600} \,
(D \!-\!4)^2 \; . \label{choice}
\end{equation}
With this choice the renormalized spin two quotient is,
\begin{eqnarray}
\lefteqn{\mathcal{Q}_{2R} = \frac{\kappa^2 H^4}{(4\pi)^4} \Biggl\{
\frac{\square}{H^2} \Biggl[ \frac1{240} \!\times\! \frac{4}{y}
\ln(\Bigl( \frac{y}4\Bigr) \Biggr] \!-\! \frac1{120} \!\times\!
\frac{4}{y} \ln\Bigl( \frac{y}{4}\Bigr) \!+\! \frac1{240} \!\times\!
\frac{4}{y} } \nonumber \\
& & \hspace{2cm} +\frac1{12}\!\times\!\frac{4}{y} \ln\Bigl( \frac{y}{4}\Bigr)
- \frac7{30}\!\times\!\frac{4}{y}     
+ \frac14 \ln^2\Bigl(\frac{y}{4}\Bigr) 
- \frac{119}{60} \ln\Bigl(\frac{y}{4}\Bigr) \Biggr\} . \qquad
\end{eqnarray}
Choosing (\ref{choice}) also produces a finite result for the spin two
remainder term,
\begin{eqnarray}
\lefteqn{\mathcal{R}_{2R} = \frac{\kappa^2 H^4}{(4\pi)^4} \Biggl\{
\frac{17}{10} \!\times\!\frac{4}{y} \ln\Bigl( \frac{y}{4}\Bigr) 
- \frac{149}{30} \!\times\!\frac{4}{y}
- \frac{41}{10}\ln^2\Bigl(\frac{y}{4}\Bigr) + \frac{193}{6} \ln\Bigl(\frac{y}{4}\Bigr)
+ \frac{359}{20}} 
\nonumber \\
& & \hspace{1.5cm} 
+\frac{32}{15(4-y)^3}\biggl[90 \Bigl(\frac{y}{4}\Bigr)^4 - 291 \Bigl(\frac{y}{4}\Bigr)^3 + 333 \Bigl(\frac{y}{4}\Bigr)^2 - 152 \Bigl(\frac{y}{4}\Bigr)
\nonumber \\
& & \hspace{1.5cm}
+ 21\biggr]  \Bigl(\frac{4}{y}\Bigr)\ln(\frac{y}{4})
+\frac{4}{45(4-y)^3}\biggl[432 \Bigl(\frac{y}{4}\Bigr)^3 - 792 \Bigl(\frac{y}{4}\Bigr)^2 - 288 \Bigl(\frac{y}{4}\Bigr) 
 \nonumber \\
& & \hspace{1.5cm}
+ 991  - 474 \Bigl(\frac{4}{y}\Bigr) 
- 84\Bigl(\frac{4}{y}\Bigr)^2\biggr]
-\frac{7}{60}(\frac{4}{y})^3\ln(1-\frac{y}{4})
-\frac{9}{10} \ln^2(\frac{y}{4})\Biggr\} \;.
\nonumber \\
\end{eqnarray}
Acting the $D=4$ Green's function (\ref{G2}) on the remainder and adding
the result to the quotient gives our final result for the renormalized 
spin two structure function (recall the definition (\ref{dilog}) of the 
dilogarithm function),
\begin{eqnarray}
\lefteqn{\mathcal{F}_{2R} = \frac{\kappa^2 H^4}{(4\pi)^4} \Biggl\{
\frac{\square}{H^2} \Biggl[ \frac1{240} \!\times\! \frac{4}{y}
\ln(\Bigl( \frac{y}4\Bigr) \Biggr] \!+\! \frac3{40} \!\times\!
\frac{4}{y} \ln\Bigl( \frac{y}{4}\Bigr) \!-\! \frac{11}{48} \!\times\!
\frac{4}{y} 
+ \frac14 \ln^2\Bigl(\frac{y}{4}\Bigr) } 
\nonumber \\
& & \hspace{1cm}     
- \frac{119}{60} \ln\Bigl(\frac{y}{4}\Bigr) 
+\frac{4096}{(4y-y^2 -8)^4}\Biggl[
\biggl[-\frac{47}{15}\Bigl(\frac{y}{4}\Bigr)^8 + \frac{141}{10} \Bigl(\frac{y}{4}\Bigr)^7 
\nonumber \\
& & \hspace{1cm} 
- \frac{2471}{90}\Bigl(\frac{y}{4}\Bigr)^6 
+ \frac{34523}{720}\Bigl(\frac{y}{4}\Bigr)^5 
-\frac{132749}{1440}\Bigl(\frac{y}{4}\Bigr)^4 + \frac{38927}{320}\Bigl(\frac{y}{4}\Bigr)^3 
\nonumber \\
& & \hspace{1cm} 
- \frac{10607}{120} \Bigl(\frac{y}{4}\Bigr)^2 
+ \frac{22399}{720}\Bigl(\frac{y}{4}\Bigr) 
- \frac{3779}{960} \biggr]\frac{4}{4-y}
+ \biggl[\frac{193}{30}\Bigl(\frac{y}{4}\Bigr)^4
-\frac{131}{10}\Bigl(\frac{y}{4}\Bigr)^3 
\nonumber \\
& & \hspace{1cm} 
+ \frac{7}{20} \Bigl(\frac{y}{4}\Bigr)^2
+\frac{379}{60} \Bigl(\frac{y}{4}\Bigr)- \frac{193}{120}\biggr]\ln(2-\frac{y}{2})
+ \biggl[-\frac{14}{15} \Bigl(\frac{y}{4}\Bigr)^5 -\frac{1}{5}\Bigl(\frac{y}{4}\Bigr)^4 
\nonumber \\
& & \hspace{1cm} 
+\frac{19}{2} \Bigl(\frac{y}{4}\Bigr)^3-\frac{889}{60}\Bigl(\frac{y}{4}\Bigr)^2 + \frac{143}{20}\Bigl(\frac{y}{4}\Bigr) -\frac{13}{20} 
-\frac{7}{60}\Bigl(\frac{4}{y}\Bigr)\biggr]\ln(1-\frac{y}{4})
\nonumber \\
& &  \hspace{1cm} 
+\biggl[-\frac{476}{15}\Bigl(\frac{y}{4}\Bigr)^9 + 160 \Bigl(\frac{y}{4}\Bigr)^8 -\frac{5812}{15} \Bigl(\frac{y}{4}\Bigr)^7+\frac{8794}{15} \Bigl(\frac{y}{4}\Bigr)^6 
\nonumber \\
& & \hspace{1.5cm}
-\frac{18271 }{30} \Bigl(\frac{y}{4}\Bigr)^5
+\frac{54499}{120} \Bigl(\frac{y}{4}\Bigr)^4 -\frac{59219}{240} \Bigl(\frac{y}{4}\Bigr)^3 +\frac{1917}{20}\Bigl(\frac{y}{4}\Bigr)^2 
\nonumber \\
& & \hspace{1.5cm}
-\frac{1951}{80} \Bigl(\frac{y}{4}\Bigr)+\frac{367}{120}\biggr]\frac{4}{4-y}\ln(\frac{y}{4})
+\biggl[4 \Bigl(\frac{y}{4}\Bigr)^7-12 \Bigl(\frac{y}{4}\Bigr)^6
+20 \Bigl(\frac{y}{4}\Bigr)^5
\nonumber \\
& & \hspace{1.5cm}
-20 \Bigl(\frac{y}{4}\Bigr)^4+15 \Bigl(\frac{y}{4}\Bigr)^3-7 \Bigl(\frac{y}{4}\Bigr)^2+\Bigl(\frac{y}{4}\Bigr)\biggr]\frac{4-y}{4}\ln^2(\frac{y}{4})
\nonumber \\
& & \hspace{1cm}
+\biggl[\frac{367}{30} \Bigl(\frac{y}{4}\Bigr)^4 -\frac{4121}{120}\Bigl(\frac{y}{4}\Bigr)^3 +\frac{237}{16}\Bigl(\frac{y}{4}\Bigr)^2 +\frac{1751}{240}\Bigl(\frac{y}{4}\Bigr)-\frac{367}{120} \biggr] \ln(\frac{y}{2})
\nonumber \\
& & \hspace{1cm}
+\frac{1}{64}(y^2 -8) \Bigl[4(2 - y) -(4y - y^2)\Bigr]\biggl[\frac{1}{5} \mbox{Li}_2(1-\frac{y}{4})
+\frac{7}{10}\mbox{Li}_2(\frac{y}{4})\biggr]
\Biggr]\Biggr\}\;.
\nonumber \\ 
\label{F2renprm}
\end{eqnarray}

\section{Discussion}

We have derived two forms for the one loop contribution to the
graviton self-energy from a massless, minimally coupled scalar on de
Sitter background. The first form (\ref{3-point-A}) is fully
dimensionally regulated, with the ultraviolate divergences neither
localized nor subtracted off with counterterms. This version of the
result agrees with the stress tensor correlator recently computed by
Perez-Nadal, Roura and Verdaguer \cite{PNRV}. Our second form is
fully renormalized, with the unregulated limit taken,
\begin{eqnarray}
\lefteqn{-i \Bigl[\mbox{}^{\mu\nu} \Sigma_{\rm
ren}^{\rho\sigma}\Bigr](x;x') = \sqrt{-\overline{g}(x)} \,
\mathcal{P}^{\mu\nu}(x) \sqrt{-\overline{g}(x')} \,
\mathcal{P}^{\rho\sigma}(x') \Bigl[
\mathcal{F}_{1R}(y) \Bigr] } \nonumber \\
& & \hspace{.5cm} + \sqrt{-\overline{g}(x)} \,
\mathcal{P}^{\mu\nu}_{\alpha\beta\gamma\delta}(x)
\sqrt{-\overline{g}(x')}
\mathcal{P}^{\rho\sigma}_{\alpha\beta\gamma\delta}(x') \Bigl[
\mathcal{T}^{\alpha\kappa} \mathcal{T}^{\beta\lambda}
\mathcal{T}^{\gamma\theta} \mathcal{T}^{\delta\phi}
\mathcal{F}_{2R}(y) \Bigr] \; . \qquad \label{finalform}
\end{eqnarray}
In this expression the spin zero operator $\mathcal{P}^{\mu\nu}$ was
defined in (\ref{P0}), the spin two operator
$\mathcal{P}^{\mu\nu}_{\alpha\beta\gamma\delta}$ was defined in
(\ref{P2}), and the bitensor $\mathcal{T}^{\alpha\kappa}$ was given
in (\ref{Tak}). Our results for the renormalized spin zero and spin
two structure functions are expressions (\ref{F1renprm}) and
(\ref{F2renprm}), respectively.

Our final form (\ref{finalform}) is manifestly transverse, as
required by gauge invariance. It is also de Sitter invariant,
despite the fact that the massless, minimally coupled propagator
breaks de Sitter invariance \cite{AF}, because the de Sitter
breaking term drops out of mixed second derivatives (\ref{double}).
Our result agrees with the flat space limit \cite{FW}. And the
divergent parts of the counterterms we used to subtract off the
divergences agree with those found long ago by `t Hooft and Veltman
\cite{HV}. We actually included finite renormalizations of Newton's
constant and of the cosmological constant. Such renormalizations are
presumably necessary when considering the effective field equations
of quantum gravity if the parameters $\Lambda$ and $G$ are to have
their correct physical meanings.

The point of this exercise has been to quantum correct the
linearized Einstein equation,
\begin{equation}
\sqrt{-\overline{g}} \, \mathcal{D}^{\mu\nu\rho\sigma}
h_{\rho\sigma}(x) - \int \!\! d^4x' \Bigl[ \mbox{}^{\mu\nu}
\Sigma_{\rm ren}^{\rho\sigma}\Bigr](x;x') h_{\rho\sigma}(x') =
\frac12 \kappa \sqrt{-\overline{g}} \, T^{\mu\nu}_{\mbox{\tiny
lin}}(x) \; , \label{lineqn}
\end{equation}
where $\mathcal{D}^{\mu\nu\rho\sigma}$ is the Lichnerowicz operator
(\ref{Lich}) specialized to de Sitter background. In a future paper
we will employ this effective field equation to work out the one
loop quantum corrections to mode functions for dynamical gravitons
and to the gravitational response to a stationary point mass. It is
worthwhile closing this paper with an adumbration of the procedure
and some of the physical considerations.

Our first comment is that one must use the Schwinger-Keldysh
formalism \cite{Schwinger,new} to correctly describe the quantum response 
from a prepared initial state. This amounts to replacing the in-out 
self-energy in (\ref{lineqn}) by the sum of two of the four Schwinger-Keldysh
self-energies,
\begin{equation}
\Bigl[\mbox{}^{\mu\nu} \Sigma_{\rm ren}^{\rho\sigma}\Bigr](x;x')
\longrightarrow \Bigl[\mbox{}^{\mu\nu} \Sigma_{\rm
ren}^{\rho\sigma}\Bigr]_{\scriptscriptstyle ++}\!\!(x;x') +
\Bigl[\mbox{}^{\mu\nu} \Sigma_{\rm
ren}^{\rho\sigma}\Bigr]_{\scriptscriptstyle +-}\!\!(x;x') \; .
\end{equation}
At the one loop order we are working $[\mbox{}^{\mu\nu} \Sigma_{\rm
ren}^{\rho\sigma}]_{\scriptscriptstyle ++}\!(x;x')$ agrees exactly
with the in-out result (\ref{finalform}) we have derived. To get
$[\mbox{}^{\mu\nu} \Sigma_{\rm
ren}^{\rho\sigma}]_{\scriptscriptstyle +-}(x;x')$, at this order,
one simply adds a minus sign and replaces the de Sitter length
function $y(x;x')$ everywhere with,
\begin{equation}
y(x;x') \longrightarrow y_{\scriptscriptstyle +-}\!(x;x') \equiv H^2
a(\eta) a(\eta') \Bigl[\Vert \vec{x} \!-\! \vec{x}'\Vert^2 - (\eta
\!-\! \eta' \!+\! i \epsilon)^2 \Bigr] \; .
\end{equation}
It will be seen that the $++$ and $+-$ self-energies cancel unless
the point $x^{\prime \mu}$ is on or inside the past light-cone of
$x^{\mu}$. That makes the effective field equation (\ref{lineqn})
causal. When $x^{\prime \mu}$ is on or inside the past light-cone of
$x^{\mu}$ the $+-$ self-energy is the complex conjugate of the $++$
one, which makes the effective field equation (\ref{lineqn}) real.

Our second comment concerns the various derivative operators in
expression (\ref{finalform}). Because the second order operators
$\mathcal{P}^{\mu\nu}(x)$ and
$\mathcal{P}^{\mu\nu}_{\alpha\beta\gamma\delta}(x)$ act on
$x^{\mu}$, they can be pulled outside of the integration over
$x^{\prime \mu}$. The same is true for the covariant scalar
d'Alembertian acting on the most singular terms of the two structure
functions. The second order operators $\mathcal{P}^{\rho\sigma}(x')$
and $\mathcal{P}^{\rho\sigma}_{\kappa\lambda\theta\phi}(x')$ can be
partially integrated to act on the graviton field
$h_{\rho\sigma}(x')$. This will give no spatial surface terms because
the two self-energies cancel for $x^{\prime \mu}$ outside the past
light-cone of $x^{\mu}$. Nor will there be any temporal surface terms 
at the upper limit, because the integrand vanishes like $(\eta - \eta')^3$. 
There {\it will} be temporal surface terms at the lower limit. We 
conjecture that these are all absorbed by perturbative corrections to 
the initial state \cite{KOW}.

Our third comment is that, because we only know the self-energy at
order $\kappa^2$, all we can do is to solve (\ref{lineqn})
perturbatively by expanding the graviton field and the self-energy
in powers of $\kappa^2$,
\begin{eqnarray}
h_{\mu\nu}(x) &=& h^{(0)}_{\mu\nu}(x) + \kappa^2 h^{(1)}_{\mu\nu}(x)
+ O(\kappa^4) \; , \\
\Bigl[\mbox{}^{\mu\nu} \Sigma_{\rm ren}^{\rho\sigma}\Bigr](x;x') & =
& \kappa^2 \Bigl[\mbox{}^{\mu\nu} \Sigma_1^{\rho\sigma}\Bigr](x;x')
+ O(\kappa^4) \; .
\end{eqnarray}
Of course $h^{(0)}_{\mu\nu}(x)$ obeys the classical, linearized
Einstein equation. Given this solution, the corresponding one loop
correction is defined by the equation,
\begin{equation}
\sqrt{-\overline{g}(x)} \, \mathcal{D}^{\mu\nu\rho\sigma}
h^{(1)}_{\rho\sigma}(x) = \int \!\! d^4x' \, \Bigl[\mbox{}^{\mu\nu}
\Sigma_1^{\rho\sigma}\Bigr](x;x') h^{(0)}_{\rho\sigma}(x') \; .
\label{oneloop}
\end{equation}

We are interested in the one loop corrections to two sorts of
classical solutions. The first is a dynamical graviton of wave
vector $\vec{k}$. The classical solution for this takes the form
\cite{TW3},
\begin{equation}
h^{(0)}_{\rho\sigma}(x) = \epsilon_{\rho\sigma}(\vec{k}) u(\eta,k)
e^{i \vec{k} \cdot \vec{x}} \; ,
\end{equation}
where the tree order mode function is,
\begin{equation}
u(\eta,k) = \frac{H}{\sqrt{2 k^3}} \Bigl[1 - \frac{i k}{H a}\Bigr]
\exp\Bigl[ \frac{ik}{H a}\Bigr] \; ,
\end{equation}
and the polarization tensor obeys all the same relations as in flat
space,
\begin{equation}
0 = \epsilon_{0\mu} = k_i \epsilon_{ij} = \epsilon_{jj} \quad {\rm
and} \quad \epsilon_{ij} \epsilon_{ij}^* = 1 \; .
\end{equation}
The second classical solution we wish to correct is the linearized
response to a stationary point mass $M$ \cite{TW3},
\begin{equation}
h^{(0)}_{00}(x) = a^2 \times \frac{2 G M}{a \Vert \vec{x}\Vert} \; ,
\; h^{(0)}_{0i}(x) = 0 \; , \; h^{(0)}_{ij}(x) = a^2 \times \frac{2 G
M}{a \Vert \vec{x} \Vert} \times \delta_{ij} \; .
\end{equation}

The one loop corrections we seek to compute represent the response
(of either dynamical gravitons or the force of gravity) to the vast
ensemble of infrared scalars which are produced by inflation. It is
simple to show that the occupation number for {\it each mode} with
wave number $\vec{k}$ grows like \cite{PW1},
\begin{equation}
N(k,\eta) = \bigg(\frac{H a(\eta)}{2k} \bigg)^2
\end{equation}
This growth is balanced by expansion of the 3-volume so that the
number density of infrared particles with $0 < k < H a$ remains
fixed,
\begin{equation}
n(\eta) = \int \!\! \frac{d^3 k}{(2 \pi a)^3} \, \theta(Ha \!-\! k)
N(k,\eta) = \frac{H^3}{8 \pi^2} \; .
\end{equation}
The constant density of virtual scalars in flat space background has 
no effect at all on dynamical gravitons (after field strength 
renormalization) \cite{HV}, so we expect that dynamical gravitons on
de Sitter will likewise suffer no important quantum corrections.
The virtual scalars of flat space do induce a correction to the 
classical potential \cite{EMGR,us} and we expect one as well on de 
Sitter background. On dimensional grounds the flat space result must 
(and does) take the form,
\begin{equation}
\Phi_{\rm flat} = -\frac{GM}{r} \Biggl\{1 + {\rm constant} \times  
\frac{G}{r^2} + O(G^2) \Biggr\} \; .
\end{equation}
On de Sitter background there is a dimensionally consistent
alternative provided by the Hubble parameter $H$ and the secular 
growth driven by continuous particle production,
\begin{equation}
\Phi_{\rm dS} = -\frac{GM}{r} \Biggl\{1 + {\rm constant} \times  G H^2 
\ln(a) + O(G^2) \Biggr\} \; .
\end{equation}
If such a correction were to occur its natural interpretation 
would be as a time dependent renormalization of the Newton constant.
The physical origin of the effect (if it is present) would be that virtual
infrared quanta which emerge near the source tend to collapse to it,
leading to a progressive increase in the source.

\vskip .5cm

\centerline{\bf Acknowledgements}

We are grateful to S. Deser and A. Roura for discussions on this
topic. This work was partially supported by NSF grant PHY-0855021 and 
by the Institute for Fundamental Theory at the University of Florida.

\end{document}